\documentclass[12pt,a4paper]{article}
\pdfoutput=1
\usepackage{jheppub, amsthm,amsbsy,amsfonts,mathrsfs,enumerate,float,wrapfig,amsmath,amssymb,mathtools}
\usepackage{tikz}
\usepackage{tkz-euclide}
\usetikzlibrary{patterns}
\tikzset{every picture/.style={line width=0.75pt}} 
\usetikzlibrary{positioning}
\usetikzlibrary{arrows, arrows.meta}
\usetikzlibrary{decorations, decorations.markings, decorations.pathmorphing}

\allowdisplaybreaks

\makeatletter
\gdef\@fpheader{}
\makeatother

\title{Thermodynamic limit for SO(2N) gauge theories with spinors/conjugate spinors}
\author[a]{Xiaobin Li,}
\author[b]{Futoshi Yagi}
\affiliation[a]{School of Mathematics, Southwest Jiaotong University,
West zone, High-tech district, Chengdu, Sichuan 611756, China}
\affiliation[b]{School of Science, Huzhou Normal University, No.~759 East 2nd Road, Huzhou, Zhejiang, 313000, China}
\emailAdd{lixiaobin@home.swjtu.edu.cn}
\emailAdd{60714@huznu.edu.cn}

\abstract{In this paper, we investigate five-dimensional $\mathcal{N} = 1$  supersymmetric $SO(2N)$ gauge theories coupled to hypermultiplets in spinor/conjugate spinor representation based on 5-brane web constructions with O5-planes. Based on the topological vertex formalism with O5-plane, we find new expressions for the partition functions for these theories, emphasizing the difference between the case with two spinors and that with one spinor and one conjugate spinor. Through the thermodynamic limit of these partition functions, we derive the Cameral and Seiberg-Witten curves. We show that the difference between the spinor and the conjugate spinor appears as a difference in the boundary conditions of the Seiberg-Witten curves at the orientifold planes.}

\begin{document}
\maketitle
\allowdisplaybreaks
\section{Introduction}\label{sec:intro}

Supersymmetric gauge theories with eight supercharges have been extensively studied using the brane constructions initiated by Hanany and Witten \cite{Hanany:1996ie}. In particular, the type IIB 5-brane web construction developed in \cite{Aharony:1997ju, Aharony:1997bh} is used to engineer 5d $\mathcal{N} = 1$ gauge theories, offering a geometric perspective on non-perturbative aspects of these theories, such as UV fixed points, duality, and symmetry enhancement. 

\begin{figure}
\centering
\begin{tikzpicture}
\begin{scope}[shift={(-5,0)}]
\draw[dashed, thick](-1.8,0)--(1.8,0);
\draw (-1.5,0.1) node[below] {\scriptsize O5$^+$};
\draw (0.1,0.1) node[below] {\scriptsize O5$^-$};
\draw (1.5,0.1) node[below] {\scriptsize O5$^+$};
\draw [thick](-1.3, 1.4) -- (-0.7,1.1) -- (-0.4,0.8) -- (-0.4,0.5) -- (-0.7,0.2) -- (-1.1,0); 
\draw [thick](1.3, 1.4) -- (0.7,1.1) -- (0.4,0.8) -- (0.4,0.5) -- (0.7,0.2) -- (1.1,0); 
\draw [thick](-0.7,1.1) -- (0.7,1.1);
\draw [thick](-0.4,0.8) -- (0.4,0.8);
\draw [thick](-0.4,0.5) -- (0.4,0.5);
\draw [thick](-0.7,0.2) -- (0.7,0.2);
\end{scope}
\begin{scope}[shift={(-0.7,0)}]
\draw[dashed, thick](-1.8,0)--(2.5,0);
\draw (-1.4,0.1) node[below] {\scriptsize O5$^+$};
\draw (0.1,0.1) node[below] {\scriptsize O5$^-$};
\draw (1.5,0.1) node[below] {\scriptsize O5$^+$};
\draw (2.2,0.1) node[below] {\scriptsize O5$^-$};
\draw [thick](-1.3, 1.4) -- (-0.7,1.1) -- (-0.4,0.8) -- (-0.4,0.5) -- (-0.7,0.2) -- (-1.1,0); 
\draw [thick](1.3, 1.4) -- (0.7,1.1) -- (0.4,0.8) -- (0.4,0.5) -- (0.7,0.2) -- (1.1,0); 
\draw [thick](-0.7,1.1) -- (0.7,1.1);
\draw [thick](-0.4,0.8) -- (0.4,0.8);
\draw [thick](-0.4,0.5) -- (0.4,0.5);
\draw [thick](-0.7,0.2) -- (0.7,0.2);
\draw [thick](1.8,0) -- (2.4,0.3);
\end{scope}
\begin{scope}[shift={(5,0)}]
\draw[dashed, thick](-2.5,0)--(2.5,0);
\draw (-2.2,0.1) node[below] {\scriptsize O5$^-$};
\draw (-1.4,0.1) node[below] {\scriptsize O5$^+$};
\draw (0.1,0.1) node[below] {\scriptsize O5$^-$};
\draw (1.5,0.1) node[below] {\scriptsize O5$^+$};
\draw (2.2,0.1) node[below] {\scriptsize O5$^-$};
\draw [thick](-1.3, 1.4) -- (-0.7,1.1) -- (-0.4,0.8) -- (-0.4,0.5) -- (-0.7,0.2) -- (-1.1,0); 
\draw [thick](1.3, 1.4) -- (0.7,1.1) -- (0.4,0.8) -- (0.4,0.5) -- (0.7,0.2) -- (1.1,0); 
\draw [thick](-0.7,1.1) -- (0.7,1.1);
\draw [thick](-0.4,0.8) -- (0.4,0.8);
\draw [thick](-0.4,0.5) -- (0.4,0.5);
\draw [thick](-0.7,0.2) -- (0.7,0.2);
\draw [thick](-1.8,0) -- (-2.4,0.3);
\draw [thick](1.8,0) -- (2.4,0.3);
\end{scope}
\end{tikzpicture}
\caption{Left: 5-brane web diagram for pure SO($2N$) gauge theory. Middle: 5-brane web diagram for SO($2N$) gauge theory with spinor or conjugate spinor representation. Right: 5-brane web diagram for SO($2N$) gauge theory with either two spinors, two conjugate spinors or one spinor plus one conjugate spinor. All of the 5-brane web diagrams are drawn for $N=4$ for concreteness.}
\label{fig:5-brane-SO}
\end{figure}
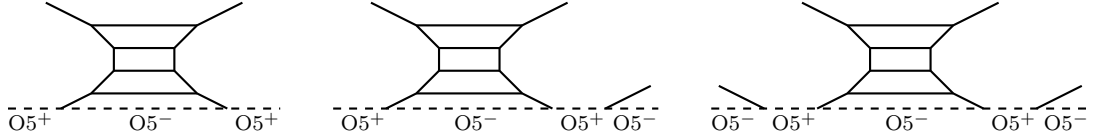

The incorporation of orientifold planes, such as O5 and O7 planes, enables us to realize the orthogonal (SO) and the symplectic (Sp) gauge groups \cite{Giveon:1998sr}. As an example, the 5-brane web diagram for 5d $\mathcal{N}=1$ SO($2N$) gauge theory is depicted on the left of Figure \ref{fig:5-brane-SO}. 

It is remarkable that the orientifold planes also enable us to realize hypermultiplets in various representations. One significant example is the realization of spinor or conjugate spinor representation in SO gauge theories suggested in \cite{Zafrir:2015ftn}. As depicted in the middle of Figure \ref{fig:5-brane-SO}, it is realized by adding a $(2,1)$ 5-brane attached to the O5-plane. Whether the representation of the hypermultiplet is either the spinor representation or the conjugate spinor representation is just a matter of convention, and there is no essential difference between them. Interestingly, in this brane setup, the spinor matter has a non-perturbative origin: the instanton particles in the Sp(1) gauge group reduce to the spinor matter in the Higgsing process. Thus, the large distance between the two external 5-branes on the right indicates a large mass of the spinor.

One of the ways to add one more hypermultiplet in spinor or conjugate spinor representation is to add a $(2,-1)$ 5-brane attached to the O5-plane as depicted on the right of Figure \ref{fig:5-brane-SO}. Strangely enough, if the masses of the spinor and/or conjugate spinor are large enough, we cannot distinguish whether this 5-brane web diagram corresponds to two spinors, two conjugate spinors, or one spinor plus one conjugate spinor only from this figure. The difference between the theory with two spinors and the one with two conjugate spinors is again a matter of convention. However, the theory with one spinor plus one conjugate spinor is essentially different from the other two theories.

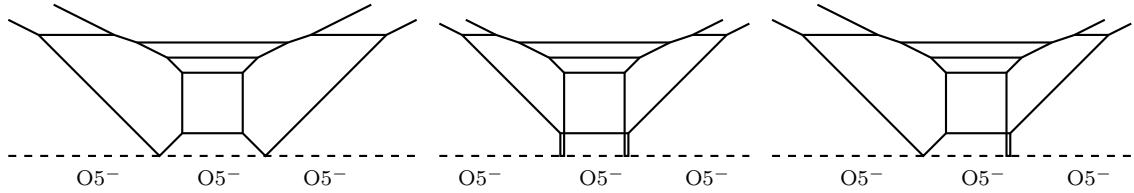
\begin{figure}
\centering
\begin{tikzpicture}
\begin{scope}[shift={(-5.05,0)}]
\draw[dashed, thick](-2.7,0)--(2.7,0);
\draw (-1.5,0) node[below] {\scriptsize O5$^-$};
\draw (0.1,0) node[below] {\scriptsize O5$^-$};
\draw (1.5,0) node[below] {\scriptsize O5$^-$};
\draw [thick] (-2.1,2) -- (-1.3, 1.6) -- (-1, 1.5) -- (-0.6,1.3) -- (-0.4,1.1) -- (-0.4,0.3) -- (-0.7,0) -- (-2.3,1.6) -- (-2.7,1.8); 
\draw [thick] (2.1,2) -- (1.3, 1.6) -- (1, 1.5) -- (0.6,1.3) -- (0.4,1.1) -- (0.4,0.3) -- (0.7,0) -- (2.3,1.6) -- (2.7,1.8); 
\draw [thick] (-1, 1.5) -- (1, 1.5);
\draw [thick] (-0.6,1.3) -- (0.6,1.3);
\draw [thick] (-0.4,1.1) -- (0.4,1.1);
\draw [thick] (-0.4,0.3) -- (0.4,0.3);
\draw [thick] (-2.3,1.6) -- (-1.3, 1.6);
\draw [thick] (2.3,1.6) -- (1.3, 1.6);
\end{scope}
\begin{scope}[shift={(0,0)}]
\draw[dashed, thick](-2.05,0)--(2.05,0);
\draw (-1.5,0) node[below] {\scriptsize O5$^-$};
\draw (0.1,0) node[below] {\scriptsize O5$^-$};
\draw (1.5,0) node[below] {\scriptsize O5$^-$};
\draw [thick]  (-1.7,1.8) -- (-1.3, 1.6) -- (-1, 1.5) -- (-0.6,1.3) -- (-0.4,1.1) -- (-0.4,0) -- (-0.45,0)-- (-0.45,0.3) -- (-1.75,1.6) -- (-2.05,1.75); 
\draw [thick]  (1.7,1.8) -- (1.3, 1.6) -- (1, 1.5) -- (0.6,1.3) -- (0.4,1.1) -- (0.4,0) -- (0.45,0)-- (0.45,0.3) -- (1.75,1.6) -- (2.05,1.75); 
\draw [thick] (-1, 1.5) -- (1, 1.5);
\draw [thick] (-0.6,1.3) -- (0.6,1.3);
\draw [thick] (-0.4,1.1) -- (0.4,1.1);
\draw [thick] (-0.45,0.3) -- (0.45,0.3);
\draw [thick] (-1.75,1.6) -- (-1.3, 1.6);
\draw [thick] (1.75,1.6) -- (1.3, 1.6);
\end{scope}
\begin{scope}[shift={(5.05,0)}]
\draw[dashed, thick](-2.7,0)--(2.05,0);
\draw (-1.5,0) node[below] {\scriptsize O5$^-$};
\draw (0.1,0) node[below] {\scriptsize O5$^-$};
\draw (1.5,0) node[below] {\scriptsize O5$^-$};
\draw [thick] (-2.1,2) -- (-1.3, 1.6) -- (-1, 1.5) -- (-0.6,1.3) -- (-0.4,1.1) -- (-0.4,0.3) -- (-0.7,0) -- (-2.3,1.6) -- (-2.7,1.8); 
\draw [thick]  (1.7,1.8) -- (1.3, 1.6) -- (1, 1.5) -- (0.6,1.3) -- (0.4,1.1) -- (0.4,0) -- (0.45,0)-- (0.45,0.3) -- (1.75,1.6) -- (2.05,1.75); 
\draw [thick] (-1, 1.5) -- (1, 1.5);
\draw [thick] (-0.6,1.3) -- (0.6,1.3);
\draw [thick] (-0.4,1.1) -- (0.4,1.1);
\draw [thick] (-0.4,0.3) -- (0.45,0.3);
\draw [thick] (1.75,1.6) -- (1.3, 1.6);
\draw [thick] (-2.3,1.6) -- (-1.3, 1.6);
\end{scope}
\end{tikzpicture}
\caption{Left: 5-brane web diagram for SO($2N$) gauge theory with two spinors. Middle: 5-brane web diagram for SO($2N$) gauge theory with two conjugate spinors. Right: 5-brane web diagram for SO($2N$) gauge theory with one spinor plus one conjugate spinor. All of the 5-brane web diagrams are drawn for $N=4$ for concreteness. The differences among these three diagrams lie in the structure of the 5-brane web around the O5-plane.}
\label{fig:2S-2C-1S+1C}
\end{figure}

It is discussed in \cite{Hayashi:2018bkd, Hayashi:2017btw}, that 5-brane web diagrams distinguish the difference between these theories in the phase where the masses of the spinor and/or conjugate spinor are small enough. Figure \ref{fig:2S-2C-1S+1C} shows 5-brane web diagrams for SO($2N$) gauge theories with two spinors, with two conjugate spinors, and with one spinor and one conjugate spinor, respectively. 
As expected, we find that the 5-brane web diagrams on the left and in the middle of Figure \ref{fig:2S-2C-1S+1C} are transformed into each other by assuming the ``generalized flop transition'' discussed in \cite{Hayashi:2017btw}. On the contrary, the 5-brane web diagram on the right of Figure \ref{fig:2S-2C-1S+1C} is not related to the other two by the generalized flop transition. 
In this paper, we clarify how the differences between these 5-brane web diagrams appear at the level of unrefined partition functions and Seiberg-Witten curves \cite{Seiberg:1994rs, Seiberg:1994aj} by explicitly computing them. 

The Nekrasov's instanton partition function \cite{Nekrasov:2002qd, Nekrasov:2004vw} for 5d $\mathcal{N}=1$ SO($2N$) gauge theory with a spinor or a conjugate spinor compactified on $S^1$ is computed \cite{Chen:2023smd} using the ADHM approach. In this approach, the spinor and the conjugate spinor are distinguished whether we sum or subtract the contribution from the plus sector and the minus sector of the dual $O(j)$ group realized on the world volume of $j$ D-strings. This is analogous to the partition function for the 5d $\mathcal{N}=1$ Sp($N$) gauge theories with different discrete theta angles, $\theta=0, \pi$ \cite{Kim:2012gu, Bergman:2013ala}. 

In this paper, however, we take an alternative approach. When a 5-brane web diagram is given, the topological vertex formalism \cite{Aganagic:2003db} provides a systematic method to compute the unrefined topological string partition function, which is known to reproduce Nekrasov's instanton partition function. Originally, this method was introduced to apply only to the toric Calabi-Yau 3-folds, whose toric diagram has the same shape as the corresponding 5-brane web diagram \cite{Leung:1997tw}. Later, it turns out that this method can be generalized to more general 5-brane web diagrams, such as those with an O5-plane \cite{Kim:2017jqn}. 

By applying the topological vertex formalism to the different 5-brane web diagrams in Figure \ref{fig:2S-2C-1S+1C}, we can compute the corresponding partition functions directly. As a consequence, we obtain new expressions for the partition functions of 5d $\mathcal{N}=1$ SO($2N$) gauge theoreis with spinors and/or conjugate spinors, which have different expressions from the ones known in \cite{Chen:2023smd} 
but are expected to be equivalent.

This new expression allows us to express the partition function in terms of profile functions corresponding to Young diagrams, facilitating the thermodynamic limit using the technique developed in \cite{Nekrasov:2003rj, Nekrasov:2012xe}. By applying saddle-point methods, we derive the equations governing the amplitude functions and cycle integrals, culminating in the Seiberg-Witten curve. Through this process, we find that two distinct boundary conditions for the Seiberg-Witten curves arise at the positions of the O5-planes.
This difference in boundary conditions characterizes the relative difference between the spinors and the conjugate spinors at the level of the Seiberg-Witten curves.

The organization of this paper is as follows:
In section \ref{sec:par}, we compute the partition function for 5d $\mathcal{N}=1$ SO($2N$) gauge theory with two spinors and for the one with one spinor plus one conjugate spinor, respectively, finding new expressions for them. We perform consistency checks on these new expressions by reproducing the expected perturbative contributions. In section \ref{sec:thermo}, we take the thermodynamic limit of the partition functions obtained in the previous section and derive the corresponding Seiberg-Witten curves. We clarify the difference in the boundary conditions satisfied by the two different curves. Section \ref{sec:concl} is devoted to the conclusion and discussion.

\section{Partition function via topological vertex}\label{sec:par}

In this section, we compute the unrefined topological string partition function for 5d $\mathcal{N}=1$ SO($2N$) gauge theories with either two spinors, two conjugate spinors, or one spinor and one conjugate spinor for $N=2,3,4$. We use the topological vertex formalism \cite{Aganagic:2003db}, which is generalized in a way to be applicable to 5-brane web diagrams with O5-planes \cite{Kim:2017jqn}. 

\subsection{Spinor/conjugate spinor strips}\label{subsec:par-strip}

The idea of topological vertex formalism is to construct the partition function by reading off contributions from each edge and vertex in the 5-brane web diagram. Because of this construction, it is often convenient to decompose into components that are easier to compute. Then, the total partition function can be obtained by gluing them. Since the strip diagram is computed in \cite{Iqbal:2004ne}, the computation would simplify if we could decompose the considered diagram into several strip diagrams. In this paper, we also take this strategy.

When a 5-brane web diagram contains an orientifold plane, we must consider the contribution only from the fundamental domain of the diagram so that we avoid double-counting the edges or vertices that are identified by the orientifold action. One natural approach would be to simply select the upper half as the fundamental domain of the 5-brane web diagram, as is already shown in Figure \ref{fig:2S-2C-1S+1C}. Or, more generally, we can first choose the fundamental domain of the spacetime, and then consider only the part of the diagram that is included in the chosen fundamental domain of the spacetime. However, for actual computation, it is more convenient to start with the whole diagram, including the mirror image, and to choose half of them in such a way that the strip diagram appears as a sub-diagram. 
In other words, we can obtain the fundamental domain of the 5-brane web diagram by reflecting some part of the upper half diagram in terms of the O5-plane. The rule for topological vertex formalism for the O5-plane is constructed in a way that the resulting partition function does not depend on such a reflection \cite{Kim:2017jqn, Li:2021rqr}.

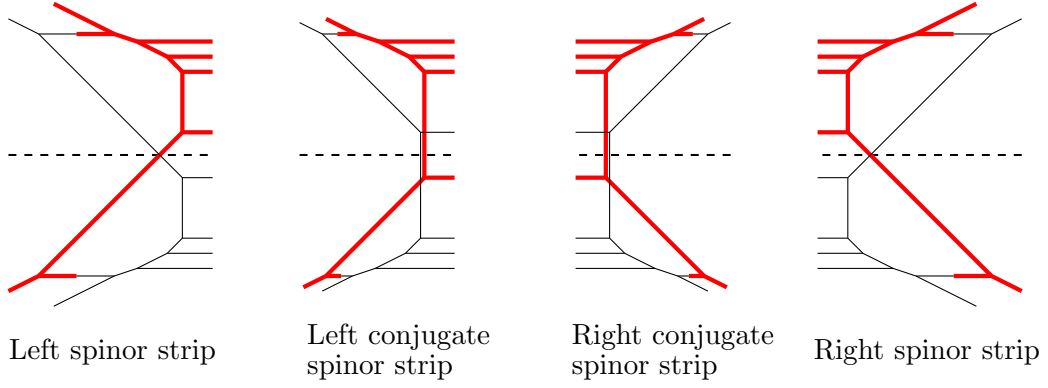
\begin{figure}
\centering
\begin{tikzpicture}
\begin{scope}[shift={(-4,0)}]
\draw (0.2,-2.6) node[left] {\small Left spinor strip};
\draw[dashed](-2.7,0)--(0,0);
\draw [ultra thick,red] (-2.1,2) -- (-1.3, 1.6) -- (-1, 1.5) -- (-0.6,1.3) -- (-0.4,1.1) -- (-0.4,0.3) -- (-0.7,0) -- (-2.3,-1.6) -- (-2.7,-1.8);  
\draw [thin] (-2.1,-2) -- (-1.3, -1.6) -- (-1, -1.5) -- (-0.6,-1.3) -- (-0.4,-1.1) -- (-0.4,-0.3) -- (-0.7,0) -- (-2.3,1.6) -- (-2.7,1.8);  
\draw [ultra thick, red] (-1, 1.5) -- (0, 1.5);
\draw [ultra thick, red] (-0.6,1.3) -- (0,1.3);
\draw [ultra thick, red] (-0.4,1.1) -- (0,1.1);
\draw [ultra thick, red] (-0.4,0.3) -- (0,0.3);
\draw [thin] (-1, -1.5) -- (0, -1.5);
\draw [thin] (-0.6,-1.3) -- (0,-1.3);
\draw [thin] (-0.4,-1.1) -- (0,-1.1);
\draw [thin] (-0.4,-0.3) -- (0,-0.3);
\draw [ultra thick, red] (-1.8,1.6) -- (-1.3, 1.6);
\draw [ultra thick, red] (-2.3,-1.6) -- (-1.8, -1.6);
\draw [thin] (-1.8,-1.6) -- (-1.3, -1.6);
\draw [thin] (-2.3,1.6) -- (-1.8, 1.6);
\end{scope}
%
\begin{scope}[shift={(-0.8,0)}]
\draw (-2.1,-2.4) node[right]  {\small Left conjugate};
\draw (-2.1,-2.8) node[right]  {\small spinor strip};
\draw[dashed](-2.05,0)--(0,0);
\draw [ultra thick, red]  (-1.7,1.8) -- (-1.3, 1.6) -- (-1, 1.5) -- (-0.6,1.3) -- (-0.4,1.1) -- (-0.4,0) -- (-0.4,-0.3) -- (-1.7,-1.6) -- (-2,-1.75); 
\draw [thin]  (-1.75,-1.8) -- (-1.35, -1.6) -- (-1.05, -1.5) -- (-0.65,-1.3) -- (-0.45,-1.1) -- (-0.45,0) -- (-0.45,0.3) -- (-1.75,1.6) -- (-2.05,1.75); 
\draw [ultra thick, red] (-1, 1.5) -- (0, 1.5);
\draw [ultra thick, red] (-0.6,1.3) -- (0,1.3);
\draw [ultra thick, red] (-0.4,1.1) -- (0,1.1);
\draw [ultra thick, red] (-0.4,-0.3) -- (0,-0.3);
\draw [thin] (-1.05, -1.5) -- (0, -1.5);
\draw [thin] (-0.65,-1.3) -- (0,-1.3);
\draw [thin] (-0.45,-1.1) -- (0,-1.1);
\draw [thin] (-0.45,0.3) -- (0,0.3);
\draw [ultra thick, red] (-1.55,1.6) -- (-1.3, 1.6);
\draw [ultra thick, red] (-1.7, -1.6) -- (-1.5,-1.6);
\draw [thin]  (-1.75, 1.6) -- (-1.55,1.6);
\draw [thin] (-1.5, - 1.6) -- (-1.35, - 1.6);
\end{scope}
\begin{scope}[shift={(0.8,0)}]
\draw (-0.2,-2.4) node[right]  {\small Right conjugate};
\draw (-0.2,-2.8) node[right]  {\small spinor strip};
\draw[dashed](2.05,0)--(0,0);
\draw [ultra thick, red]  (1.7,1.8) -- (1.3, 1.6) -- (1, 1.5) -- (0.6,1.3) -- (0.4,1.1) -- (0.4,0) -- (0.4,-0.3) -- (1.7,-1.6) -- (2,-1.75); 
\draw [thin]  (1.75,-1.8) -- (1.35, -1.6) -- (1.05, -1.5) -- (0.65,-1.3) -- (0.45,-1.1) -- (0.45,0) -- (0.45,0.3) -- (1.75,1.6) -- (2.05,1.75); 
\draw [ultra thick, red] (1, 1.5) -- (0, 1.5);
\draw [ultra thick, red] (0.6,1.3) -- (0,1.3);
\draw [ultra thick, red] (0.4,1.1) -- (0,1.1);
\draw [ultra thick, red] (0.4,-0.3) -- (0,-0.3);
\draw [thin] (1.05, -1.5) -- (0, -1.5);
\draw [thin] (0.65,-1.3) -- (0,-1.3);
\draw [thin] (0.45,-1.1) -- (0,-1.1);
\draw [thin] (0.45,0.3) -- (0,0.3);
\draw [ultra thick, red] (1.55,1.6) -- (1.3, 1.6);
\draw [ultra thick, red] (1.7, -1.6) -- (1.5,-1.6);
\draw [thin]  (1.75, 1.6) -- (1.55,1.6);
\draw [thin] (1.5, - 1.6) -- (1.35, - 1.6);
\end{scope}
\begin{scope}[shift={(4,0)}]
\draw (-0.2,-2.6) node[right] {\small Right spinor strip};
\draw[dashed](2.7,0)--(0,0);
\draw [ultra thick,red] (2.1,2) -- (1.3, 1.6) -- (1, 1.5) -- (0.6,1.3) -- (0.4,1.1) -- (0.4,0.3) -- (0.7,0) -- (2.3,-1.6) -- (2.7,-1.8);  
\draw [thin] (2.1,-2) -- (1.3, -1.6) -- (1, -1.5) -- (0.6,-1.3) -- (0.4,-1.1) -- (0.4,-0.3) -- (0.7,0) -- (2.3,1.6) -- (2.7,1.8);  
\draw [ultra thick, red] (1, 1.5) -- (0, 1.5);
\draw [ultra thick, red] (0.6,1.3) -- (0,1.3);
\draw [ultra thick, red] (0.4,1.1) -- (0,1.1);
\draw [ultra thick, red] (0.4,0.3) -- (0,0.3);
\draw [thin] (1, -1.5) -- (0, -1.5);
\draw [thin] (0.6,-1.3) -- (0,-1.3);
\draw [thin] (0.4,-1.1) -- (0,-1.1);
\draw [thin] (0.4,-0.3) -- (0,-0.3);
\draw [ultra thick, red] (1.8,1.6) -- (1.3, 1.6);
\draw [ultra thick, red] (2.3,-1.6) -- (1.8, -1.6);
\draw [thin] (1.8,-1.6) -- (1.3, -1.6);
\draw [thin] (2.3,1.6) -- (1.8, 1.6);
\end{scope}
\end{tikzpicture}
\caption{Reflecting part of the 5-brane web diagram to obtain the left/right spinor/conjugate spinor strip.}
\label{fig:reflect-SC}
\end{figure}

In order to demonstrate this point in our case, we first focus on the left half of the leftmost diagram in Figure \ref{fig:2S-2C-1S+1C}, which includes the contribution from the hypermultiplet in spinor representation as well as part of the contribution from the vector multiplet. The same part also appears in the rightmost diagram in Figure \ref{fig:2S-2C-1S+1C}. On the far left in Figure \ref{fig:reflect-SC}, the sub-diagram under consideration is depicted, along with its mirror image. Among them, we select the part depicted in thick red lines as the fundamental domain of the diagram, from which we take into account the contribution to the amplitude. We observe that this chosen part is a strip diagram of the type whose amplitude is already computed in \cite{Iqbal:2004ne}. In this paper, we denote this strip as ``left spinor strip'' for convenience.

Next, we also consider the left half of the middle diagram in Figure \ref{fig:2S-2C-1S+1C}, which corresponds to the contribution from the hypermultiplet in conjugate spinor representation as well as the part of the contribution from the vector multiplet. In this case, we select the fundamental domain of the diagram as depicted in thick red lines on the second from the left in Figure \ref{fig:reflect-SC}. Again, the chosen part is the strip diagram. We denote this as ``left conjugate spinor strip''. The right half of each diagram in Figure \ref{fig:2S-2C-1S+1C} can also be done in an analogous way, as depicted on the right in Figure \ref{fig:reflect-SC}, which we denote ``right spinor strip'' and ``right conjugate spinor strip'', respectively. We observe that all the diagrams in Figure \ref{fig:2S-2C-1S+1C} can be constructed by combining the left spinor/conjugate spinor strip and the right spinor/conjugate spinor strip.

Once we choose the fundamental domain of the 5-brane web diagram in such a way that it can be decomposed into strip diagrams, the next step is to compute each strip diagram amplitude following the method in \cite{Iqbal:2004ne}. Here, instead of repeating the same computation, we just use the known result for the strip diagram amplitude, which is explicitly written down, for example, in \cite{Li:2021rqr}. 

In order to write down the strip-diagram amplitude concretely according to the topological vertex formalism, we need to assign distinct Young diagrams to the edges, with their orientations specified. Changing the orientation is equivalent to taking the transpose of the assigned Young diagram. In this paper, we denote Young diagrams by bold Greek letters as $\boldsymbol{\mu}, \boldsymbol{\nu}, \cdots$, sometimes with lower indices to distinguish different Young diagrams. Each Young diagram $\boldsymbol{\mu}$ is specified by a non-negative monotonically decreasing sequence of integers, $\mu_1 \ge \mu_2 \ge \cdots \ge 0$, which are denoted with non-bold Greek letters with indices. We assume $\mu_i=0$ for large enough $i$. We also denote the transpose of the Young diagram $\boldsymbol{\mu}$ as $\boldsymbol{\mu}^T$. We also introduce the notation $|  \boldsymbol{\mu}  | : = \sum_{i} \mu_i$ and $\|  \boldsymbol{\mu}  \|^2 : = \sum_{i} \mu_i^2$.

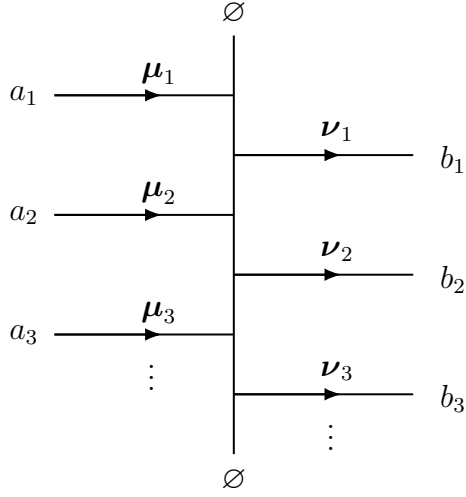
\begin{figure}
    \centering
    \begin{tikzpicture}[x=0.75pt,y=0.75pt,yscale=-0.6,xscale=0.6]%
\draw    (300,350) -- (450,350) ;
\draw [->, >={Latex[scale=1.0]}] (300,350)--(390,350) ;
\draw    (300,250) -- (450,250) ;
\draw [->, >={Latex[scale=1.0]}] (300,250)--(390,250) ;
\draw    (300,150) -- (450,150) ;
\draw [->, >={Latex[scale=1.0]}] (300,150)--(390,150) ;
\draw    (300,300) -- (150,300) ;
\draw [->, >={Latex[scale=1.0]}] (150,300)--(240,300) ;
\draw    (300,200) -- (150,200) ;
\draw [->, >={Latex[scale=1.0]}] (150,200)--(240,200) ;
\draw    (300,100) -- (150,100) ;
\draw [->, >={Latex[scale=1.0]}] (150,100)--(240,100) ;
\draw (300,50) -- (300,400) ;
\draw (300,20) node [anchor=north][inner sep=0.75pt]    {$\boldsymbol{\varnothing}$};
\draw (300,410) node [anchor=north][inner sep=0.75pt]    {$\boldsymbol{\varnothing}$};

\draw (470,140) node [anchor=north west][inner sep=0.75pt]    {$b_{1}$};
\draw (470,240) node [anchor=north west][inner sep=0.75pt]    {$b_{2}$};
\draw (470,340) node [anchor=north west][inner sep=0.75pt]    {$b_{3}$};
\draw (375,360) node [anchor=north west][inner sep=0.75pt]    {$\vdots$};

\draw (110,90) node [anchor=north west][inner sep=0.75pt]    {$a_{1}$};
\draw (110,190) node [anchor=north west][inner sep=0.75pt]    {$a_{2}$};
\draw (110,290) node [anchor=north west][inner sep=0.75pt]    {$a_{3}$};
\draw (225,310) node [anchor=north west][inner sep=0.75pt]    {$\vdots$};
\draw (370,120) node [anchor=north west][inner sep=0.75pt]    {$ \boldsymbol{\nu}_{1}$};
\draw (370,220) node [anchor=north west][inner sep=0.75pt]    {$\boldsymbol{\nu}_{2}$};
\draw (370,320) node [anchor=north west][inner sep=0.75pt]    {$\boldsymbol{\nu}_{3}$};

\draw (220,70) node [anchor=north west][inner sep=0.75pt]    {$\boldsymbol{\mu}_{1}$};
\draw (220,170) node [anchor=north west][inner sep=0.75pt]    {$\boldsymbol{\mu}_{2}$};
\draw (220,270) node [anchor=north west][inner sep=0.75pt]    {$\boldsymbol{\mu}_{3}$};

\end{tikzpicture}
\caption{A strip diagram. Here, $(p,1)$ 5-branes at the center are all written vertically for simplicity. $a_i, \boldsymbol{\mu}_i(i=1,2,3, \ldots)$ are the heights and the Young diagrams corresponding to the D5-branes on the left, respectively. $b_i, \boldsymbol{\nu}_i(i=1,2,3, \ldots)$ are the heights and the Young diagrams corresponding to the D5-branes on the right, respecitvely.
}
 \label{fig:generalstrip}
\end{figure}

In Figure \ref{fig:generalstrip}, a generic strip diagram is depicted.
We denote the height of the left horizontal lines as $a_i$ ($i=1,2,3,\cdots $) while the height of the right horizontal lines as $b_j$ ($j=1,2,3,\cdots $).
We assume $a_1 > a_2 > a_3 > \cdots $ and $b_1 > b_2 > b_3 > \cdots $ without loss of generality. We do not impose any inequality relation between $a_i$ and $b_j$.
The Young diagrams $ \boldsymbol{\mu} _i$ are assigned to the left horizontal lines, while $ \boldsymbol{\nu} _j$ are assigned to the right horizontal lines. We assume that the orientation of the horizontal lines is all to the right. Then, the amplitude for the strip diagram given in Figure \ref{fig:generalstrip} is known to be given by \cite{Iqbal:2004ne, Li:2021rqr}
\footnote{The dependence on the circumference $\beta$ of the circle on which five-dimensional gauge theory is compactified, which appeared in \cite{Li:2021rqr}, has been absorbed into the redefinition of $\hbar$, $a_i$, $b_i$ in this paper.}
\begin{align}\label{eq:strip-general}
Z_{\text{strip}} =
& \prod_{i} g^{\frac{\|  \boldsymbol{\mu} _i^T \|^2}{2}} \tilde{Z}_{ \boldsymbol{\mu} _i}
\prod_{j} g^{\frac{\|  \boldsymbol{\nu} _j \|^2}{2}} \tilde{Z}_{ \boldsymbol{\nu} _j}
\cr
&\prod_{i<j} R^{-1}_{ \boldsymbol{\mu} _i  \boldsymbol{\mu} _j^T} ( e^{- (a_i - a_j)} ) 
\prod_{i<j}  R^{-1}_{ \boldsymbol{\nu} _i  \boldsymbol{\nu} _j^T}  (e^{- (b_i - b_j)} )
\prod_{i, j} X_{ \boldsymbol{\mu} _i,  \boldsymbol{\nu} _j}  \left( a_i,  b_j \right), 
\end{align}
where we denote the exponential of the Omega deformation parameters $\epsilon_1 = - \epsilon_2:= \hbar$ as $g:=e^{-\hbar}$. 
Here, we have introduced building blocks, which typically appear in the computation of the strip amplitude:
\begin{align}
X_{ \boldsymbol{\mu} ,  \boldsymbol{\nu} } \left( a,  b  \right)
:= \left\{
\begin{array}{ll}
R_{ \boldsymbol{\mu}   \boldsymbol{\nu} ^T}\left( e^{- (a - b)} \right) & \text{if }  a > b  \\
R_{ \boldsymbol{\nu}   \boldsymbol{\mu} {}^T} \left( e^{- (b - a)} \right) & \text{if }  a < b \\
R_{ \boldsymbol{\mu}   \boldsymbol{\mu} {}^T} ( 1 ) \delta_{ \boldsymbol{\mu}   \boldsymbol{\nu} } & \text{if }  a = b \
\end{array}
\right. 
\end{align}
with
\begin{align}\label{eq:defofR}
R_{ \boldsymbol{\lambda}  \boldsymbol{\mu} } (Q)
:= & 
\prod_{i=1}^{\infty} \prod_{j=1}^{\infty} \left(1 - Q g^{i+j - \lambda_i - \mu_j - 1} \right)
\end{align}
and 
\begin{align}
\widetilde{Z}_{ \boldsymbol{\nu} }
:= & \prod_{(i,j) \in \boldsymbol{\nu}}\frac1{1-g^{\nu_i+\nu^T_j-i-j+1}} .
\end{align}

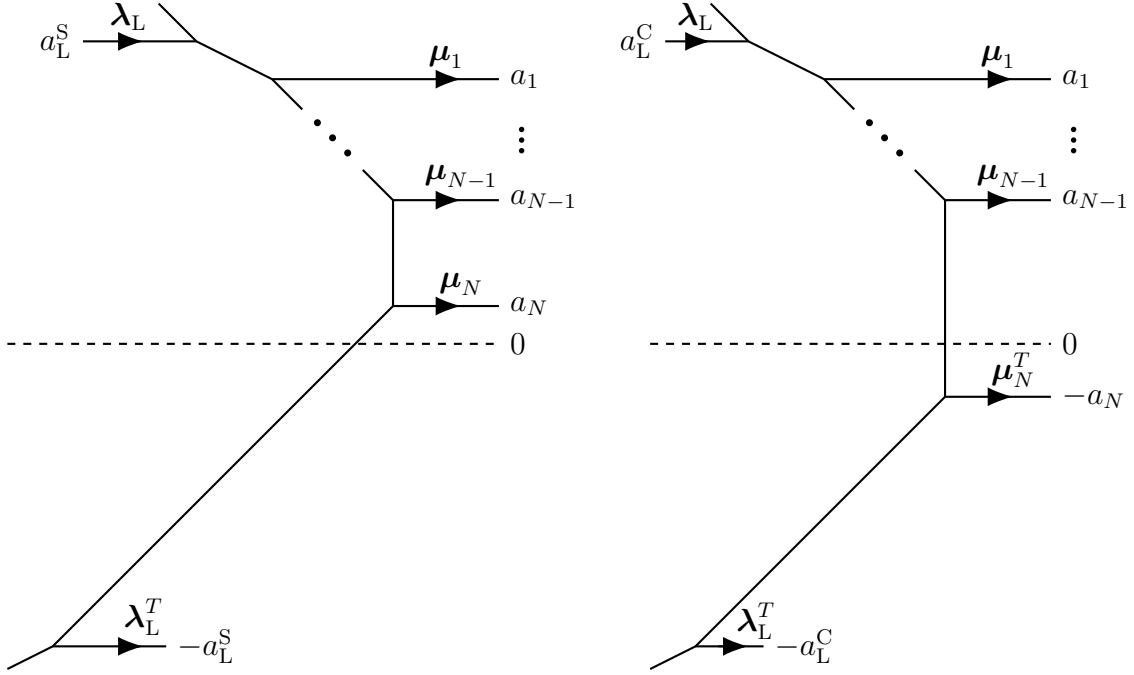
\begin{figure}
\centering
\begin{tikzpicture}
\begin{scope}[shift={(-3.5,0)}]
\draw[thick, dashed](-5.5,0)--(1,0);
\draw (1,0) node[right] {$0$};
\draw [thick] (-3.5,4.5) -- (-3, 4) -- (-2,3.5) -- (-1.6,3.1);
\draw (-1.4,2.9) node {\huge $\cdot$};
\draw (-1.2,2.7) node {\huge $\cdot$};
\draw (-1.0,2.5) node {\huge $\cdot$};
\draw [thick] (-0.8, 2.3) -- (-0.4, 1.9) -- (-0.4,0.5) -- (-0.9,0) -- (-4.9, -4) -- (-5.5, -4.3);
\draw [thick](-2,3.5) -- (1,3.5);
\draw (1,3.5) node[right] {$a_1$};
\draw [->, >={Latex[scale=1.5]}] (0.4,3.5) -- (0.5,3.5);
\draw (0.3,3.5) node[above] {$\boldsymbol{\mu}_1$};
\draw (1.3,2.8) node {\huge $\vdots$};
\draw [thick](-0.4, 1.9) -- (1,1.9);
\draw (1,1.9) node[right] {$a_{N-1}$};
\draw [->, >={Latex[scale=1.5]}] (0.4,1.9) -- (0.5,1.9);
\draw (0.5,1.9) node[above] {$\boldsymbol{\mu}_{N-1}$};
\draw [thick](-0.4, 0.5) -- (1,0.5);
\draw (1,0.5) node[right] {$a_{N}$};
\draw [->, >={Latex[scale=1.5]}] (0.4,0.5) -- (0.5,0.5);
\draw (0.5,0.5) node[above] {$\boldsymbol{\mu}_{N}$};
%
\draw [thick] (-4.5, 4)  -- (-3, 4) ;
\draw (-4.5,4) node[left] {$a_{\mathrm{L}}^{\mathrm{S}}$};
\draw [->, >={Latex[scale=1.5]}] (-3.9,4) -- (-3.7,4);
\draw (-3.9,4) node[above] {$\boldsymbol{\lambda}_{\mathrm{L}}$};
\draw [thick] (-4.9, -4)  -- (-3.4, -4) ;
\draw (-3.4,-4) node[right] {$-a_{\mathrm{L}}^{\mathrm{S}}$};
\draw [->, >={Latex[scale=1.5]}] (-3.8,-4) -- (-3.7,-4);
\draw (-3.7,-4) node[above] {$\boldsymbol{\lambda}_{\mathrm{L}}^T$};
\end{scope}

\begin{scope}[shift={(3.8,0)}]
\draw[thick, dashed](-4.3,0)--(1,0);
\draw (1,0) node[right] {$0$};
\draw [thick] (-3.5,4.5) -- (-3, 4) -- (-2,3.5) -- (-1.6,3.1);
\draw (-1.4,2.9) node {\huge $\cdot$};
\draw (-1.2,2.7) node {\huge $\cdot$};
\draw (-1.0,2.5) node {\huge $\cdot$};
\draw [thick] (-0.8, 2.3) -- (-0.4, 1.9) -- (-0.4,-0.7) -- (-3.7, -4) -- (-4.3, -4.3);
\draw [thick](-2,3.5) -- (1,3.5);
\draw (1,3.5) node[right] {$a_1$};
\draw [->, >={Latex[scale=1.5]}] (0.4,3.5) -- (0.5,3.5);
\draw (0.3,3.5) node[above] {$\boldsymbol{\mu}_1$};
\draw (1.3,2.8) node {\huge $\vdots$};
\draw [thick](-0.4, 1.9) -- (1,1.9);
\draw (1,1.9) node[right] {$a_{N-1}$};
\draw [->, >={Latex[scale=1.5]}] (0.4,1.9) -- (0.5,1.9);
\draw (0.5,1.9) node[above] {$\boldsymbol{\mu}_{N-1}$};
\draw [thick](-0.4, -0.7) -- (1,-0.7);
\draw (1,-0.7) node[right] {$-a_{N}$};
\draw [->, >={Latex[scale=1.5]}] (0.4,-0.7) -- (0.5,-0.7);
\draw (0.5,-0.72) node[above] {$\boldsymbol{\mu}_{N}^T$};
%
\draw [thick] (-4.1, 4)  -- (-3, 4) ;
\draw (-4.1,4) node[left] {$a_{\mathrm{L}}^{\mathrm{C}}$};
\draw [->, >={Latex[scale=1.5]}] (-3.8,4) -- (-3.5,4);
\draw (-3.7,4) node[above] {$\boldsymbol{\lambda}_{\mathrm{L}}$};
\draw [thick] (-3.7, -4)  -- (-2.8, -4) ;
\draw (-2.8,-4) node[right] {$-a_{\mathrm{L}}^{\mathrm{C}}$};
\draw [->, >={Latex[scale=1.5]}] (-3.4,-4) -- (-3,-4);
\draw (-2.9,-4.05) node[above] {$\boldsymbol{\lambda}_{\mathrm{L}}^T$};
\end{scope}
\end{tikzpicture}
\caption{Parametrization and assignment of Young diagram for the left spinor strip (left figure) and the left conjugate spinor strip (right figure). Only the fundamental domains of the 5-brane web diagrams are depicted, respectively.}
\label{fig:Young-assign}
\end{figure}

In order to compute the strip amplitude given in Figure \ref{fig:reflect-SC} using this result, we need to fix the parametrization. We first focus on the left spinor/conjugate spinor strips for explanation. As depicted in Figure \ref{fig:Young-assign}, we denote the height of the horizontal lines corresponding to the color branes in the upper half as $a_i$ ($i=1,2,\cdots, N$), which are identified as the Coulomb moduli of SO($2N$) gauge theory. In our notation, the heights of the horizontal lines are measured from the O5-plane. Note that, in this notation, the height of the color brane in the lower half, which is obtained by reflecting part of the upper half diagram in terms of the O5-plane, is given with a minus sign, such as $-a_N$ for the left conjugate spinor strip. We also denote the height of the remaining horizontal line in the upper half as $a_{\mathrm{L}}^{\mathrm{S}}$ and $a_{\mathrm{L}}^{\mathrm{C}}$ for the spinor strip and the conjugate spinor strip, respectively. Here, the labels S and C stand for the spinor and conjugate spinor, respectively, and the label L stands for the left strip. For the parametrization to be consistent with reflection, the height of the remaining horizontal line in the lower half should be $-a_{\mathrm{L}}^{\mathrm{S}}$ and $-a_{\mathrm{L}}^{\mathrm{C}}$. 

As discussed in \cite{Akhond2026}, the parameters $a_{\mathrm{L}}^{\mathrm{S}}$ and $a_{\mathrm{L}}^{\mathrm{C}}$ are related to the mass of the spinor or conjugate spinor mass $m_{\mathrm{L}}$ as 
\begin{align}\label{eq:param-b}
a_{\mathrm{L}}^{\mathrm{S}} = \frac12 \sum_{i=1}^N a_i - m_{\mathrm{L}}, 
\quad
a_{\mathrm{L}}^{\mathrm{C}} = \left. a_{\mathrm{L}}^{\mathrm{S}} \right|_{a_N \to -a_N}.
\end{align}
This parametrization is fixed in such a way that the area computed from the 5-brane web diagram agrees with the first derivative of the prepotential \cite{Intriligator:1997pq} in terms of the Coulomb branch parameter.

The assignment of the Young diagrams to the horizontal lines is also given in Figure \ref{fig:Young-assign}. We assign $\boldsymbol{\mu}_i$ to the horizontal line with height $a_i$. Following the convention in Figure \ref{fig:generalstrip}, we suppose the orientation to be to the right. We note that we assign $\boldsymbol{\mu}_N^T$ instead of $\boldsymbol{\mu}_N$ to the color brane at height $-a_{N}$, which was obtained by reflecting the horizontal line with height $a_N$ in terms of the O5-plane. This assignment is based on the property that the orientation changes under the orientifold action. This property can be explained from the identity that the topological vertex satisfies, as explained in \cite{Kim:2017jqn}. Since the horizontal line at height $-a_{\mathrm{L}}^{\mathrm{S}}$ or $-a_{\mathrm{L}}^{\mathrm{C}}$ was connected to the horizontal line at height $a_{\mathrm{L}}^{\mathrm{S}}$ or $a_{\mathrm{L}}^{\mathrm{C}}$ before reflection, respectively, we should assign Young diagrams to them that are related by transposition. Thus, we assign $\boldsymbol{\lambda}_{\mathrm{L}}^T$ and $\boldsymbol{\lambda}_{\mathrm{L}}$ to them, respectively, where again, the label L stands for the left strip.

Applying \eqref{eq:strip-general} to the setup given in Figure \ref{fig:Young-assign}, the amplitude for the left spinor strip is given by
\begin{align}\label{eq:leftS}
Z_{\text{left}}^{\text{S}}
&= g^{||\boldsymbol{\lambda}_{\mathrm{L}}^T||^2 } \tilde{Z}_{\boldsymbol{\lambda}_{\mathrm{L}}}^2 
\left(
\prod_{j=1}^{N} 
g^{\frac{||\boldsymbol{\mu}_{j}||^2}{2} } 
\tilde{Z}_{\boldsymbol{\mu}_{j}} 
\right) 
\left(
\prod_{1 \le i < j \le N} R^{-1}_{\boldsymbol{\mu}_i \boldsymbol{\mu}_j^T}\left( e^{-(a_i-a_j)} \right) 
\right)
\cr
& 
\left(
\prod_{i=1}^{N}  
R^{-1}_{\boldsymbol{\mu}_i \boldsymbol{\lambda}_{\mathrm{L}}}\left( e^{-(a_i+a_{\mathrm{L}}^{\mathrm{S}})} \right)
\right)
\left(
\prod_{i=1}^{N}  
R_{\boldsymbol{\lambda}_{\mathrm{L}} \boldsymbol{\mu}_i^T}\left( e^{-(a_{\mathrm{L}}^{\mathrm{S}} - a_i)} \right)
\right) 
R_{\boldsymbol{\lambda}_{\mathrm{L}}\boldsymbol{\lambda}_{\mathrm{L}}}\left( e^{- 2 a_{\mathrm{L}}^{\mathrm{S}}} \right).
\end{align}

Comparing the left spinor strip and the left conjugate spinor strip in Figure \ref{fig:Young-assign}, we observe that the difference between them is whether the sign in front of $a_N$ is plus or minus, whether we use $a_{\mathrm{L}}^{\mathrm{S}}$ or $a_{\mathrm{L}}^{\mathrm{C}}$, and whether the Young diagram $\boldsymbol{\mu}_N$ is transposed or not. 
Moreover, we find from the parametrization \eqref{eq:param-b} that the exchange of $a_{\mathrm{L}}^{\mathrm{S}}$ and $a_{\mathrm{L}}^{\mathrm{C}}$ is automatically induced from the replacement $a_N \to -a_N$. Thus, the amplitude for the conjugate spinor strip is given just by replacing $a_N \to -a_N$, and $\boldsymbol{\mu}_N \to \boldsymbol{\mu}_N^T$ in the amplitude for the spinor strip result: 
\begin{align}\label{eq:leftC}
Z_{\text{left}}^{\text{C}}
= 
\left. Z_{\text{left}}^{\text{S}} \right|_{a_N \to -a_N, \,\, \boldsymbol{\mu}_{N} \to \boldsymbol{\mu}_{N}^T}.
\end{align}

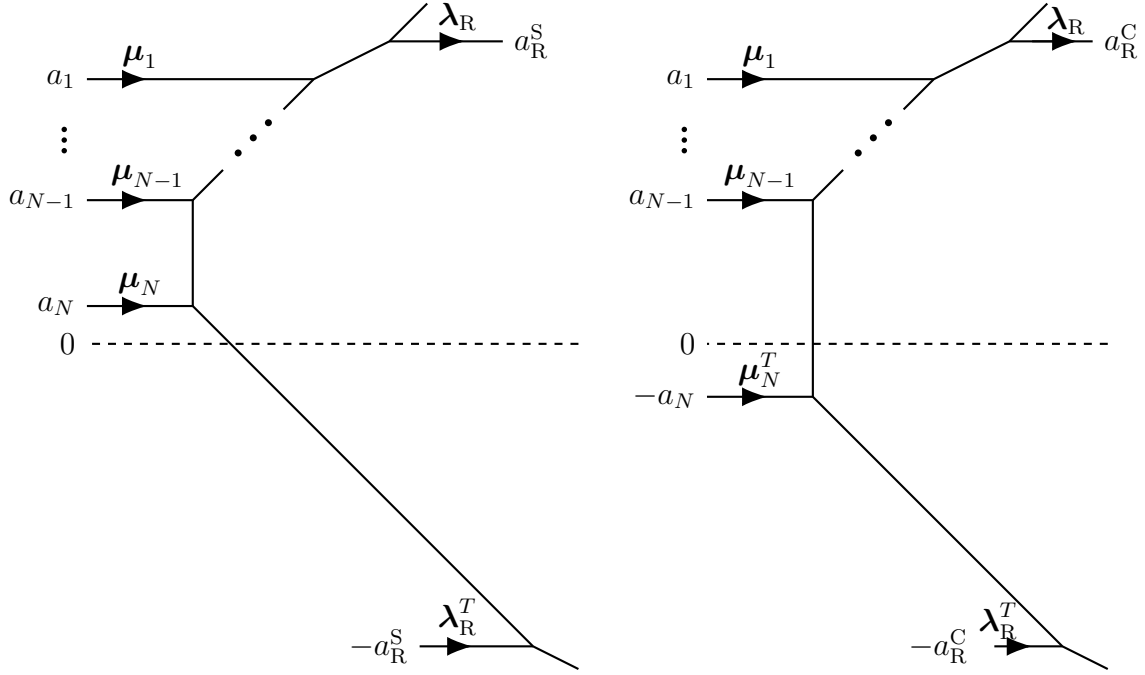
\begin{figure}
\centering
\begin{tikzpicture}
\begin{scope}[shift={(-4,0)}]
\draw[thick, dashed](5.5,0)--(-1,0);
\draw (-1,0) node[left] {$0$};
\draw [thick] (3.5,4.5) -- (3, 4) -- (2,3.5) -- (1.6,3.1);
\draw (1.4,2.9) node {\huge $\cdot$};
\draw (1.2,2.7) node {\huge $\cdot$};
\draw (1.0,2.5) node {\huge $\cdot$};
\draw [thick] (0.8, 2.3) -- (0.4, 1.9) -- (0.4,0.5) -- (0.9,0) -- (4.9, -4) -- (5.5, -4.3);
\draw [thick](2,3.5) -- (-1,3.5);
\draw (-1,3.5) node[left] {$a_1$};
\draw [->, >={Latex[scale=1.5]}] (-0.4,3.5) -- (-0.2,3.5);
\draw (-0.3,3.5) node[above] {$\boldsymbol{\mu}_1$};
\draw (-1.3,2.8) node {\huge $\vdots$};
\draw [thick](0.4, 1.9) -- (-1,1.9);
\draw (-1,1.9) node[left] {$a_{N-1}$};
\draw [->, >={Latex[scale=1.5]}] (-0.4,1.9) -- (-0.2,1.9);
\draw (-0.2,1.9) node[above] {$\boldsymbol{\mu}_{N-1}$};
\draw [thick](0.4, 0.5) -- (-1,0.5);
\draw (-1,0.5) node[left] {$a_{N}$};
\draw [->, >={Latex[scale=1.5]}] (-0.4,0.5) -- (-0.2,0.5);
\draw (-0.3,0.5) node[above] {$\boldsymbol{\mu}_{N}$};
%
\draw [thick] (4.5, 4)  -- (3, 4) ;
\draw (4.5,4) node[right] {$a_{\mathrm{R}}^{\mathrm{S}}$};
\draw [->, >={Latex[scale=1.5]}] (3.9,4) -- (4,4);
\draw (3.9,4) node[above] {$\boldsymbol{\lambda}_{\mathrm{R}}$};
\draw [thick] (4.9, -4)  -- (3.4, -4) ;
\draw (3.4,-4) node[left] {$-a_{\mathrm{R}}^{\mathrm{S}}$};
\draw [->, >={Latex[scale=1.5]}] (3.8,-4) -- (4.1,-4);
\draw (3.9,-4) node[above] {$\boldsymbol{\lambda}_{\mathrm{R}}^T$};
\end{scope}

\begin{scope}[shift={(4.2,0)}]
\draw[thick, dashed](4.3,0)--(-1,0);
\draw (-1,0) node[left] {$0$};
\draw [thick] (3.5,4.5) -- (3, 4) -- (2,3.5) -- (1.6,3.1);
\draw (1.4,2.9) node {\huge $\cdot$};
\draw (1.2,2.7) node {\huge $\cdot$};
\draw (1.0,2.5) node {\huge $\cdot$};
\draw [thick] (0.8, 2.3) -- (0.4, 1.9) -- (0.4,-0.7) -- (3.7, -4) -- (4.3, -4.3);
\draw [thick](2,3.5) -- (-1,3.5);
\draw (-1,3.5) node[left] {$a_1$};
\draw [->, >={Latex[scale=1.5]}] (-0.4,3.5) -- (-0.2,3.5);
\draw (-0.3,3.5) node[above] {$\boldsymbol{\mu}_1$};
\draw (-1.3,2.8) node {\huge $\vdots$};
\draw [thick](0.4, 1.9) -- (-1,1.9);
\draw (-1,1.9) node[left] {$a_{N-1}$};
\draw [->, >={Latex[scale=1.5]}] (-0.4,1.9) -- (-0.2,1.9);
\draw (-0.3,1.9) node[above] {$\boldsymbol{\mu}_{N-1}$};
\draw [thick](0.4, -0.7) -- (-1,-0.7);
\draw (-1,-0.7) node[left] {$-a_{N}$};
\draw [->, >={Latex[scale=1.5]}] (-0.4,-0.7) -- (-0.2,-0.7);
\draw (-0.3,-0.72) node[above] {$\boldsymbol{\mu}_{N}^T$};
%
\draw [thick] (4.1, 4)  -- (3, 4) ;
\draw (4.1,4) node[right] {$a_{\mathrm{R}}^{\mathrm{C}}$};
\draw [->, >={Latex[scale=1.5]}] (3.3,4) -- (3.9,4);
\draw (3.75,3.95) node[above] {$\boldsymbol{\lambda}_{\mathrm{R}}$};
\draw [thick] (3.7, -4)  -- (2.8, -4) ;
\draw (2.6,-4) node[left] {$-a_{\mathrm{R}}^{\mathrm{C}}$};
\draw [->, >={Latex[scale=1.5]}] (3,-4) -- (3.3,-4);
\draw (2.85,-4.05) node[above] {$\boldsymbol{\lambda}_{\mathrm{R}}^T$};
\end{scope}
\end{tikzpicture}
\caption{Parametrization and assignment of Young diagram for the right spinor strip (left figure) and the right conjugate spinor strip (right figure). Only the fundamental domains of the 5-brane web diagrams are depicted, respectively. }
\label{fig:Young-assign-right}
\end{figure}

The parametrization and the assignment of Young diagrams for the right spinor/ conjugate spinor are given analogously in Figure \ref{fig:Young-assign-right}. Here, the parameters $a_{\mathrm{R}}^{\mathrm{S}}$ and $a_{\mathrm{R}}^{\mathrm{C}}$ are related to the spinor or conjugate spinor mass $m_{\mathrm{R}}$ analogous to \eqref{eq:param-b} as 
\begin{align}\label{eq:param-b-R}
a_{\mathrm{R}}^{\mathrm{S}} = \frac12 \sum_{i=1}^N a_i - m_{\mathrm{R}}, 
\quad
a_{\mathrm{R}}^{\mathrm{C}} = \left. a_{\mathrm{R}}^{\mathrm{S}} \right|_{a_N \to - a_N}
\end{align}

The amplitude for the right spinor strip is also computed based on this in an analogous way:
\begin{align}\label{eq:rightS}
Z_{\text{right}}^{\text{S}}
&= g^{||\boldsymbol{\lambda}_{\mathrm{R}}||^2 } \tilde{Z}_{\boldsymbol{\lambda}_{\mathrm{R}}}^2 
\left(
\prod_{j=1}^{N} 
g^{\frac{||\boldsymbol{\mu}_{j}^T||^2}{2} } 
\tilde{Z}_{\boldsymbol{\mu}_{j}} 
\right) 
\left(
\prod_{1 \le i < j \le N} R^{-1}_{\boldsymbol{\mu}_i \boldsymbol{\mu}_j^T}\left( e^{-(a_i-a_j)} \right) 
\right)
\cr
&  
\left(
\prod_{i=1}^{N}  
R^{-1}_{\boldsymbol{\mu}_i \boldsymbol{\lambda}_{\mathrm{R}}}\left( e^{-(a_i+a_{\mathrm{R}}^{\mathrm{S}})} \right)
\right)
\left(
\prod_{i=1}^{N}  
R_{\boldsymbol{\lambda}_{\mathrm{R}} \boldsymbol{\mu}_i^T}\left( e^{-(a_{\mathrm{R}}^{\mathrm{S}} - a_i)} \right)
\right) 
R_{\boldsymbol{\lambda}_{\mathrm{R}}\boldsymbol{\lambda}_{\mathrm{R}}}\left( e^{-2a_{\mathrm{R}}^{\mathrm{S}}} \right) .
\end{align}
We observe that the right spinor strip amplitude is obtained from the left spinor strip amplitude after the replacement $m_{\mathrm{L}} \to m_{\mathrm{R}}$ and $\boldsymbol{\lambda}_{\mathrm{L}} \to \boldsymbol{\lambda}_{\mathrm{R}}$ up to  framing factors as follows:
\begin{align}\label{eq:rightC}
Z_{\text{right}}^{\text{S}}
= 
\left. Z_{\text{left}}^{\text{S}}\right|_{m_{\mathrm{L}} \to m_{\mathrm{R}}, \boldsymbol{\lambda}_{\mathrm{L}} \to \boldsymbol{\lambda}_{\mathrm{R}}}
\times
\mathfrak{f}_{\boldsymbol{\lambda}_{\mathrm{R}}}^{-2}
\prod_{i=1}^{N} (-1)^{|\boldsymbol{\mu}_{i}|} \mathfrak{f}_{\boldsymbol{\mu}_{i}}.
\end{align}
Here, 
the framing factor is defined as \footnote{This should be distinguished from the profile function $f_{\boldsymbol{\mu}}$, which appear in later section.}
\begin{align}\label{eq:def-framing}
\mathfrak{f}_{\boldsymbol{\mu}} := 
(-1)^{|\boldsymbol{\mu}|} g^{\frac12 (\| \boldsymbol{\mu}^T \|^2 -\| \boldsymbol{\mu} \|^2 ) }.
\end{align}

The right conjugate spinor strip is, again, obtained by replacing $a_N \to -a_N$ and $\boldsymbol{\mu}_N \to \boldsymbol{\mu}_N^T$ in the right spinor strip amplitude.
\begin{align}\label{eq:Zr-C-S}
Z_{\text{right}}^{\text{C}}
= 
\left. Z_{\text{right}}^{\text{S}} \right|_{a_N \to -a_N, \,\, \boldsymbol{\mu}_{N} \to \boldsymbol{\mu}_{N}^T}.
\end{align}

\subsection{Gluing strips}\label{subsec:par-gluing}

In this subsection, we compute the unrefined topological string partition function using the amplitudes for left/right spinor/conjugate spinor computed in the previous subsection. 

According to the topological vertex formalism, the total topological string partition function is obtained by gluing the strip diagram amplitudes. This gluing is implemented by multiplying the edge factor corresponding to the edge where the strips are connected, and then summing over the Young diagrams assigned to those edges.

The edge factor is given in the form 
\begin{align}\label{eq:edge-fac}
(-Q)^{|\boldsymbol{\mu}|} \, \mathfrak{f}_{\boldsymbol{\mu}}^{\, n},
\end{align}
where we denote the Young diagram assigned to the edge as $\boldsymbol{\mu}$ and the corresponding K\"ahler parameter as $Q$, which can be read off from the 5-brane web diagram as
\begin{align}
Q = \exp \left( - \frac{ (\text{Length}) }{ \sqrt{p^2+q^2} }\right).
\end{align}
Here, ``(Length)'' denotes the length of the edge and $p,q$ are the charge of the 5-brane, which can be read off from the slope of the edge.
The power $n$ for the framing factor in \eqref{eq:edge-fac} is an integer determined by the $(p,q)$ charges of the neighboring 5-branes, as explained in detail in the following.

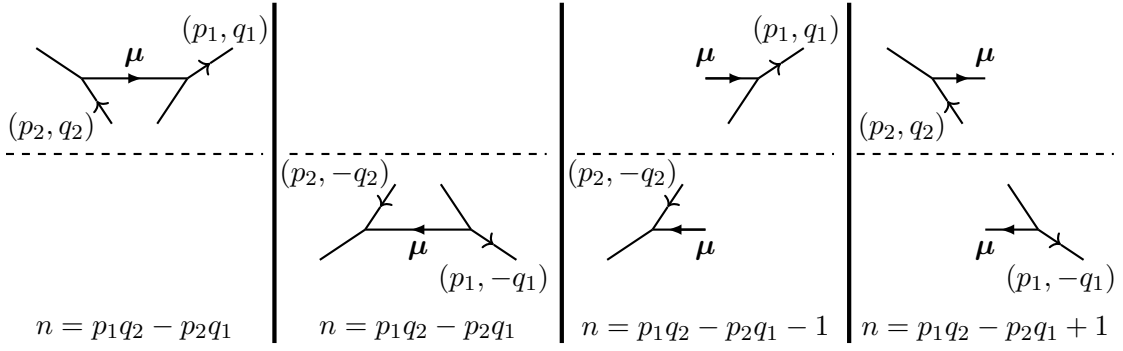
\begin{figure}
\centering
\begin{tikzpicture}
\begin{scope}[shift={(-5.65,0)}]
\draw [dashed,thick] (-1.7,0) -- (1.7,0);
\draw [thick](-0.7,1) -- (0.7,1) ;
\draw [->,>=latex, thick] (0,1) -- (0.1,1) ;
\draw (0,1) node [above] {\small $\boldsymbol{\mu}$};
\draw [->,thick] (-0.3,0.4) -- (-0.5,0.7) ;
\draw [thick](-0.5,0.7) -- (-0.7,1) ;
\draw (-1.1,0.7) node [below] {\small $(p_2,q_2)$};
\draw [thick](-0.7,1) -- (-1.3,1.4) ;
\draw [->,thick]  (0.7,1) -- (1,1.2);
\draw [thick](1,1.2) -- (1.3,1.4);
\draw (1.2,1.3) node [above] {\small $(p_1,q_1)$};
\draw [thick](0.7,1) -- (0.3,0.4) ;
\draw (0,-2) node [below] {\small $n=p_1 q_2 - p_2 q_1$};
\end{scope}
\draw [ultra thick](-3.8,-2.5) -- (-3.8,2);
\begin{scope}[shift={(-1.9,0)}]
\draw [dashed,thick] (-1.7,0) -- (1.7,0);
\draw [thick](-0.7,-1) -- (0.7,-1) ;
\draw [->,>=latex, thick] (0,-1) -- (-0.1,-1) ;
\draw (0,-1) node [below] {\small $\boldsymbol{\mu}$};
\draw [->,thick] (-0.3,-0.4) -- (-0.5,-0.7) ;
\draw [thick](-0.5,-0.7) -- (-0.7,-1) ;
\draw (-1.1,-0.6) node [above] {\small $(p_2,-q_2)$};
\draw [thick](-0.7,-1) -- (-1.3,-1.4) ;
\draw [->,thick]  (0.7,-1) -- (1,-1.2);
\draw [thick](1,-1.2) -- (1.3,-1.4);
\draw (1,-1.3) node [below] {\small $(p_1,-q_1)$};
\draw [thick](0.7,-1) -- (0.3,-0.4) ;
\draw (0,-2) node [below] {\small $n=p_1 q_2 - p_2 q_1$};
\end{scope}
\draw [ultra thick] (0,-2.5) -- (0,2);
\begin{scope}[shift={(1.9,0)}]
\draw [dashed,thick] (-1.7,0) -- (1.7,0);
\draw [thick](0,1) -- (0.7,1) ;
\draw [->,>=latex, thick] (0,1) -- (0.5,1) ;
\draw (0,1) node [above] {\small $\boldsymbol{\mu}$};
\draw [thick](-0.7,-1) -- (0,-1) ;
\draw [->,>=latex, thick] (0,-1) -- (-0.5,-1) ;
\draw (0,-1) node [below] {\small $\boldsymbol{\mu}$};
\draw [->,thick] (-0.3,-0.4) -- (-0.5,-0.7) ;
\draw [thick](-0.5,-0.7) -- (-0.7,-1) ;
\draw (-1.1,-0.6) node [above] {\small $(p_2,-q_2)$};
\draw [thick](-0.7,-1) -- (-1.3,-1.4) ;
\draw [->,thick]  (0.7,1) -- (1,1.2);
\draw [thick](1,1.2) -- (1.3,1.4);
\draw (1.2,1.3) node [above] {\small $(p_1,q_1)$};
\draw [thick](0.7,1) -- (0.3,0.4) ;
\draw (0,-2) node [below] {\small $n=p_1 q_2 - p_2 q_1-1$};
\end{scope}
\draw [ultra thick](3.8,-2.5) -- (3.8,2);
\begin{scope}[shift={(5.6,0)}]
\draw [dashed,thick] (-1.7,0) -- (1.7,0);
\draw [thick](-0.7,1) -- (0,1) ;
\draw [->,>=latex, thick] (-0.3,1) -- (-0.2,1) ;
\draw (0,1) node [above] {\small $\boldsymbol{\mu}$};
\draw [thick](0,-1) -- (0.7,-1) ;
\draw [->,>=latex, thick] (0.3,-1) -- (0.2,-1) ;
\draw (0,-1) node [below] {\small $\boldsymbol{\mu}$};
\draw [->,thick] (-0.3,0.4) -- (-0.5,0.7) ;
\draw [thick](-0.5,0.7) -- (-0.7,1) ;
\draw (-1.1,0.7) node [below] {\small $(p_2,q_2)$};
\draw [thick](-0.7,1) -- (-1.3,1.4) ;
\draw [->,thick]  (0.7,-1) -- (1,-1.2);
\draw [thick](1,-1.2) -- (1.3,-1.4);
\draw (1,-1.3) node [below] {\small $(p_1,-q_1)$};
\draw [thick](0.7,-1) -- (0.3,-0.4) ;
\draw (0,-2) node [below] {\small $n=p_1 q_2 - p_2 q_1+1$};
\end{scope}
\end{tikzpicture}
\caption{Four different types of gluing patterns, where we glue with the horizontal edge with Young diagram $\boldsymbol{\mu}$ assigned. The dashed line represents the O5-plane. }
\label{fig:gluing-pattern}
\end{figure}

Since we consider the strip diagram obtained by reflecting part of the diagram across the O5-plane, we have four patterns of gluing. That is, the edge is not reflected, the whole edge is reflected, the left half of the edge is reflected, or the right half of the edge is reflected, as depicted from the left in Figure \ref{fig:gluing-pattern}, respectively. 

If the considered edge is not reflected, the rule is identical to the original topological vertex formalism \cite{Aganagic:2003db}. It is well known that the power $n$ is given by
\begin{align}\label{eq:framing-power}
n = p_1 q_2 - p_2 q_1\ ,
\end{align}
where $(p_1, q_1)$ and $(p_2, q_2)$ are the 5-brane charges of the upper-right and lower-left 5-branes, which are defined according to the arrows in the leftmost figure in Figure \ref{fig:gluing-pattern}. 

The second left figure in \ref{fig:gluing-pattern} is obtained by reflecting the whole edge across the O5-plane. In this process, the edge orientation is reversed, as described in the previous subsection. Note that the assignment of the Young diagram $\boldsymbol{\mu}$ with the orientation to the left is equivalent to the assignment of the Young diagram $\boldsymbol{\mu}^T$ with the orientation to the right. The identical rule \eqref{eq:framing-power} applies also for this case. 

More concern is required when half of the edge is reflected across the O5-plane. The second figure from the right and the rightmost figure in Figure \ref{fig:gluing-pattern} are both obtained by reflecting the left half and the right half, respectively. In this case, we need to shift by $-1$ and $+1$, respectively as
\begin{align}\label{eq:framing-power-exotic}
n = p_1 q_2 - p_2 q_1 \pm 1\ .
\end{align}
This rule is obeyed by requiring that the final result for the unrefined partition function be independent of which part of the 5-brane web is reflected. This rule is proposed, discussed, and/or justified, explicitly or implicitly, in various papers including \cite{Kim:2017jqn, Hayashi:2018bkd, Hayashi:2020hhb, Kim:2021cua, Li:2021rqr, Kim:2022dbr}.

\begin{figure}
\centering
\begin{tikzpicture}[scale=0.4, thick, font=\small]
\draw [dashed] (-18,0) -- (18,0);
\draw (-15,13) -- (-13,11) -- (-11,10) -- (-9,9);
\draw (-8.4,8.7) node {$\cdot$};
\draw (-8.0,8.5) node {$\cdot$};
\draw (-7.6,8.3) node {$\cdot$};
\draw (-7,8) -- (-4,6) -- (-3,4.5);
\draw [->] (-3,4.5) -- (-3.5,5.25);
\draw (-3.4,4.8) node [left]{$(-N+i+1,1)$};
\draw (-3,4.0) node {$\cdot$};
\draw (-3,3.7) node {$\cdot$};
\draw (-3,3.4) node {$\cdot$};
\draw (-3,3) -- (-3,2) -- (-5,0) -- (-16,-11) -- (-18,-12);
\draw (15,13) -- (13,11) -- (11,10) -- (9,9);
\draw (8.4,8.7) node {$\cdot$};
\draw (8.0,8.5) node {$\cdot$};
\draw (7.6,8.3) node {$\cdot$};
\draw (7,8) -- (4,6) -- (3,4.5);
\draw [->] ((4,6) -- (5.8,7.2);
\draw (5.8,7) node [right]{$(N-i,1)$};
\draw (3,4.0) node {$\cdot$};
\draw (3,3.7) node {$\cdot$};
\draw (3,3.4) node {$\cdot$};
\draw (3,3) -- (3,2) -- (5,0) -- (16,-11) -- (18,-12);
\draw (-11,10) -- (-1,10);
\draw [->, >=Latex] (-2,10) -- (-1.6,10);
\draw (0,10) node {$\boldsymbol{\mu}_1$};
\draw (1,10) -- (11,10);
\draw [->, >=Latex] (2,10) -- (2.1,10);
\draw [dotted] (-10,10) -- (-11.5,10);
\draw (-11.2,10) node [left]{$a_1$};
\draw (-4,6) -- (-1,6);
\draw [->, >=Latex] (-2,6) -- (-1.6,6);
\draw (0,6) node {$\boldsymbol{\mu}_i$};
\draw (1,6) -- (4,6);
\draw [->, >=Latex] (2,6) -- (2.1,6);
\draw[dotted] (-4,6) -- (-4,8);
\draw[dotted] (4,6) -- (4,8);
\draw [->] (0,7) -- (-4,7);
\draw [->] (0,7) -- (4,7);%
\draw (0,7) node [above] {$\ell_i^{\text{SS}}$};
\draw [dotted] (-4,6) -- (-5.5,6) node [left]{$a_i$};
\draw (-3,2) -- (-1,2);
\draw [->, >=Latex] (-2,2) -- (-1.6,2);
\draw (0,2) node {$\boldsymbol{\mu}_N$};
\draw (1,2) -- (3,2);
\draw [->, >=Latex] (2,2) -- (2.1,2);
\draw [dotted] (-3,2) -- (-3,-1.5);
\draw [dotted] (3,2) -- (3,-1.5);
\draw [->] (0,-0.5) -- (-3,-0.5);
\draw [->] (0,-0.5) -- (3,-0.5);
\draw (0,-1.2) node {$\ell_N^{\text{SS}}$};
\draw [dotted] (-3,2) -- (-4.5,2) node [left]{$a_N$};
\draw (-14.8,11) -- (-13,11);
\draw [->, >=Latex] (-14,11) -- (-13.5,11);
\draw (-14.5,11) node [left]{$\boldsymbol{\lambda}_{\mathrm{L}}$};
%
\draw (-16,-11) -- (-14.2,-11);
\draw [->, >=Latex] (-15,-11) -- (-14.6,-11);
\draw (-14.5,-11) node [right]{$\boldsymbol{\lambda}_{\mathrm{L}}^T$};
%
\draw [dotted] (-16,-12) -- (-16,12);
\draw [dotted] (-13,9) -- (-13,12);
\draw [->] (-14.5,10) -- (-13,10);
\draw [->] (-14.5,10) -- (-16,10);
\draw (-14.5,10) node [below]{$\ell_{\mathrm{L}}^{\text{S}}$};
\draw (14.8,11) -- (13,11);
\draw [->, >=Latex] (14,11) -- (14.4,11);
\draw (14.5,11) node [right]{$\boldsymbol{\lambda}_{\mathrm{R}}$};
\draw (16,-11) -- (14.2,-11);
\draw [->, >=Latex] (15,-11) -- (15.5,-11);
\draw (14.5,-11) node [left]{$\boldsymbol{\lambda}_{\mathrm{R}}^T$};
%
\draw [dotted] (16,-12) -- (16,12);
\draw [dotted] (13,9) -- (13,12);
\draw [->] (14.5,10) -- (13,10);
\draw [->] (14.5,10) -- (16,10);
\draw (14.5,10) node [below]{$\ell_{\mathrm{R}}^{\text{S}}$};
\draw [red, ultra thick] (-7,-2) -- (7,-2) -- (7,3) -- (-7,3) -- (-7,-2);
\end{tikzpicture}
\caption{Gluing the strip diagrams for SO($2N$) gauge theory with two spinors.}
\label{fig:gluing}
\end{figure}
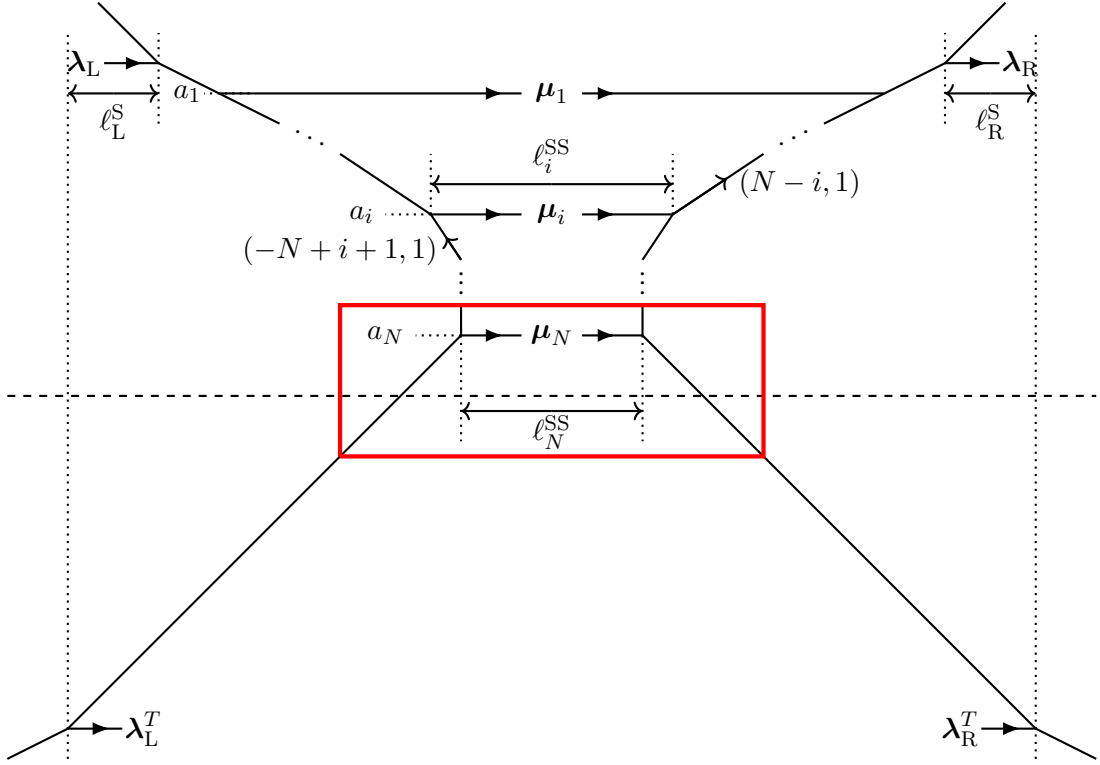

\begin{figure}
\centering
\begin{tikzpicture}[thick]
\begin{scope}[shift={(-4,0)}]
\draw [red, ultra thick] (-3,-1.7) -- (3,-1.7) -- (3,1.2) -- (-3,1.2) -- (-3,-1.7);
\draw [dashed] (-3,0) -- (3,0);
\draw (-1.5,0.8) -- (-1.5,-0.5) -- (-2.5,-1.5);
\draw (1.5,0.8) -- (1.5,-0.5) -- (2.5,-1.5);
\draw (-1.5,-0.5) -- (-0.5,-0.5);
\draw [->, >=Latex] (-1,-0.5) -- (-0.7,-0.5);
\draw (0,-0.5) node {$\boldsymbol{\mu}_N^T$};
\draw (0.5,-0.5) -- (1.5,-0.5);
\draw [->, >=Latex] (1,-0.5) -- (1.1,-0.5);
\draw [->] (0,0.3) -- (-1.5,0.3);
\draw [->] (0,0.3) -- (1.5,0.3);
\draw (0,0.6) node {$\ell_N^{\text{CC}}$};
\draw [dotted] (-1.5,-0.5) -- (-2,-0.5) node [left]{$-a_N$};
\end{scope}
\begin{scope}[shift={(4,0)}]
\draw [red, ultra thick]  (-3.5,-1.7) -- (3,-1.7) -- (3,1.2) -- (-3.5,1.2) -- (-3.5,-1.7);
\draw [dashed] (-3.5,0) -- (3,0);
\draw (-1.5,1) -- (-1.5,0.5) -- (-3.2,-1.2);
\draw (1.5,1) -- (1.5,-0.5) -- (2.5,-1.5);
\draw (-1.5,0.5) -- (-0.5,0.5);
\draw [->, >=Latex] (-1,0.5) -- (-0.7,0.5);
\draw (0,0.5) node {$\boldsymbol{\mu}_N$};
\draw (0,-0.5) node {$\boldsymbol{\mu}_N^T$};
\draw (0.5,-0.5) -- (1.5,-0.5);
\draw [->, >=Latex] (1,-0.5) -- (1.1,-0.5);
\draw [dotted] (-1.5,0.5) -- (-2,0.5) node [left]{$a_N$};
\draw [dotted] (1.5,-0.5) -- (2,-0.5) node [right]{$-a_N$};
\draw [dotted] (-1.5,0.5) -- (-1.5,-1.5);
\draw [dotted] (1.5,-0.5) -- (1.5,-1.5);
\draw [->] (0,-1) -- (-1.5,-1);
\draw [->] (0,-1) -- (1.5,-1);
\draw (0,-1.35) node {$\ell_N^{\text{SC}}$};
\end{scope}
\end{tikzpicture}
\caption{Structure of gluing the strip diagrams around O5-plane for SO($2N$) gauge theory with two conjugate spinors (left) and with one spinor and one conjugate spinor (right). }
\label{fig:gluing-SC-CC}
\end{figure}
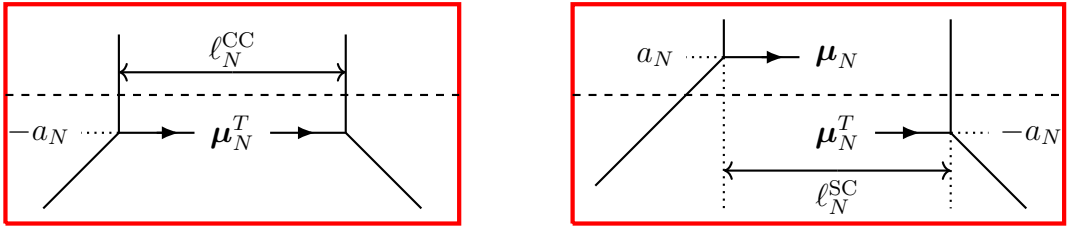

To compute the edge factors in our cases explicitly using the rule described above, we first need to fix the parameterization of the horizontal edges.
The length $\ell_i$ of the $i$-th color brane from the top, 
as given in Figure \ref{fig:gluing}, is computed to be given in terms of the instanton mass $m_0$ and the spinor/conjugate spinor masses $m_{\mathrm{L}}, m_{\mathrm{R}}$ as \cite{Akhond2026} 
\begin{align}\label{eq:param-li}
\ell_i^{\text{SS}} 
&= m_0 - (2^{N-4}-1) (m_{\mathrm{L}}+m_{\mathrm{R}}) + (2N-2i-1)a_i + \sum_{j=1}^{i-1} a_j - \sum_{j=i+1}^{N} a_j 
\cr
\ell_i^{\text{CC}} 
&=\left. \ell_i^{\text{SS}} \right|_{a_N \to -a_N}, 
\qquad
\ell_i^{\text{SC}} 
= \frac12(\ell_i^{\text{SS}}+\ell_i^{\text{CC}}) .
\end{align}
The mass dependence in \eqref{eq:param-li} is understood by considering the decoupling limit of the spinor/conjugate spinors exactly in the same way as the Appendix in \cite{Hayashi:2018lyv}, which discussed for SO$(2N+1)$ gauge theory. 
That is, if we consider the limit where $m_1$ and $m_2$ are large enough, the prepotential reduces to that for pure SO gauge theory, with the instanton mass identified as $m_0^{\text{pure}} = m_0 - 2^{N-5} (m_{\mathrm{L}}+m_{\mathrm{R}})$.  

Also, the length of the remaining horizontal lines at height $a_{\mathrm{S}}^{\mathrm{L}}$ or $a_{\mathrm{C}}^{\mathrm{L}}$ in the left spinor/conjugate spinor strip and the corresponding one at the right spinor/conjugate spinor are given respectively as
\begin{align}\label{eq:l-SC-LR}
&\ell_{\mathrm{L}}^{\mathrm{S}} 
= (N-2) m_{\mathrm{L}} - \frac12(N-4) \sum_{i=1}^N a_i
= (2-N) a_{\mathrm{L}}^S + \sum_{i=1}^N a_i, 
\cr
&\ell_{\mathrm{L}}^{\mathrm{C}} 
= \left. \ell_{\mathrm{L}}^{\mathrm{S}}\right|_{a_N \to - a_N},
\quad 
\ell_{\mathrm{R}}^{\mathrm{S}} 
= \left. \ell_{\mathrm{L}}^{\mathrm{S}}\right|_{m_{\mathrm{L}} \to m_{\mathrm{R}}},
\quad
\ell_{\mathrm{R}}^{\mathrm{C}}  
= \left. \ell_{\mathrm{L}}^{\mathrm{S}}\right|_{\substack{m_{\mathrm{L}} \to m_{\mathrm{R}} \\ a_N \to - a_N}}.
\end{align}
Here, the labels L, R stand for left and right, while S, C stand for spinor and conjugate spinor.

Using all these notations, we find that the topological string partition functions for the case of two spinors, two conjugate spinors, and one spinor plus one conjugate spinor are given by gluing the left strip and the right strip, respectively, as
\begin{align}\label{eq:top}
Z^{\mathrm{SS}} =& \sum_{\{ \boldsymbol{\mu} \}, \boldsymbol{\lambda}_{\mathrm{L}}, \boldsymbol{\lambda}_{\mathrm{R}}}
\left( \prod_{i=1}^{N} \left( - e^{-\ell_i^{\text{SS}}} \right)^{|\mu_i|}
\mathfrak{f}_{\boldsymbol{\mu}_i}^{\, 2N-2i-1} \right)
\cr
&
\qquad\qquad
\times (- e^{\ell_{\mathrm{L}}^{\mathrm{S}}})^{|\boldsymbol{\lambda}_{\mathrm{L}}|}  \mathfrak{f}_{\boldsymbol{\lambda}_{\mathrm{L}}}^{\, 2-N} 
\times (- e^{\ell_{\mathrm{R}}^{\mathrm{S}}})^{|\boldsymbol{\lambda}_{\mathrm{R}}|} \mathfrak{f}_{\boldsymbol{\lambda}_{\mathrm{R}}}^{\, 4-N} 
\times Z_{\text{left}}^{\text{S}} Z_{\text{right}}^{\text{S}},
\cr
Z^{\mathrm{CC}} =& \sum_{\{ \boldsymbol{\mu} \}, \boldsymbol{\lambda}_{\mathrm{L}}, \boldsymbol{\lambda}_{\mathrm{R}}}
\left( \prod_{i=1}^{N-1} \left( - e^{-\ell_i^{\text{CC}}} \right)^{|\mu_i|}
\mathfrak{f}_{\boldsymbol{\mu}_i}^{\, 2N-2i-1} \right)
\times 
\left( - e^{-\ell_N^{\text{CC}}} \right)^{|\mu_N|}
\mathfrak{f}_{\boldsymbol{\mu}_N}
\cr
& \qquad\qquad
\times (- e^{\ell_{\mathrm{L}}^{\mathrm{C}}})^{|\boldsymbol{\lambda}_{\mathrm{L}}|}  \mathfrak{f}_{\boldsymbol{\lambda}_{\mathrm{L}}}^{\, 2-N} 
\times (- e^{\ell_{\mathrm{R}}^{\mathrm{C}}})^{|\boldsymbol{\lambda}_{\mathrm{R}}|}  \mathfrak{f}_{\boldsymbol{\lambda}_{\mathrm{R}}}^{\, 4-N} 
\times
Z_{\text{left}}^{\text{C}} Z_{\text{right}}^{\text{C}},
\cr
Z^{\mathrm{S}\mathrm{C}} =& \sum_{\{ \boldsymbol{\mu} \}, \boldsymbol{\lambda}_{\mathrm{L}}, \boldsymbol{\lambda}_{\mathrm{R}}}
\left( \prod_{i=1}^{N-1} \left( - e^{-\ell_i^{\text{SC}}} \right)^{|\mu_i|}
\mathfrak{f}_{\boldsymbol{\mu}_i}^{\, 2N-2i-1} \right)
\times 
\left( - e^{-\ell_N^{\text{SC}}} \right)^{|\mu_N|} \mathfrak{f}_{\boldsymbol{\mu}_N}
\cr
&
\qquad \qquad
\times (- e^{\ell_{\mathrm{L}}^{\mathrm{S}}})^{|\boldsymbol{\lambda}_{\mathrm{L}}|}  \mathfrak{f}_{\boldsymbol{\lambda}_{\mathrm{L}}}^{\, 2-N} 
\times (- e^{\ell_{\mathrm{R}}^{\mathrm{C}}})^{|\boldsymbol{\lambda}_{\mathrm{R}}|}  \mathfrak{f}_{\boldsymbol{\lambda}_{\mathrm{R}}}^{\, 4-N} 
\times
Z_{\text{left}}^{\text{S}} Z_{\text{right}}^{\text{C}},
\end{align}
where, $Z_{\text{left}}^{\text{S}}$, $Z_{\text{left}}^{\text{C}}$, $Z_{\text{right}}^{\text{S}}$, $Z_{\text{right}}^{\text{C}}$,  are given in \eqref{eq:leftS}, \eqref{eq:leftC}, \eqref{eq:rightS}, \eqref{eq:Zr-C-S}, respectively.

For later convenience, we absorb these non-trivial factors, originated from the contribution from horizontal edges, into the left/right spinor/conjugate spinor strip amplitudes as
\begin{align}
\tilde{Z}_{\text{left}}^{\text{S}} 
:= &
\left( \prod_{i=1}^{N} e^{- \frac12 
\ell_i^{\text{SS}} |\boldsymbol{\mu}_i|} \,
\left( (-1)^{|\boldsymbol{\mu}_i|} \mathfrak{f}_{\boldsymbol{\mu}_i} \right)^{\, N-i} \right)
\times (- e^{-\ell_{\mathrm{L}}^{\mathrm{S}}})^{|\boldsymbol{\lambda}_{\mathrm{L}}|}  \mathfrak{f}_{\boldsymbol{\lambda}_{\mathrm{L}}}^{\, 2-N} 
\times Z_{\text{left}}^{\text{S}} 
\cr
\tilde{Z}_{\text{left}}^{\text{C}} 
:= &
\left( \prod_{i=1}^{N-1} e^{- \frac12 
\ell_i^{\text{CC}} |\boldsymbol{\mu}_i|} \,
\left( (-1)^{|\boldsymbol{\mu}_i|} \mathfrak{f}_{\boldsymbol{\mu}_i} \right)^{\, N-i} \right)
\times (- e^{\ell_{\mathrm{L}}^{\mathrm{C}}})^{|\boldsymbol{\lambda}_{\mathrm{L}}|}  \mathfrak{f}_{\boldsymbol{\lambda}_{\mathrm{L}}}^{\, 2-N} 
\times Z_{\text{left}}^{\text{C}} 
\cr
\tilde{Z}_{\text{right}}^{\text{S}} 
:= &
\left( \prod_{i=1}^{N} 
e^{- \frac12 \ell_i^{\text{SS}} |\boldsymbol{\mu}_i|} \, 
\left( (-1)^{|\boldsymbol{\mu}_i|} \mathfrak{f}_{\boldsymbol{\mu}_i} \right)^{\, N-i-1} \right)
\times (- e^{\ell_{\mathrm{R}}^{\mathrm{S}}})^{|\boldsymbol{\lambda}_{\mathrm{R}}|}  \mathfrak{f}_{\boldsymbol{\lambda}_{\mathrm{R}}}^{\, 4-N} 
\times Z_{\text{right}}^{\text{S}} 
\cr
\tilde{Z}_{\text{right}}^{\text{C}} 
:= &
\left( \prod_{i=1}^{N-1} e^{- \frac12 
\ell_i^{\text{CC}} |\boldsymbol{\mu}_i|}
\,
\left( (-1)^{|\boldsymbol{\mu}_i|} \mathfrak{f}_{\boldsymbol{\mu}_i} \right)^{\, N-i-1} \right)
\cr
& \qquad\qquad
\times e^{-\frac12  \ell_N^{\text{CC}} |\boldsymbol{\mu}_N|} 
(-1)^{|\boldsymbol{\mu}_N|} \mathfrak{f}_{\boldsymbol{\mu}_N}
\times (- e^{\ell_{\mathrm{R}}^{\mathrm{C}}})^{|\boldsymbol{\lambda}_{\mathrm{R}}|}  \mathfrak{f}_{\boldsymbol{\lambda}_{\mathrm{R}}}^{\, 4-N} 
\times Z_{\text{right}}^{\text{C}}  
\end{align}
so that the topological string partition function is written simply as
\begin{align}\label{eq:Top-result}
Z^{\mathrm{SS}} =& \sum_{\{ \boldsymbol{\mu} \}, \boldsymbol{\lambda}_{\mathrm{L}}, \boldsymbol{\lambda}_{\mathrm{R}}}
\tilde{Z}_{\text{left}}^{\text{S}} \tilde{Z}_{\text{right}}^{\text{S}} ,
\cr
Z^{\mathrm{CC}} =& \sum_{\{ \boldsymbol{\mu} \}, \boldsymbol{\lambda}_{\mathrm{L}}, \boldsymbol{\lambda}_{\mathrm{R}}}
\tilde{Z}_{\text{left}}^{\text{C}} \tilde{Z}_{\text{right}}^{\text{C}} ,
\cr
Z^{\mathrm{S}\mathrm{C}} =& \sum_{\{ \boldsymbol{\mu} \}, \boldsymbol{\lambda}_{\mathrm{L}}, \boldsymbol{\lambda}_{\mathrm{R}}}
\tilde{Z}_{\text{left}}^{\text{S}} \tilde{Z}_{\text{right}}^{\text{C}}.
\end{align}
Here, each amplitude is given explicitly as
\begin{align}\label{eq:def-Z-tilde}
\tilde{Z}_{\text{left}}^{\text{S}} 
= &
\left( \prod_{i=1}^{N} e^{- 
\frac12 \left( m_0 - (2^{N-4}-1)(m_{\mathrm{L}}+m_{\mathrm{R}}) + (2N-2i-1)a_i + \sum_{j=1}^{i-1} a_j - \sum_{j=i+1}^{N} a_j \right)
|\boldsymbol{\mu}_i|} \,
\left( (-1)^{\boldsymbol{\mu}_i} \mathfrak{f}_{\boldsymbol{\mu}_i} \right)^{\, N-i} \right)
\cr
&
\times (- e^{ (2-N) a_{\mathrm{L}}^S + \sum_{i=1}^N a_i})^{|\boldsymbol{\lambda}_{\mathrm{L}}|}  \mathfrak{f}_{\boldsymbol{\lambda}_{\mathrm{L}}}^{\, 2-N} 
g^{||\boldsymbol{\lambda}_{\mathrm{L}}^T||^2 } \tilde{Z}_{\boldsymbol{\lambda}_{\mathrm{L}}}^2 
\left(
\prod_{j=1}^{N} 
g^{\frac{||\boldsymbol{\mu}_{j}||^2}{2} } 
\tilde{Z}_{\boldsymbol{\mu}_{j}} 
\right) 
\cr
&
\left(
\prod_{1 \le i < j \le N} R^{-1}_{\boldsymbol{\mu}_i \boldsymbol{\mu}_j^T}\left( e^{-(a_i-a_j)} \right) 
\right)
\left(
\prod_{i=1}^{N}  
R^{-1}_{\boldsymbol{\mu}_i \boldsymbol{\lambda}_{\mathrm{L}}}\left( e^{-(a_i+a_{\mathrm{L}}^{\mathrm{S}})} \right)
\right)
\cr
& 
\left(
\prod_{i=1}^{N}  
R_{\boldsymbol{\lambda}_{\mathrm{L}} \boldsymbol{\mu}_i^T}\left( e^{-(a_{\mathrm{L}}^{\mathrm{S}} - a_i)} \right)
\right) 
R_{\boldsymbol{\lambda}_{\mathrm{L}}\boldsymbol{\lambda}_{\mathrm{L}}}\left( e^{- 2 a_{\mathrm{L}}^{\mathrm{S}}} \right) ,
\cr
\tilde{Z}_{\text{left}}^{\text{C}} 
= &\left. \tilde{Z}_{\text{left}}^{\text{S}}  
\right|_{a_N \to -a_N, \boldsymbol{\mu}_N \to \boldsymbol{\mu}_N^T} \, ,
\cr
\tilde{Z}_{\text{right}}^{\text{S}} 
= 
& 
\left. \tilde{Z}_{\text{left}}^{\text{S}}  
\right|_{m_{\mathrm{L}} \to m_{\mathrm{R}}, \boldsymbol{\lambda}_{\mathrm{L}} \to \boldsymbol{\lambda}_{\mathrm{R}}} 
\, ,
\cr
\tilde{Z}_{\text{right}}^{\text{C}} 
=
& 
\left. \tilde{Z}_{\text{right}}^{\text{S}}  
\right|_{a_N \to -a_N, \boldsymbol{\mu}_N \to \boldsymbol{\mu}_N^T} 
=
\left. \tilde{Z}_{\text{left}}^{\text{S}}  
\right|_{a_N \to -a_N, m_{\mathrm{L}} \to m_{\mathrm{R}}, \boldsymbol{\mu}_N \to \boldsymbol{\mu}_N^T, \boldsymbol{\lambda}_{\mathrm{L}} \to \boldsymbol{\lambda}_{\mathrm{R}}} \, ,
\end{align}
with 
$
a_{\mathrm{L}}^{\mathrm{S}} = \frac12 \sum_{i=1}^N a_i - m_{\mathrm{L}}, 
$
as given in \eqref{eq:param-b}.

We observe that
\begin{align}\label{eq:SS-CC}
Z^{\mathrm{CC}} = \left. Z^{\mathrm{SS}} \right|_{a_N \to - a_N}. 
\end{align}
Thus, it is enough to focus only on $Z^{\mathrm{SS}}$ and $Z^{\mathrm{SC}}$ in the following.

\subsection{Perturbative part}

The unrefined topological string partition function is expected to include both the perturbative part and the instanton part of the corresponding Nekrasov partition function. In this subsection, we check that the expected perturbative part is reproduced from the topological string partition function obtained in the previous subsection. 

The perturbative part is given by the weak coupling limit where the gauge coupling constant vanishes, which means the instanton mass goes to infinity $m_0 \to \infty$. Due to the factor of the form $e^{-\frac12 (m_0 + \cdots) |\boldsymbol{\mu}_i|}$ in \eqref{eq:def-Z-tilde}, only the term with Young diagram $\boldsymbol{\mu}_i = \varnothing$ contribute in \eqref{eq:Top-result}, which factorizes the partition function into the contribution from each strip in the following form
\begin{align}\label{eq:SS-SC-m0-inf}
\lim_{m_0 \to \infty} Z^{\text{SS}}
 = Z_{\text{left, pert}}^{\text{S}}  
Z_{\text{right, pert}}^{\text{S}} \, ,
\quad
\lim_{m_0 \to \infty} Z^{\text{SC}}
 = Z_{\text{left, pert}}^{\text{S}} 
Z_{\text{right, pert}}^{\text{C}} .
\end{align}
where we put
\begin{align}\label{eq:pert-half}
Z_{\text{left, pert}}^{\text{S}} 
&:=
\sum_{\boldsymbol{\lambda}_{\mathrm{L}}}
 \left. \tilde{Z}_{\text{left}}^{\text{S}} \right|_{\boldsymbol{\mu}_i = \varnothing},
\cr
Z_{\text{right, pert}}^{\text{S}} 
&:= 
\sum_{\boldsymbol{\lambda}_{\mathrm{R}}}
 \left. 
 \tilde{Z}_{\text{right}}^{\text{S}} \right|_{\boldsymbol{\mu}_i = \varnothing}, 
\cr
Z_{\text{right, pert}}^{\text{C}} 
&:= 
\sum_{\boldsymbol{\lambda}_{\mathrm{R}}}
 \left. 
 \tilde{Z}_{\text{right}}^{\text{C}} \right|_{\boldsymbol{\mu}_i = \varnothing}. 
\end{align}

We find that the second quantity in \eqref{eq:pert-half} is essentially the same as the first one: 
\begin{align}\label{eq:pert-right-S}
Z_{\text{right, pert}}^{\text{S}} 
= \left. Z_{\text{left, pert}}^{\text{S}} \right|_{m_{\mathrm{L}} \to m_{\mathrm{R}}} ,
\end{align}
where we have also used 
the relation between $\ell_{\mathrm{L}}^{\mathrm{S}}$ and $\ell_{\mathrm{R}}^{\mathrm{S}}$ given in \eqref{eq:l-SC-LR}. Also, from the relation between $Z_{\text{right}}^{\text{S}}$ and $Z_{\text{right}}^{\text{C}}$ given in \eqref{eq:Zr-C-S} as well as the relation between $\ell_{\mathrm{R}}^{\mathrm{S}}$ and $\ell_{\mathrm{R}}^{\mathrm{C}}$ given in \eqref{eq:l-SC-LR}, we also find that the third quantity in \eqref{eq:pert-half} is also simply related to the second one as
\begin{align}\label{eq:pert-right-C}
Z_{\text{right, pert}}^{\text{C}} 
= \left. Z_{\text{right, pert}}^{\text{S}} \right|_{a_N \to -a_N} .
\end{align}
Thus, it is enough to consider only the first quantity in \eqref{eq:pert-half}. 
From \eqref{eq:leftS}, \eqref{eq:def-framing} and \eqref{eq:l-SC-LR}, the first one in \eqref{eq:pert-half} is given explicitly as
\begin{align}\label{eq:left-pert-1}
Z_{\text{left, pert}}^{\text{S}} 
=& \sum_{\boldsymbol{\lambda} }
\left( - e^{-(2 a_{\mathrm{L}}^S + \sum_{i=1}^N (a_i-a_{\mathrm{L}}^S))}  
\right)^{|\boldsymbol{\lambda}|}
\mathfrak{f}_{\boldsymbol{\lambda}_{\mathrm{L}}}^{\, 2-N} 
g^{||\boldsymbol{\lambda}^T||^2 } \tilde{Z}_{\boldsymbol{\lambda}}^2 
\cr
& 
\left(
\prod_{1 \le i < j \le N} R^{-1}_{\varnothing \varnothing}\left( e^{-(a_i-a_j)} \right) 
\right)
\left(
\prod_{i=1}^{N}  
R^{-1}_{\varnothing \boldsymbol{\lambda} }\left( e^{-(a_i+a_{\mathrm{L}}^{\mathrm{S}})} \right)
\right)
\cr
&
\left(
\prod_{i=1}^{N}  
R_{\boldsymbol{\lambda} \varnothing}\left( e^{-(a_{\mathrm{L}}^{\mathrm{S}} - a_i)} \right)
\right) 
R_{\boldsymbol{\lambda} \boldsymbol{\lambda} }\left( e^{- 2 a_{\mathrm{L}}^{\mathrm{S}}} \right),
\end{align}
where we relabeled the dummy index $\boldsymbol{\lambda}_{\mathrm{L}} \to \boldsymbol{\lambda}$ for simplicity.

In order to proceed the computation, it is convenient to decompose the building block $R_{\boldsymbol{\mu} \boldsymbol{\nu}} (Q)$ into the simple infinite product part and non-trivial finite part as
\begin{align}\label{eq:R-RN}
R_{\boldsymbol{\lambda} \boldsymbol{\nu}} (Q)
= R_{\varnothing \varnothing} (Q)
\boldsymbol{N}_{\boldsymbol{\lambda}^T \boldsymbol{\nu}} (Q).
\end{align}
Here, the simple infinite product part can be rewritten in the Plethystic exponential form as
\begin{align}\label{eq:R-PE}
R_{\varnothing \varnothing}  (Q)
= & 
\prod_{i=1}^{\infty} \prod_{j=1}^{\infty} \left(1 - Q g^{i+j- 1} \right)
= \text{PE} \left( - \frac{g}{(1-g)^2} Q \right).
\end{align}
The non-trivial finite part is the Nekrasov factor given as
\begin{align}\label{def:Nekra}
\boldsymbol{N}_{\boldsymbol{\lambda}\boldsymbol{\nu}}(Q):=
\prod_{(i,j)\in \boldsymbol{\lambda}} \left(1-Qg^{1-i-j+\lambda_i+\nu^T_j}\right)\prod_{(m,n)\in \boldsymbol{\nu}}
\left(1-Qg^{m+n-\lambda^T_n-\nu_m-1}\right).
\end{align}
The factor $\tilde{Z}_{\boldsymbol{\lambda}}^2$ can also be rewritten in terms of this Nekrasov factor as
\begin{align}\label{eq:Z-tilde-lambda}
\tilde{Z}_{\boldsymbol{\lambda}}^2 
= g^{ - ||\boldsymbol{\lambda}^T||^2 } 
(-1)^{|\boldsymbol{\lambda}|} \mathfrak{f}_{\boldsymbol{\lambda}}
\boldsymbol{N}^{-1}_{\boldsymbol{\lambda}\boldsymbol{\lambda}}(1)
= g^{ - ||\boldsymbol{\lambda}||^2 } 
(-1)^{|\boldsymbol{\lambda}|} \mathfrak{f}_{\boldsymbol{\lambda}}^{\, -1}
\boldsymbol{N}^{-1}_{\boldsymbol{\lambda}\boldsymbol{\lambda}}(1).
\end{align}
Also, by using the identity
\begin{align}
\boldsymbol{N}_{\boldsymbol{\lambda} \boldsymbol{\nu}}(Q)
&=
Q^{|\boldsymbol{\lambda}|+|\boldsymbol{\nu}|}
\mathfrak{f}_{\boldsymbol{\lambda}}^{-1}
\mathfrak{f}_{\boldsymbol{\nu}} \boldsymbol{N}_{\boldsymbol{\nu} \boldsymbol{\lambda} } \left( Q^{-1} \right)\ ,
\end{align}
to the factor $\boldsymbol{N}_{\boldsymbol{\lambda}^T \varnothing} \left( e^{-(a_{\mathrm{L}}^{\mathrm{S}} - a_i)} \right)$, which appears in the process of computation, the expression \eqref{eq:left-pert-1} is rewritten as
\begin{align}\label{eq:pert-intermed2}
Z_{\text{left, pert}}^{\text{S}} 
=& 
\text{PE} 
\left[ \frac{g}{(1-g)^2} \left( 
\sum_{1 \le i <j \le N} e^{-(a_i-a_j)} 
+ \sum_{i=1}^N (e^{-a_i} - e^{a_i}) e^{-a_{\mathrm{L}}^{\mathrm{S}}}
- e^{-2a_{\mathrm{L}}^{\mathrm{S}}}
\right) \right]
\cr
&\sum_{\boldsymbol{\lambda} }
e^{- 2 a_{\mathrm{L}}^S |\boldsymbol{\lambda}|} 
\mathfrak{f}_{\boldsymbol{\lambda}}^{\, 3}
\frac{
\boldsymbol{N}_{\boldsymbol{\lambda}^T \boldsymbol{\lambda} }\left( e^{- 2 a_{\mathrm{L}}^{\mathrm{S}}} \right)
}{
\boldsymbol{N}_{\boldsymbol{\lambda}\boldsymbol{\lambda}} (1) 
} 
\prod_{i=1}^{N}  
\frac{
\boldsymbol{N}_{\varnothing \boldsymbol{\lambda}^T }
\left( e^{-(a_i - a_{\mathrm{L}}^{\mathrm{S}})} \right)
}{ 
\boldsymbol{N}_{\varnothing \boldsymbol{\lambda} }
\left( e^{-(a_i + a_{\mathrm{L}}^{\mathrm{S}})} \right)
} .
\end{align}

In order to compare this result with the known perturbative result, 
we need to perform the summation for the Young diagram $\boldsymbol{\lambda}$ exactly. We conjecture the following identity:
\begin{align}\label{eq:conjecture-id}
&\sum_{\boldsymbol{\lambda}}
Q^{|\boldsymbol{\lambda}|} 
\mathfrak{f}_{\boldsymbol{\boldsymbol{\lambda}}}^{\, 3}
\frac{\boldsymbol{N}_{\boldsymbol{\lambda}^T \boldsymbol{\lambda}}\left( Q \right)}{\boldsymbol{N}_{\boldsymbol{\lambda}\boldsymbol{\lambda}}\left( 1 \right)}
\prod_{i=1}^N \frac{\boldsymbol{N}_{\varnothing \boldsymbol{\lambda}^T}( A_i)}{\boldsymbol{N}_{\varnothing \boldsymbol{\lambda}}\left( A_i Q \right)} 
\cr
&= \mathrm{PE}
\left( 
\frac{g \, Q}{(1-g)^2} \sum_{k=0}^{N} (-1)^k
\sum_{1 \le i_1 < \cdots < i_k \le N}  
A_{i_1} \cdots A_{i_k}
\right).
\end{align}
The $N=0$ case was noted and confirmed \cite{Hayashi:2020hhb} up to the order $\mathcal{O}(Q^{11})$. We also confirm this identity up to the order $\mathcal{O}(Q^{8})$ for $N=1,2,3,4$, respectively. 
Although no proof of this identity is currently available to the best of the authors' knowledge, these checks imply that the conjecture is plausible. 
Using this identity by identifying
\begin{align}\label{eq:paramAQ}
A_i = e^{-(a_i-a_{\mathrm{L}}^{\mathrm{S}})}, \qquad Q = e^{-2a_{\mathrm{L}}^{\mathrm{S}}},
\end{align}
we find \eqref{eq:pert-intermed2} becomes
\begin{align}\label{eq:S-lert-pert-PE}
Z_{\text{left, pert}}^{\text{S}} 
&= 
\text{PE} \left[ \frac{g}{(1-g)^2} 
\left(
\sum_{1 \le i <j \le N} e^{-(a_i-a_j)}
- \sum_{i=1}^N e^{ - ( a_{\mathrm{L}}^{\mathrm{S}} - a_i )}
\right.
\right.
\cr 
& \left. \left. \qquad \qquad \qquad
+ \sum_{k=2}^{N} (-1)^k
\sum_{1 \le i_1 < \cdots < i_k \le N}  
e^{-( \sum_{j=1}^{k} a_{i_j} - (k-2) a_{\mathrm{L}}^{\mathrm{S}})}
\right) \right].
\end{align}
We note that the contributions from the terms corresponding to $k=0, 1$ in \eqref{eq:conjecture-id} are cancelled with some of the terms in the first line in \eqref{eq:pert-intermed2}.

In order to check that it reproduces the expected perturbative part, we use the following replacement 
\begin{align}\label{eq:replace}
\text{PE} \left( - \frac{g}{(1-g)^2} Q \right)
\to 
\text{PE} \left( - \frac{g}{(1-g)^2} Q^{-1} \right),
\end{align}
which often appears in the context of flop invariance \cite{Konishi:2006ev, Taki:2008hb} of the topological string partition function and is often used to check that the topological string partition function computed via topological vertex formalism reproduces the perturbative part of the corresponding gauge theory \cite{Iqbal:2004ne, Mitev:2014jza, Kim:2015jba, Hayashi:2016abm, Hayashi:2018bkd}. This is understood as an analytic continuation from the region $|Q| \ll 1$ to $|Q| \gg 1$ with the zeta function regularization, as discussed in \cite{Iqbal:2004ne}.
\footnote{More rigorously, we also need to include the polynomial contribution of the classical part of the topological string partition function, which we omit in this paper.}

With this replacement, we find \eqref{eq:S-lert-pert-PE} becomes
\begin{align}\label{eq:S-left-pert-to}
Z_{\text{left, pert}}^{\text{S}} 
& \to 
\text{PE} \left[ \frac{g}{(1-g)^2} 
\left( - \frac{N}{2}
+ \frac12 \sum_{i=1}^N \sum_{j=1}^N e^{-(a_i-a_j)}
- \sum_{i=1}^N e^{ - ( a_i - a_{\mathrm{L}}^{\mathrm{S}} )}
\right.
\right.
\cr 
& \qquad \qquad \qquad
+ \frac12 \sum_{1 \le i < j \le N}  
\left(
e^{ - (a_i + a_j) } + e^{a_i + a_j} 
\right)
\cr
& \left. \left. \qquad \qquad \qquad
+ \sum_{k=3}^{N} (-1)^k
\sum_{1 \le i_1 < \cdots < i_k \le N}  
e^{-( \sum_{j=1}^{k} a_{i_j} - (k-2) a_{\mathrm{L}}^{\mathrm{S}})}
\right) \right]
\cr
& = 
\text{PE} \left[ \frac{g}{(1-g)^2} 
\left( \frac12 \chi_{\text{Adj}} (a) - e^{-m_{\mathrm{L}}} \chi_{\text{S}} (a)\right)
\right] \times Z_{\text{extra}},
\end{align}
where we put
\begin{align}\label{eq:extra}
Z_{\text{extra}} := \text{PE} 
\left[ \frac{g}{(1-g)^2} 
\left( - \frac12 (N-1) + \delta_{N,4} e^{-2m_{\mathrm{L}}} \right)
\right].
\end{align}
Note that the first line in \eqref{eq:S-left-pert-to} comes from the first line in \eqref{eq:S-lert-pert-PE} while the second line in \eqref{eq:S-left-pert-to} originates from the $k=2$ contribution in the second line in \eqref{eq:S-lert-pert-PE}.
Here, $\chi_{\text{Adj}} (a)$ and $\chi_{\text{S}} (a)$ are characters of adjoint representation and spinor representation of SO($2N$), respectively. These are given explicitly as follows:
\begin{align}
\chi_{\text{Adj}} (a) 
:= & \sum_{i=1}^N \sum_{j=1}^N e^{-(a_i-a_j)} 
+ \sum_{1 \le i < j \le N} \left( e^{-(a_i+a_j)} + e^{a_i+a_j} \right),
\cr
\chi_{\text{S}} (a)
:= &
\frac12 \left( \prod_{i=1}^N (e^{- \frac{a_i}{2} } + e^{\frac{a_i}{2}  }) - \prod_{i=1}^N (e^{- \frac{a_i}{2} } - e^{\frac{a_i}{2}  }) \right),
\cr
\chi_{\text{C}} (a)
:= &
\frac12 \left( \prod_{i=1}^N (e^{- \frac{a_i}{2} } + e^{\frac{a_i}{2}  }) + \prod_{i=1}^N (e^{- \frac{a_i}{2} } - e^{\frac{a_i}{2}  }) \right),
\end{align}
where we also wrote the character for the conjugate spinor representation for later convenience.
The factor $Z_{\text{extra}}$ given in \eqref{eq:extra} is the factor that does not depend on the Coulomb moduli $a_i$.
As discussed in various works including \cite{Bao:2013pwa, Hayashi:2013qwa}, such an extra factor should be discarded when compared with Nekrasov partition function.

Combining the results above with \eqref{eq:SS-SC-m0-inf}, \eqref{eq:pert-right-S}, and \eqref{eq:pert-right-C}, we find that our partition function in the limit $m_0 \to \infty$ gives
\begin{align}\label{eq:limpert}
&\lim_{m_0 \to \infty} Z^{\text{SS}}
\to \text{PE} \left[ \frac{g}{(1-g)^2} 
\left( \chi_{\text{Adj}} (a) - e^{-m_{\mathrm{L}}} \chi_{\text{S}} (a) - e^{-m_{\mathrm{R}}} \chi_{\text{S}} (a)\right)
\right],
\cr
&\lim_{m_0 \to \infty} Z^{\text{SC}}
\to \text{PE} \left[ \frac{g}{(1-g)^2} 
\left( \chi_{\text{Adj}} (a) - e^{-m_{\mathrm{L}}} \chi_{\text{S}} (a) - e^{-m_{\mathrm{R}}} \chi_{\text{C}} (a)\right)
\right],
\end{align}
up to the replacement \eqref{eq:replace} and up to the extra factors of the form \eqref{eq:extra}.
Thus, we have confirmed that our unrefined topological string partition function correctly reproduces the expected perturbative part of the corresponding gauge theories.

\subsection{Instanton partition function}

In this subsection, we discuss the instanton part of the partition function. This part can be extracted by normalizing the partition function by the $m_0 \to \infty$ limit, from which we obtained the perturbative part in the previous subsection. Thus, we define
\begin{align}\label{eq:norm-p}
\hat{Z}^{\text{SS}} 
= &\frac{Z^{\text{SS}}}{\displaystyle \lim_{m_0 \to \infty} Z^{\text{SS}}}
= 
 \frac{\sum_{\{ \boldsymbol{\mu} \}, \boldsymbol{\lambda}_{\mathrm{L}}, \boldsymbol{\lambda}_{\mathrm{R}}} \tilde{Z}_{\text{left}}^{\text{S}} \tilde{Z}_{\text{right}}^{\text{S}}}{Z_{\text{left, pert}}^{\text{S}}  
Z_{\text{right, pert}}^{\text{S}}} 
= \sum_{\{ \boldsymbol{\mu} \}} 
\left( 
\sum_{\boldsymbol{\lambda}_{\mathrm{L}}}
\hat{Z}_{\text{left}}^{\text{S}} 
\right)
\left( 
\sum_{\boldsymbol{\lambda}_{\mathrm{R}}}
\hat{Z}_{\text{right}}^{\text{S}} 
\right),
\cr
\hat{Z}^{\text{SC}} 
= & 
\frac{Z^{\text{SC}}}{\displaystyle\lim_{m_0 \to \infty} Z^{\text{SC}}} \, 
= 
 \frac{\sum_{\{ \boldsymbol{\mu} \}, \boldsymbol{\lambda}_{\mathrm{L}}, \boldsymbol{\lambda}_{\mathrm{R}}} \tilde{Z}_{\text{left}}^{\text{S}} \tilde{Z}_{\text{right}}^{\text{C}}}{Z_{\text{left, pert}}^{\text{S}}  
Z_{\text{right, pert}}^{\text{C}}} 
= \sum_{\{ \boldsymbol{\mu} \}} 
\left( 
\sum_{\boldsymbol{\lambda}_{\mathrm{L}}}
\hat{Z}_{\text{left}}^{\text{S}} 
\right)
\left( 
\sum_{\boldsymbol{\lambda}_{\mathrm{R}}}
\hat{Z}_{\text{right}}^{\text{C}} 
\right) ,
\end{align}
where we define the normalized strip
\begin{align}\label{eq:def-norm-strip-LS}
\hat{Z}_{\text{left}}^{\text{S}} := \frac{\tilde{Z}_{\text{left}}^{\text{S}}}{Z_{\text{left, pert}}^{\text{S}}}, 
\quad
\hat{Z}_{\text{right}}^{\text{S}} := \frac{\tilde{Z}_{\text{right}}^{\text{S}}}{Z_{\text{right, pert}}^{\text{S}}}, 
\quad
\hat{Z}_{\text{right}}^{\text{C}} := \frac{\tilde{Z}_{\text{right}}^{\text{C}}}{Z_{\text{right, pert}}^{\text{C}}}.
\end{align}

The instanton partition function is obtained from this normalized partition function by removing the extra factors 
\begin{align}\label{eq:InstantonP}
Z_{\text{inst}}^{\text{SS}} = \frac{\hat{Z}^{\text{SS}}}{\hat{Z}_{\text{extra}}^{\text{SS}}}, 
\qquad 
Z_{\text{inst}}^{\text{SC}} = \frac{\hat{Z}^{\text{SC}}}{\hat{Z}_{\text{extra}}^{\text{SC}}},
\end{align}
where the extra factor is the factor that does not depend on the Coulomb moduli. Typically, it comes from the contribution of the parallel external lines \cite{Bergman:2013ala, Bergman:2013aca, Bao:2013pwa, Hayashi:2013qwa}. In our case, there are non-trivial contributions only for SO(4).
\begin{align}\label{eq:ExtraF}
Z_{\text{extra}}^{\text{SS}} 
= Z_{\text{extra}}^{\text{SC}}
=
\left\{ 
\begin{array}{cl}
\text{PE} \left( \frac{g}{(1-g)^2} e^{-(m_0-\frac12(m_{\mathrm{L}}+m_{\mathrm{R}}))} \right) & \text{ for } N=2 
\\
1 & \text{ for } N=3,4
\end{array}
\right.
\end{align}

In the following, we rewrite the normalized strip amplitude \eqref{eq:def-norm-strip-LS} by decomposing the building block $R_{\boldsymbol{\mu} \boldsymbol{\nu}} (Q)$ included in the strip amplitudes \eqref{eq:def-Z-tilde} according to the Plethystic exponential and the Nekrasov factor as in \eqref{eq:R-RN}, \eqref{eq:R-PE}, analogous to the computation in the previous subsection. Then, divide by the $m_0 \to \infty$ limit, which is given by \eqref{eq:S-lert-pert-PE}. Writing explicitly, we find
\begin{align}
\hat{Z}_{\text{left}}^{\text{S}}
= &\text{PE} \left[ - \frac{g}{(1-g)^2} 
\sum_{k=0}^{N} (-1)^k
e^{(k-2) a_{\mathrm{L}}^{\mathrm{S}}} 
\sum_{1 \le i_1 < \cdots < i_k \le N}  
e^{-\sum_{j=1}^{k} a_{i_j} }
\right]
\cr
&
\prod_{i=1}^{N}  
e^{- 
\frac12 \left( m_0 - (2^{N-4}-1) (m_{\mathrm{L}}+m_{\mathrm{R}}) + (2N-2i-1)a_i + \sum_{j=1}^{i-1} a_j - \sum_{j=i+1}^{N} a_j \right)
|\boldsymbol{\mu}_i|} \,
\left( (-1)^{|\boldsymbol{\mu}_i|}
\mathfrak{f}_{\boldsymbol{\mu}_i}
\right)^{\, N-i-\frac12} 
\cr
&
e^{-((2-N) a_{\mathrm{L}}^S + \sum_{i=1}^N a_i))|\boldsymbol{\lambda}_{\mathrm{L}}|} 
\mathfrak{f}_{\boldsymbol{\lambda}_{\mathrm{L}}}^{\, 3-N} 
\left( 
\prod_{i=1}^{N}  
\boldsymbol{N}^{-\frac12}_{\boldsymbol{\mu}_i \boldsymbol{\mu}_i}(1)
\right)
\boldsymbol{N}^{-1}_{\boldsymbol{\lambda}_{\mathrm{L}}\boldsymbol{\lambda}_{\mathrm{L}}}(1)
\cr
&
\left(
\prod_{1 \le i < j \le N} \boldsymbol{N}^{-1}_{\boldsymbol{\mu}^T_i \boldsymbol{\mu}_j^T}\left( e^{-(a_i-a_j)} \right) 
\right)
\left(
\prod_{i=1}^{N}  
\boldsymbol{N}^{-1}_{\boldsymbol{\mu}^T_i \boldsymbol{\lambda}_{\mathrm{L}}}\left( e^{-(a_i+a_{\mathrm{L}}^{\mathrm{S}})} \right)
\right)
\cr
&
\left(
\prod_{i=1}^{N}  
\boldsymbol{N}_{\boldsymbol{\lambda}_{\mathrm{L}}^T \boldsymbol{\mu}_i^T}\left( e^{-(a_{\mathrm{L}}^{\mathrm{S}} - a_i)} \right)
\right)  
\boldsymbol{N}_{\boldsymbol{\lambda}_{\mathrm{L}}^T \boldsymbol{\lambda}_{\mathrm{L}}}\left( e^{- 2 a_{\mathrm{L}}^{\mathrm{S}}} \right).
\end{align}

Here, we introduce a Nekrasov factor given in terms of the sinh functions as
\begin{align}\label{eq:new-nek}
\tilde{\boldsymbol{n}}_{\boldsymbol{\lambda} \boldsymbol{\nu}} (a) 
:= &
\prod_{(i,j) \in \boldsymbol{\lambda}} 
2 \sinh \frac12 \left( 
a + \hbar (1-i-j+\lambda_i+\nu^T_j)
\right)
\cr
& \qquad  \times
\prod_{(m,n) \in \boldsymbol{\nu}} 
2 \sinh \frac12 \left( 
a + \hbar (m+n-\lambda^T_n-\nu_m-1)
\right),
\end{align}
which is related to the original Nekrasov factor \eqref{def:Nekra} as 
\begin{align}
\boldsymbol{N}_{\boldsymbol{\lambda} \boldsymbol{\nu}} (e^{-a})
= &
\left( (-e^{a})^{|\boldsymbol{\lambda}|}\mathfrak{f}_{\boldsymbol{\lambda}} \right)^{\, - \frac12} 
\left( (-e^{-a})^{|\boldsymbol{\nu}|} \mathfrak{f}_{\boldsymbol{\nu}} \right) ^{\, \frac12}
\tilde{\boldsymbol{n}}_{\boldsymbol{\lambda} \boldsymbol{\nu}} (a)
\end{align}
and satisfy the identities
\begin{align}
\tilde{\boldsymbol{n}}_{\boldsymbol{\lambda} \boldsymbol{\nu}} (a) 
= (-1)^{|\boldsymbol{\lambda}| - |\boldsymbol{\nu}|} \tilde{\boldsymbol{n}}_{\boldsymbol{\nu} \boldsymbol{\lambda}} (-a) .
\end{align}
We also note that this Nekrasov factor can be constructed as a geometric mean of the original Nekrasov factor as
\begin{align}\label{eq:geom-mean-nN}
\tilde{\boldsymbol{n}}_{\boldsymbol{\lambda} \boldsymbol{\nu}} (a) 
= (-1)^{\frac12(-|\boldsymbol{\lambda}| + |\boldsymbol{\nu}|)} 
\boldsymbol{N}^{\frac12}_{\boldsymbol{\lambda} \boldsymbol{\nu}} (e^{-a})
\boldsymbol{N}^{\frac12}_{\boldsymbol{\nu} \boldsymbol{\lambda} } (e^{a}).
\end{align}

Rewriting the strip amplitude in terms of this Nekrasov factors, we have 
\begin{align}\label{eq:inst-strip}
\hat{Z}_{\text{left}}^{\mathrm{S}}
= &\text{PE} \left[ - \frac{g}{(1-g)^2} 
\sum_{k=0}^{N} (-1)^k
e^{(k-2) a_{\mathrm{L}}^{\mathrm{S}}} 
\sum_{1 \le i_1 < \cdots < i_k \le N}  
e^{-\sum_{j=1}^{k} a_{i_j} }
\right]
\cr
&
(- 1 )^{|\boldsymbol{\lambda}_{\mathrm{L}}|}
\prod_{i=1}^N 
\left( - e^{ - \left( m_0 - (2^{N-4}-1) (m_{\mathrm{L}}+m_{\mathrm{R}})  \right) } \right) ^{\frac12 |\boldsymbol{\mu}_i|}
\cr
& 
\left( (-e^{-a_{\mathrm{L}}^S})^{|\boldsymbol{\lambda}_{\mathrm{L}}|}
\mathfrak{f}_{\boldsymbol{\lambda}_{\mathrm{L}}} \right)^{\, 4-N}
\left( \prod_{i=1}^N (-e^{-a_i})^{|\boldsymbol{\mu}_i|} \mathfrak{f}_{\boldsymbol{\mu}_i} 
 \right) ^{\,\frac12 (N-4)} 
\cr
&
\left(
\prod_{i=1}^N \prod_{j=1}^N \boldsymbol{n}^{-\frac12}_{\boldsymbol{\mu}^T_i \boldsymbol{\mu}_j^T}\left( a_i-a_j \right) 
\right) 
\cr
& 
\left(
\prod_{i=1}^{N}  
\boldsymbol{n}^{-1}_{\boldsymbol{\mu}^T_i \boldsymbol{\lambda}_{\mathrm{L}}}\left( a_i+a_{\mathrm{L}}^{\mathrm{S}} \right)
\right)
\left(
\prod_{i=1}^{N}  
\boldsymbol{n}_{\boldsymbol{\lambda}_{\mathrm{L}}^T \boldsymbol{\mu}_i^T}\left( a_{\mathrm{L}}^{\mathrm{S}} - a_i \right)
\right)  
\boldsymbol{n}_{\boldsymbol{\lambda}_{\mathrm{L}}^T \boldsymbol{\lambda}_{\mathrm{L}}}\left( 2 a_{\mathrm{L}}^{\mathrm{S}} \right)
\boldsymbol{n}^{-1}_{\boldsymbol{\lambda}_{\mathrm{L}} \boldsymbol{\lambda}_{\mathrm{L}}}\left( 0 \right)
.
\cr
\end{align}
The other types of strip amplitudes are given by
\begin{align}\label{eq:inst-strip-others}
\hat{Z}_{\text{left}}^{\text{C}} 
= &\left. \hat{Z}_{\text{left}}^{\text{S}}  
\right|_{a_N \to -a_N, \boldsymbol{\mu}_N \to \boldsymbol{\mu}_N^T} \, ,
\cr
\hat{Z}_{\text{right}}^{\text{S}} 
= 
& 
\left. \hat{Z}_{\text{left}}^{\text{S}}  
\right|_{m_{\mathrm{L}} \to m_{\mathrm{R}}, \boldsymbol{\lambda}_{\mathrm{L}} \to \boldsymbol{\lambda}_{\mathrm{R}}} 
\, ,
\cr
\hat{Z}_{\text{right}}^{\text{C}} 
=
& 
\left. \hat{Z}_{\text{right}}^{\text{S}}  
\right|_{a_N \to -a_N, \boldsymbol{\mu}_N \to \boldsymbol{\mu}_N^T} 
=
\left. \hat{Z}_{\text{left}}^{\text{S}}  
\right|_{a_N \to -a_N, m_{\mathrm{L}} \to m_{\mathrm{R}}, \boldsymbol{\mu}_N \to \boldsymbol{\mu}_N^T, \boldsymbol{\lambda}_{\mathrm{L}} \to \boldsymbol{\lambda}_{\mathrm{R}}} \, .
\end{align}

In summary, the explicit expressions for the instanton partition functions \eqref{eq:InstantonP} are obtained by being combined with \eqref{eq:norm-p}, \eqref{eq:ExtraF}, \eqref{eq:inst-strip}, \eqref{eq:inst-strip-others}. 

Here, we comment on the structure of our formula for the instanton partition function. From the parameter dependence on $m_0$, we find that the instanton number is given by $\sum_{i=1}^N |\boldsymbol{\mu}_i|$. The other Young diagrams $\boldsymbol{\lambda}_{\mathrm{L}}, \boldsymbol{\lambda}_{\mathrm{R}}$ are related to the power of exponentiated masse parameters for the spinors or conjugate spinors, analogous to the structure discussed in \cite{Chen:2023smd}, in which the instanton partition function is given as the positive power expansion of $e^{-m_{\mathrm{L}}}$. On the contrary, our formula is understood as a positive power expansion of $Q:=e^{-2a_{\mathrm{L}}^{\text{S}}} = e^{-(\sum_{i=1}^N a_i - 2m_{\mathrm{L}})}$ while keeping $A_i:=e^{-(a_i-a_{\mathrm{L}})} = e^{-(a_i-\frac12 \sum_j a_j + m_{\mathrm{L}})}$ as order 1, which parametriztion is introduced in \eqref{eq:paramAQ}. Thus, they are expected to be the expansion of an identical formula in different parameters. If we can sum over the Young diagram $\boldsymbol{\lambda}_{\mathrm{L}}, \boldsymbol{\lambda}_{\mathrm{R}}$ exactly, we will be able to compare them.

\section{Seiberg-Witten curve via thermodynamic limit}\label{sec:thermo}

In this section, we take the thermodynamic limit $\hbar \to 0$ for the topological string partition function or instanton partition function to derive the corresponding Seiberg-Witten curve, based on the methods developed in \cite{Nekrasov:2003rj, Nekrasov:2012xe}.

\subsection{Profile function}

In this subsection, we introduce profile function \cite{Nekrasov:2003rj}
\begin{align}\label{eq:def-profile}
f_{\boldsymbol{\lambda}}(x) = |x| + \sum_{i=1}^{\infty} 
\Bigl[ |x +\hbar (\lambda_i-i + 1)| - |x +\hbar ( \lambda_i-i)| - |x+ \hbar(i-1)| + |x+\hbar i|  \Bigr],
\end{align}
which function is determined when a Young diagram $\boldsymbol{\lambda}$ is given. This is related to the profile function corresponding to the transposed Young diagram as
\begin{align}
f_{\boldsymbol{\lambda}^T}(x) = f_{\boldsymbol{\lambda}}(-x)
\end{align}
The integral of this function 
multiplied by a monomial $x^n$ is given as
\begin{align}\label{eq:prof-integral}
\int^{\infty}_{-\infty} f_{\boldsymbol{\lambda}}''(x) dx
&= 2 ,
\cr
\int^{\infty}_{-\infty} x f_{\boldsymbol{\lambda}}''(x) dx
&= 0 ,
\cr
\int^{\infty}_{-\infty} x^2 f_{\boldsymbol{\lambda}}''(x) dx 
&= 4 \hbar^2 |\boldsymbol{\lambda}| ,
\cr
\int^{\infty}_{-\infty} x^3 f_{\boldsymbol{\lambda}}''(x) dx
&= 6 \hbar^3  ( \| \boldsymbol{\lambda}^T \|^2 - \| \boldsymbol{\lambda} \|^2 ).
\end{align}
These formula indicates that the following combinations that appear in \eqref{eq:inst-strip} can be rewritten as 
\begin{align}
(-e^{-a})^{|\boldsymbol{\lambda}|}
\mathfrak{f}_{\boldsymbol{\lambda}} 
= &
\exp \left( 
- \frac{1}{12 \hbar^2} 
\int^{\infty}_{-\infty} x^3 f_{\boldsymbol{\lambda}}''(x-a) dx
+ \frac{a^2}{6\hbar^3}
\right),
\cr
e^{-m |\boldsymbol{\lambda}|} 
= &\exp \left( 
- \frac{m}{4 \hbar^2} 
\int^{\infty}_{-\infty} x^2 f_{\boldsymbol{\lambda}}''(x-a) dx
+ \frac{a^2}{2\hbar^2}
\right).
\end{align}

We also note the building block $R_{\boldsymbol{\lambda} \boldsymbol{\mu}} (Q)$ of the partition function is also written in terms of the profile function as
\begin{align}
R_{\boldsymbol{\lambda} \boldsymbol{\mu}} (e^{-(a-b)})
=
\exp \left[ \frac14 \int^{\infty}_{-\infty}\int^{\infty}_{-\infty} f_{\boldsymbol{\lambda}}''(x-a) 
f_{\boldsymbol{\mu}^T}''(y-b) 
\gamma_{\hbar}(x-y) dx dy
\right],
\end{align}
where $\gamma_{\hbar}(x)$ is Barnes double gamma function given by
\begin{align}
\gamma_{\hbar}(x) = \sum_{n=1} \frac{1}{n} \frac{e^{-nx}}{(1-e^{-n\hbar})(1-e^{n\hbar})}. 
\end{align}
See, for example, Appendix of \cite{Li:2021rqr} for detailed derivation. Combining this with the decomposition of the building block into the Plethystic exponential and the Nekrasov factor as in \eqref{eq:R-RN} , \eqref{eq:R-PE}, and the relation between the Nekrasov factor in sinh form in \eqref{eq:geom-mean-nN}, 
we find
\begin{align}
\tilde{n}_{\boldsymbol{\mu}\boldsymbol{\nu}} (a-b)
= &(-1)^{\frac12 ( - |\boldsymbol{\mu}| + |\boldsymbol{\nu}|)}
\text{PE} \left(
- \frac{g}{2(1-g)^2} \left( e^{-(a-b)} + e^{-(b-a)} \right)
\right)
\cr
&
\exp \left[ \frac18 \int^{\infty}_{-\infty}\int^{\infty}_{-\infty} 
\left( 
f_{\boldsymbol{\lambda}^T}''(x-a) 
f_{\boldsymbol{\mu}^T}''(y-b) 
\right. 
\right.
\cr
& \qquad\qquad
\left.
+
f_{\boldsymbol{\mu}^T}''(x-b) 
f_{\boldsymbol{\lambda}^T}''(y-a) 
\right)
\gamma_{\hbar}(x-y) dx dy
\biggl].
\end{align}

Using these identities, we rewrite the strip amplitude given in \eqref{eq:inst-strip} in terms of the profile function. For this purpose, we introduce the following combinations of profile functions, each of which corresponds to the left/right spinor/conjugate spinor amplitudes,
\begin{align}\label{eq:def-comb-prof}
\tilde{f}^{\mathrm{S}}_{\mathrm{L}}(x;\vec{\boldsymbol{\mu}},\boldsymbol{\lambda}_{\mathrm{L}}) 
&:= - \sum_{i=1}^N f_{\boldsymbol{\mu}_i}(x-a_i)  +  f_{\boldsymbol{\lambda}_{\mathrm{L}}}(x-a_{\mathrm{L}}^S) - f_{\boldsymbol{\lambda}_{\mathrm{L}}^T}(x+a_{\mathrm{L}}^S)
\cr
&= - \sum_{i=1}^N f_{\boldsymbol{\mu}_i}(x-a_i)  +  f_{\boldsymbol{\lambda}_{\mathrm{L}}}(x-a_{\mathrm{L}}^S) - f_{\boldsymbol{\lambda}_{\mathrm{L}}}(-x-a_{\mathrm{L}}^S)
\cr
\tilde{f}^{\mathrm{C}}_{\mathrm{L}}(x; \vec{\boldsymbol{\mu}},\boldsymbol{\lambda}_{\mathrm{L}}) 
& := \left. \tilde{f}^{\mathrm{S}}_{\mathrm{L}}{}(x;\vec{\boldsymbol{\mu}},\boldsymbol{\lambda}_{\mathrm{L}})  \right|_{a_N \to -a_N, \boldsymbol{\mu}_N \to \boldsymbol{\mu}_N^T}
\cr
&= - \sum_{i=1}^{N-1} f_{\boldsymbol{\mu}_i}(x-a_i) - f_{\boldsymbol{\mu}_N}(-x-a_N) +  f_{\boldsymbol{\lambda}_{\mathrm{L}}}(x-a_{\mathrm{L}}^{\mathrm{C}}) - f_{\boldsymbol{\lambda}_{\mathrm{L}}}(-x-a_{\mathrm{L}}^{\mathrm{C}})
\cr
\tilde{f}^{\mathrm{S}}_{\mathrm{R}}(x;\vec{\boldsymbol{\mu}},\boldsymbol{\lambda}_{\mathrm{R}}) 
&:= -\left. \tilde{f}^{\mathrm{S}}_{\mathrm{L}}(x;\vec{\boldsymbol{\mu}},\boldsymbol{\lambda}_{\mathrm{L}})  \right|_{m_{\mathrm{L}} \to m_{\mathrm{R}}, \boldsymbol{\lambda}_{\mathrm{L}} \to \boldsymbol{\lambda}_{\mathrm{R}}}
\cr
&= \sum_{i=1}^N f_{\boldsymbol{\mu}_i}(x-a_i) -  f_{\boldsymbol{\lambda}_{\mathrm{R}}}(x-a_{\mathrm{R}}^S) + f_{\boldsymbol{\lambda}_{\mathrm{R}}}(-x-a_{\mathrm{R}}^S)
\cr
\tilde{f}_{\mathrm{R}}^{\mathrm{C}}(x;\vec{\boldsymbol{\mu}},\boldsymbol{\lambda}_{\mathrm{R}}) 
&:= - \left. \tilde{f}^{\mathrm{S}}_{\mathrm{L}}(x;\vec{\boldsymbol{\mu}},\boldsymbol{\lambda}_{\mathrm{L}})  \right|_{a_N \to -a_N, m_{\mathrm{L}} \to m_{\mathrm{R}}, \boldsymbol{\mu}_N \to \boldsymbol{\mu}_N^T, \boldsymbol{\lambda}_{\mathrm{L}} \to \boldsymbol{\lambda}_{\mathrm{R}}}
\cr
&= \sum_{i=1}^{N-1} f_{\boldsymbol{\mu}_i}(x-a_i) + f_{\boldsymbol{\mu}_N}(-x-a_N) - f_{\boldsymbol{\lambda}_{\mathrm{R}}}(x-a_{\mathrm{R}}^{\mathrm{C}}) + f_{\boldsymbol{\lambda}_{\mathrm{R}}}(-x-a_{\mathrm{R}}^{\mathrm{C}})
\cr
\end{align}
The sign in front of each term is determined depending on whether the corresponding color brane extends to the left or to the right from the strip.

Using this function, the strip amplitudes are given in terms of the combination of profile functions as 
\begin{align}\label{eq:strip-amp-f}
\tilde{Z}_{\text{left}}^{\mathrm{X}}
=& \tilde{C}^{\mathrm{X}}_{\mathrm{L}} 
\exp \biggl[
- \frac{1}{24 \hbar^2} (4-N)
\int^{\infty}_{-\infty} x^3 
\tilde{f}^{\mathrm{S}}_{\mathrm{L}}{}''(x;\vec{\boldsymbol{\mu}},\boldsymbol{\lambda}_{\mathrm{L}}) dx
\cr
& \qquad \qquad 
+ \frac{1}{8 \hbar^2}
\bigl( (m_0 - (2^{N-4}-1) (m_{\mathrm{L}}+m_{\mathrm{R}}) \bigr) 
\int^{\infty}_{-\infty} x^2 \tilde{f}^{\mathrm{X}}_{\mathrm{L}}{}''(x;\vec{\boldsymbol{\mu}},\boldsymbol{\lambda}_{\mathrm{L}}) dx
\cr
& \qquad \qquad 
- \frac{1}{8} \int^{\infty}_{-\infty} \int^{\infty}_{-\infty} 
\tilde{f}^{\mathrm{X}}_{\mathrm{L}}{}''(x;\vec{\boldsymbol{\mu}},\boldsymbol{\lambda}_{\mathrm{L}}) \tilde{f}^{\mathrm{X}}_{\mathrm{L}}{}''(y;\vec{\boldsymbol{\mu}},\boldsymbol{\lambda}_{\mathrm{L}})
\gamma_{\hbar}(x-y) dx dy
\biggr],
\cr
\tilde{Z}_{\text{right}}^{\mathrm{X}}
=& \tilde{C}^{\mathrm{X}}_{\mathrm{R}} 
\exp \biggl[
+ \frac{1}{24 \hbar^2} (4-N)
\int^{\infty}_{-\infty} x^3 
\tilde{f}^{\mathrm{X}}_{\mathrm{R}}{}''(x;\vec{\boldsymbol{\mu}},\boldsymbol{\lambda}_{\mathrm{R}}) dx
\cr
& \qquad \qquad 
- \frac{1}{8 \hbar^2}
\bigl( m_0 - (2^{N-4}-1) (m_{\mathrm{L}}+m_{\mathrm{R}} ) \bigr) 
\int^{\infty}_{-\infty} x^2 \tilde{f}^{\mathrm{X}}_{\mathrm{R}}{}''(x;\vec{\boldsymbol{\mu}},\boldsymbol{\lambda}_{\mathrm{R}}) dx
\cr
& \qquad \qquad 
- \frac{1}{8} \int^{\infty}_{-\infty} \int^{\infty}_{-\infty} 
\tilde{f}^{\mathrm{X}}_{\mathrm{R}}{}''(x;\vec{\boldsymbol{\mu}},\boldsymbol{\lambda}_{\mathrm{R}}) \tilde{f}^{\mathrm{X}}_{\mathrm{R}}{}''(y;\vec{\boldsymbol{\mu}},\boldsymbol{\lambda}_{\mathrm{R}})
\gamma_{\hbar}(x-y) dx dy
\biggr].
\end{align}
where the upper label $\mathrm{X}$ refers either $\mathrm{X}=\mathrm{L}$ or $\mathrm{X}=\mathrm{R}$ and the equations are satisfied for both cases.  
$\tilde{C}^{\mathrm{S}}_{\mathrm{L}}$, $\tilde{C}^{\mathrm{C}}_{\mathrm{L}}$, $\tilde{C}^{\mathrm{S}}_{\mathrm{R}}$, $\tilde{C}^{\mathrm{C}}_{\mathrm{R}}$, are a product of factors written in the plethystic exponential form and does not depend on any of the Young diagrams. Here, we have also absorbed the phase factor into the redefinition of the instanton particle mass parameter $m_0$ as 
\begin{align}
m_0 \to m_0 - (N-3) \pi i.
\end{align}

\subsection{Thermodynamic limit and saddle point equation}

In the following, we discuss the thermodynamic limit $\hbar \to 0$ of the partition functions. Since $\gamma_{\hbar}(x) = \mathcal{O}(\hbar^{-2})$,  the exponent of the strip amplitudes \eqref{eq:strip-amp-f} is of order $\mathcal{O}(\hbar^{-2})$. Thus, as discussed in \cite{Nekrasov:2003rj}, the typical number of boxes of Young diagrams that contribute to the summation the most is expected to be also of order $\mathcal{O}(\hbar^{-2})$.

In this limit, the profile function defined in \eqref{eq:def-profile} can be treated as a continuous function. Since the strip amplitudes depend on the Young diagrams $\boldsymbol{\mu}_i$, $\boldsymbol{\lambda}_{\mathrm{L}}$, $\boldsymbol{\lambda}_{\mathrm{R}}$ only through the second derivative of these profile functions, we can replace the summation over these Young diagrams in the partition function \eqref{eq:Top-result} by the functional integral.

Based on this idea, we omit the dependence on Young diagrams of the profile functions and simply denote them as
\begin{align}\label{eq:drop-label}
&f_{\boldsymbol{\mu}_i}(x-a_i) \to f_i(x) \qquad (i=1,2,\cdots, N)
\cr
& f_{\boldsymbol{\lambda}_{\mathrm{L}}}(x-a^{\mathrm{X}}_{\mathrm{L}}) \to 
g_{\mathrm{L}}(x),
\qquad 
f_{\boldsymbol{\lambda}_{\mathrm{R}}}(x-a_{\mathrm{R}}^{\mathrm{X}}) \to g_{\mathrm{R}}(x). 
\qquad (\mathrm{X} = \mathrm{S}, \mathrm{C})
\end{align}
Here, we have also eliminated the labels to distinguish the spinor or the conjugate spinor, $\mathrm{X}=\mathrm{S}, \mathrm{C}$. 
The dependence on the Young diagram of the profile function
is now understood in terms of the region of the functional integral.

The region of the functional integral is specified by imposing the constraints on the functions.  Since the integral of the profile function is given by \eqref{eq:prof-integral}, the following constraints should be satisfied:
\begin{align}\label{eq:constraint-fg}
& \int^{\infty}_{-\infty} x f_i''(x) dx = 2a_i, \qquad 
\int^{\infty}_{-\infty} f_i''(x) dx = 2 ,
\qquad (i=1,2,\cdots, N)
\cr
&
\int^{\infty}_{-\infty} x g_{\mathrm{L}}''(x) dx 
= 
2 a^{\mathrm{X}}_{\mathrm{L}}
\qquad 
\int^{\infty}_{-\infty} g_{\mathrm{L}}''(x) dx = 2 ,
\qquad (\mathrm{X}=\mathrm{S}, \mathrm{C})
\cr
& 
\int^{\infty}_{-\infty} x g_{\mathrm{R}}''(x) dx 
= 
2 a^{\mathrm{X}}_{\mathrm{R}}, \qquad 
\int^{\infty}_{-\infty} g_{\mathrm{R}}^{\mathrm{X}}{}''(x) dx = 2.
\qquad (\mathrm{X}=\mathrm{S}, \mathrm{C})
\end{align}
The choice $\mathrm{X}=\mathrm{S}$ or $\mathrm{X}=\mathrm{C}$ depends on whether we consider the spinor strip or the conjugate spinor strip.

Analogous to the replacement \eqref{eq:drop-label}, we also denote the combinations of the profile functions in \eqref{eq:def-comb-prof} as 
\begin{align}
&\tilde{f}^{\mathrm{X}}_{\mathrm{L}}(x;\vec{\boldsymbol{\mu}},\boldsymbol{\lambda}_{\mathrm{L}}) 
\to 
\tilde{f}^{\mathrm{X}}_{\mathrm{L}}(x) ,
\qquad
\tilde{f}^{\mathrm{X}}_{\mathrm{R}}(x;\vec{\boldsymbol{\mu}},\boldsymbol{\lambda}_{\mathrm{R}}) 
\to \tilde{f}^{\mathrm{X}}_{\mathrm{R}} (x) 
\qquad (\mathrm{X} = \mathrm{S}, \mathrm{C}).
\end{align}
With these conventions, we now have
\begin{align}\label{eq:fLRSC}
\tilde{f}^{\mathrm{S}}_{\mathrm{L}}(x) 
=& - \sum_{i=1}^N f_i(x) + g_{\mathrm{L}}(x) - g_{\mathrm{L}}(-x) ,
\cr
\tilde{f}^{\mathrm{C}}_{\mathrm{L}}(x) 
=& - \sum_{i=1}^{N-1} f_i(x) - f_N(-x) + g_{\mathrm{L}}(x) - g_{\mathrm{L}}(-x) , 
\cr
\tilde{f}_{\mathrm{R}}^{\mathrm{S}} (x)
=& \sum_{i=1}^N f_i(x) - g_{\mathrm{R}}(x) + g_{\mathrm{R}}(-x) ,
\cr
\tilde{f}_{\mathrm{R}}^{\mathrm{C}} (x)
=& \sum_{i=1}^{N-1} f_i(x) + f_N(-x) - g_{\mathrm{R}}(x) + g_{\mathrm{R}}(-x) .
\end{align}

Taking all these into account, the partition function can be schematically written in the form
\begin{align}\label{eq:path-int}
Z^{\mathrm{X}\mathrm{Y}} 
\sim & \,
\tilde{C}^{\mathrm{X}}_{\mathrm{L}}
\tilde{C}^{\mathrm{Y}}_{\mathrm{R}}
\int \left( \prod_{i=1}^N 
\mathcal{D} f''_{i} \right)\,  
\mathcal{D} g_{\mathrm{L}}'' \,
\mathcal{D} g_{\mathrm{R}}'' \,
\left( \prod_{i=1}^{N} d \xi_i d\zeta_i \right) d\xi_{\mathrm{L}} d\zeta_{\mathrm{L}}d\xi_{\mathrm{R}} d\zeta_{\mathrm{R}}
\cr
& \qquad
\exp \left[ - 
\frac{\mathcal{F}^{\mathrm{XY}}[\{ f_i\},g_{\mathrm{L}},g_{\mathrm{R}}] ( \{ \xi_i \}, \{ \zeta_i \}, \xi_{\mathrm{L}}, \zeta_{\mathrm{L}}, \xi_{\mathrm{R}}, \zeta_{\mathrm{R}}) }{\hbar^{2}}
+ \mathcal{O} \left( \frac{1}{\hbar} \right)
\right]
\end{align}
where $\mathcal{F}^{\mathrm{XY}}$ consists of the three parts as
\begin{align}
&\mathcal{F}^{\mathrm{XY}}[\{ f_i\},g_{\mathrm{L}},g_{\mathrm{R}}]( \{ \xi_i \}, \{ \zeta_i \}, \xi_{\mathrm{L}}, \zeta_{\mathrm{L}}, \xi_{\mathrm{R}}, \zeta_{\mathrm{R}})
\cr
= &\mathcal{F}^{\mathrm{X}}_{\mathrm{L}} [\{ f_i\},g_{\mathrm{L}}]
+ \mathcal{F}^{\mathrm{Y}}_{\mathrm{R}} [\{ f_i\},g_{\mathrm{R}}]
+ \mathcal{F}_{\mathrm{LR}}^{\mathrm{XY}} [\{ f_i\},g_{\mathrm{L}},g_{\mathrm{R}}]( \{ \xi_i \}, \{ \zeta_i \}, \xi_{\mathrm{L}}, \zeta_{\mathrm{L}}, \xi_{\mathrm{R}}, \zeta_{\mathrm{R}}).
\end{align}
Here, $\mathcal{F}^{\mathrm{X}}_{\mathrm{L}}$ is the contribution from the left strip, $\mathcal{F}^{\mathrm{Y}}_{\mathrm{R}}$ is the contribution of the right strip, and $\mathcal{F}_{\mathrm{LR}}^{\mathrm{XY}}$ is the Lagrange multiplier term to impose the constraints in \eqref{eq:constraint-fg}, respectively.
Using
\begin{align}
\gamma_{\hbar}(x) = - \hbar^{-2} \mathrm{Li}_3 (e^{- x}) + 
\mathcal{O}(\hbar^{-1}),
\end{align}
the contributions from the left strip and the right strip are given from \eqref{eq:strip-amp-f} as
\begin{align}\label{eq:strip-functional}
\mathcal{F}^{\mathrm{X}}_{\mathrm{L}} [\{ f_i\},g_{\mathrm{L}}]
= & - \frac{1}{24}(N-4)
\int^{\infty}_{-\infty} x^3 \tilde{f}_{\mathrm{L}}^{\mathrm{X}}{}''(x) dx
\cr
& 
- \frac{1}{8}
\bigl( m_0 - (2^{N-4}-1) (m_{\mathrm{L}}+m_{\mathrm{R}} ) \bigr)  
\int^{\infty}_{-\infty} x^2 
\tilde{f}_{\mathrm{L}}^{\mathrm{X}}{}''(x) dx
\cr
& 
- \frac{1}{8} \int^{\infty}_{-\infty} \int^{\infty}_{-\infty} 
\tilde{f}_{\mathrm{L}}^{\mathrm{X}}{}''(x)
\tilde{f}_{\mathrm{L}}^{\mathrm{X}}{}''(y) 
\mathrm{Li}_3 (e^{-(x-y)})  dx dy 
\quad (\mathrm{X} = \mathrm{S}, \mathrm{C})
\cr
\mathcal{F}^{\mathrm{Y}}_{\mathrm{R}} [\{ f_i\},g_{\mathrm{R}}]
= &\frac{1}{24}  (N-4)
\int^{\infty}_{-\infty} x^3 \tilde{f}_{\mathrm{R}}^{\mathrm{Y}}{}''(x) dx
\cr
& 
+ \frac{1}{8}
\bigl( m_0 - (2^{N-4}-1) (m_{\mathrm{L}}+m_{\mathrm{R}} ) \bigr)  
\int^{\infty}_{-\infty} x^2 
\tilde{f}_{\mathrm{R}}^{\mathrm{Y}}{}''(x) dx
\cr
& 
- \frac{1}{8} \int^{\infty}_{-\infty} \int^{\infty}_{-\infty} 
\tilde{f}_{\mathrm{R}}^{\mathrm{Y}}{}''(x)
\tilde{f}_{\mathrm{R}}^{\mathrm{Y}}{}''(y)
\mathrm{Li}_3 (e^{-(x-y)})  dx dy.
\quad (\mathrm{Y} = \mathrm{S}, \mathrm{C})
\end{align}
Since the $\mathrm{Li}_3(z)$ diverges at $z \to 0$, the double integrals in \eqref{eq:strip-functional} are understood as principal integrals.
The Lagrange multiplier terms are given as
\begin{align}
&\mathcal{F}_{\mathrm{LR}}^{\mathrm{XY}} [\{ f_i\},g_{\mathrm{L}},g_{\mathrm{R}}]( \{ \xi_i \}, \{ \zeta_i \}, \xi_{\mathrm{L}}, \zeta_{\mathrm{L}}, \xi_{\mathrm{R}}, \zeta_{\mathrm{R}})
\cr
&
= \sum_{i=1}^N
\xi_{i} \left( \int^{\infty}_{-\infty} x f_i''(x) dx - 2a_i \right)
+ \sum_{i=1}^N
\zeta_i \left( \int^{\infty}_{-\infty} f_i''(x) dx - 2 \right)
\cr
& \qquad + \xi_{\mathrm{L}}\left( \int^{\infty}_{-\infty} x g_{\mathrm{L}}(x) dx - 2 a^{\mathrm{X}}_{\mathrm{L}} \right)
+ \zeta_{\mathrm{L}}\left( \int^{\infty}_{-\infty} g_{\mathrm{L}}(x) dx - 2  \right)
\cr
& \qquad + \xi_{\mathrm{R}}\left( \int^{\infty}_{-\infty} x g_{\mathrm{R}}(x) dx - 2 a^{\mathrm{X}}_{\mathrm{R}} \right)
+ \zeta_{\mathrm{R}}\left( \int^{\infty}_{-\infty} g_{\mathrm{R}}(x) dx - 2  \right).
\end{align}

In the limit $\hbar \to 0$, the functional integral \eqref{eq:path-int} is evaluated by the saddle point approximation. 
Then, the partition function is reduced to
\begin{align}
Z^{\text{XY}} = 
\tilde{C}^{\mathrm{X}}_{\mathrm{L}}
\tilde{C}^{\mathrm{Y}}_{\mathrm{R}} 
\exp \left[ -\frac{\mathcal{F}^{\mathrm{X}}_{\mathrm{L}} [\{ f_{i*}\},g_{\mathrm{L}*}]
+ \mathcal{F}^{\mathrm{Y}}_{\mathrm{R}} [\{ f_{i*} \},g_{\mathrm{R}*}]}{\hbar^2} + \mathcal{O}(\hbar^{-1}) \right].
\end{align}
where $f_{i*},\, g_{\mathrm{L}*}, \,g_{\mathrm{R}*}$ denote the solution of the following saddle point equations
\begin{align}\label{eq:saddle-pt-9}
&\frac{\delta \mathcal{F}^{\mathrm{XY}}}{\delta f_i''} = 0,
\qquad
\frac{\delta \mathcal{F}^{\mathrm{XY}}}{\delta g_{\mathrm{L}}''} = 0,
\qquad
\frac{\delta \mathcal{F}^{\mathrm{XY}}}{\delta g_{\mathrm{R}}''} = 0.
\cr
&\frac{\delta \mathcal{F}^{\mathrm{XY}}}{\partial \xi_i} = 0,
\qquad
\frac{\partial \mathcal{F}^{\mathrm{XY}}}{\partial \xi_{\mathrm{L}}} = 0,
\qquad
\frac{\partial \mathcal{F}^{\mathrm{XY}}}{\partial \xi_{\mathrm{R}}} = 0,
\cr
&\frac{\partial \mathcal{F}^{\mathrm{XY}}}{\partial \zeta_{\mathrm{i}}} = 0,
\qquad
\frac{\partial \mathcal{F}^{\mathrm{XY}}}{\partial \zeta_{\mathrm{L}}} = 0,
\qquad
\frac{\partial \mathcal{F}^{\mathrm{XY}}}{\partial \zeta_{\mathrm{R}}} = 0,
\end{align}

In order to present these saddle-point equations explicitly, we denote the support of the functions $f''_i(x)$, $g''_{\mathrm{L}}(x)$ and $g''_{\mathrm{R}}(x)$ as $\mathcal{C}_i$, $\mathcal{C}_L$, $\mathcal{C}_R$, respectively. Using the same symbol, we also denote the support of $f''_i(-x)$ $g''_{\mathrm{L}}(-x)$,  $g''_{\mathrm{R}}(-x)$ as 
 $-\mathcal{C}_N$, $-\mathcal{C}_L$, $-\mathcal{C}_R$, respectively. We assume that these regions are all disjoint. 

As for the first equation with $i=1,\cdots, N-1$ in \eqref{eq:saddle-pt-9}, we have, for $x \in \mathcal{C}_i$,
\begin{align}\label{eq:saddle1}
 0 = \frac{\delta \mathcal{F}^{\mathrm{XY}}}{\delta f_i''(x)}
= & 
\frac{1}{12} (N-4) x^3 + \frac14 \bigl( m_0 - (2^{N-4}-1) (m_{\mathrm{L}}+m_{\mathrm{R}} ) \bigr) x^2
\cr
&  
+ \frac14 \int^{\infty}_{-\infty} 
\left( f_{\mathrm{L}}^{\mathrm{X}}{}''(y) - f_{\mathrm{R}}^{\mathrm{Y}}{}''(y) 
\right)
\left( \mathrm{Li}_3 (e^{-(x-y)}) 
+ \mathrm{Li}_3 (e^{-(y-x)}) \right) dy 
\cr
& 
+ \xi_i x + \zeta_i .
\end{align}

As for the first one with $i=N$ in \eqref{eq:saddle-pt-9}, we have, for $x \in \mathcal{C}_N$,
\begin{align}\label{eq:saddle1-2}
 0 = \frac{\delta \mathcal{F}^{\mathrm{SS}}}{\delta f_N''(x)}
= & 
\frac{1}{12} (N-4) x^3 + \frac14 \bigl( m_0 - (2^{N-4}-1) (m_{\mathrm{L}}+m_{\mathrm{R}} ) \bigr) x^2
\cr
&  
+ \frac14 \int^{\infty}_{-\infty} 
\left( \tilde{f}_{\mathrm{L}}^{\mathrm{S}}{}''(y) - \tilde{f}_{\mathrm{R}}^{\mathrm{S}}{}''(y) 
\right)
\left( \mathrm{Li}_3 (e^{-(x-y)}) 
+ \mathrm{Li}_3 (e^{-(y-x)}) \right) dy 
\cr
& 
+ \xi_N x + \zeta_N .
\cr
 0 = \frac{\delta \mathcal{F}^{\mathrm{SC}}}{\delta f_N''(x)}
= & 
\frac14 \bigl( m_0 - (2^{N-4}-1) (m_{\mathrm{L}}+m_{\mathrm{R}} ) \bigr) x^2
\cr
&  
+ \frac14 \int^{\infty}_{-\infty} 
\left( \tilde{f}_{\mathrm{L}}^{\mathrm{S}}{}''(y) 
- \tilde{f}_{\mathrm{R}}^{\mathrm{C}}{}''(-y)
\right)
\left( \mathrm{Li}_3 (e^{-(x-y)}) 
+ \mathrm{Li}_3 (e^{-(y-x)}) \right) dy 
\cr
& 
+ \xi_N x + \zeta_N .
\end{align}

As for the second one in \eqref{eq:saddle-pt-9}, we have, for $x \in \mathcal{C}_{\mathrm{L}}$,
\begin{align}\label{eq:saddle2}
0 = \frac{\delta \mathcal{F}^{\mathrm{XY}}}{\delta g_{\mathrm{L}}''(x)}
= & 
- \frac{1}{12} (N-4) x^3 
\cr
& 
+ \frac18 \int^{\infty}_{-\infty} 
\left( \tilde{f}_{\mathrm{L}}^{\mathrm{X}}{}''(y) - \tilde{f}_{\mathrm{L}}^{\mathrm{X}}{}''(-y) \right)
\left( \mathrm{Li}_3 (e^{-(y-x)}) 
- \mathrm{Li}_3 (e^{-(y+x)}) \right) dy 
\cr
& + \xi_{\mathrm{L}} x + \zeta_{\mathrm{L}} .
\end{align}
As for the third one in \eqref{eq:saddle-pt-9}, we have, for $x \in C_{\mathrm{R}}$,
\begin{align}\label{eq:saddle3}
0 = \frac{\delta \mathcal{F}^{\mathrm{XY}}}{\delta g_{\mathrm{R}}''(x)}
= & 
- \frac{1}{12} (N-4) x^3 
\cr
& 
+\frac18 \int^{\infty}_{-\infty} 
\left( \tilde{f}_{\mathrm{R}}^{\mathrm{Y}}{}''(y) - \tilde{f}_{\mathrm{R}}^{\mathrm{Y}}{}''(-y) \right)
\left( \mathrm{Li}_3 (e^{-(y-x)}) 
- \mathrm{Li}_3 (e^{-(y+x)}) \right) dy 
\cr
&+ \xi_{\mathrm{R}} x + \zeta_{\mathrm{R}} .
\end{align}
The remaining six equations in \eqref{eq:saddle-pt-9} reproduce the constraints that we have discussed in \eqref{eq:constraint-fg}.

Considering the second derivative of \eqref{eq:saddle1} in terms of $x$, we have 
\begin{align}\label{eq:saddle-deriv-1-1}
0 
=\frac{\partial^2}{\partial x^2} \left( \frac{\delta \mathcal{F}^{\mathrm{XY}}}{\delta f_i''(x)} \right)
= & 
-2 x
+ \frac12 \bigl( m_0 - (2^{N-4}-2) (m_{\mathrm{L}}+m_{\mathrm{R}} ) \bigr) 
\cr
& \quad  
- \frac14 \int^{\infty}_{-\infty} 
\left( \tilde{f}_{\mathrm{L}}^{\mathrm{S}}{}''(y) - \tilde{f}_{\mathrm{R}}^{\mathrm{S}}{}''(y) 
\right) \log (1 - e^{-(y-x)})  dy 
\end{align}
The second derivative of \eqref{eq:saddle1-2} in terms of $x$ is
\begin{align}\label{eq:saddle-deriv-1-2}
0 
= \frac{\partial^2}{\partial x^2} \left( \frac{\delta \mathcal{F}^{\mathrm{SS}}}{\delta f_N''(x)} \right)
= & 
-2 x
+ \frac12 \bigl( m_0 - (2^{N-4}-2) (m_{\mathrm{L}}+m_{\mathrm{R}} ) \bigr) 
\cr
& \quad  
- \frac14 \int^{\infty}_{-\infty} 
\left( \tilde{f}_{\mathrm{L}}^{\mathrm{S}}{}''(y) - \tilde{f}_{\mathrm{R}}^{\mathrm{S}}{}''(y) 
\right) \log (1 - e^{-(y-x)})  dy 
\cr
0 
= \frac{\partial^2}{\partial x^2} \left( \frac{\delta \mathcal{F}^{\mathrm{SC}}}{\delta f_N''(x)} \right)
= & 
\frac12 \bigl( m_0 - (2^{N-4}-2) (m_{\mathrm{L}}+m_{\mathrm{R}} ) \bigr) 
\cr
&  
- \frac14 \int^{\infty}_{-\infty} 
\left( f_{\mathrm{L}}^{\mathrm{S}}{}''(y) - f_{\mathrm{R}}^{\mathrm{C}}{}''(-y) \right)
\log (1 - e^{-(y-x)})  dy 
\end{align}
The second derivative of \eqref{eq:saddle2} in terms of $x$ is given by 
\begin{align}\label{eq:saddle-deriv-2}
0 
=
\frac{\partial^2}{\partial x^2} \left( \frac{\delta \mathcal{F}^{\mathrm{XY}}}{\delta g_{\mathrm{L}}''(x)} \right)
= & 2 x
- \frac14 \int^{\infty}_{-\infty} 
\tilde{f}_{\mathrm{L}}^{\mathrm{X}}{}''(y) 
\left( \log (1 - e^{-(y-x)}) 
- \log ( 1 - e^{-(y+x)}) \right) dy
\end{align}
The second derivative of \eqref{eq:saddle3} is
\begin{align}\label{eq:saddle-deriv-3}
0 =& \frac{\partial^2}{\partial x^2} \left( \frac{\delta \mathcal{F}^{\mathrm{XY}}}{\delta g_{\mathrm{R}}''(x)} \right)
=
2 x
- \frac14 \int^{\infty}_{-\infty} 
\tilde{f}_{\mathrm{R}}^{\mathrm{Y}}{}''(y) 
\left( \log (1 - e^{-(y-x)}) 
- \log ( 1 - e^{-(y+x)}) \right)  dy.
\end{align}

In the computations above, we have used 
\begin{align}\label{eq:int-fL-fR}
& \int^{\infty}_{-\infty} y \tilde{f}_{\mathrm{L}}^{\mathrm{X}}{}''(y) dy 
= - 4m_{\mathrm{L}}, \quad \qquad
\int^{\infty}_{-\infty} \tilde{f}_{\mathrm{L}}^{\mathrm{X}}{}''(y) dy = - 2N,
\cr
& \int^{\infty}_{-\infty} y \tilde{f}_{\mathrm{R}}^{\mathrm{X}}{}''(y) dy 
= 4 m_{\mathrm{R}},
\qquad \qquad
\int^{\infty}_{-\infty} \tilde{f}_{\mathrm{R}}^{\mathrm{X}}{}''(y) dy = 2N .
\qquad (\mathrm{X} = \mathrm{S}, \mathrm{C})
\end{align}
which follow from \eqref{eq:param-b}, \eqref{eq:param-b-R}, \eqref{eq:constraint-fg} and \eqref{eq:fLRSC}.

\subsection{Amplitude function}
Using the solution of the saddle point equations above, we introduce the amplitude function defined on 
$z \in \mathbb{C} \backslash (\cup_I \mathcal{C}_I \cup \mathcal{C}_{\mathrm{L}})$ and 
$z \in \mathbb{C} \backslash (\cup_I \mathcal{C}_I \cup \mathcal{C}_{\mathrm{R}})$, respectively:
\begin{align}\label{eq:def-amp-fn}
\mathcal{Y}_{\mathrm{L}}^{\mathrm{X}}{}(z) 
:= &\exp \left[ 2 z
- \frac12 m_0 + (2^{N-4}-2) m_{\mathrm{L}}  
\right]
\cr
& \qquad \qquad 
\times \exp 
\left[
\frac12 \int^{\infty}_{-\infty} 
\tilde{f}_{\mathrm{L}*}^{\mathrm{X}}{}''(y) \log (1 - e^{-(y-z)}) dy
\right],
\cr
\mathcal{Y}_{\mathrm{R}}^{\mathrm{X}}{}(z) 
:= &\exp \left[ -2 z 
+ \frac12  m_0 - (2^{N-4}-2) m_{\mathrm{R}} 
\right]
\cr
& \qquad \qquad 
\times \exp 
\left[
\frac12 \int^{\infty}_{-\infty} 
\tilde{f}_{\mathrm{R}*}^{\mathrm{X}}{}''(y) \log (1 - e^{-(y-z)}) dy
\right].
\end{align}
Using \eqref{eq:int-fL-fR}, these amplitude functions can be also written as
\begin{align}\label{eq:def-amp-fn-2}
\mathcal{Y}_{\mathrm{L}}^{\mathrm{X}}{}(z) 
= &(-1)^N \exp \left[ - (N-2) z
- \frac12 m_0 + 2^{N-4} m_{\mathrm{L}} 
\right]
\cr
& \qquad \qquad 
\times \exp 
\left[
\frac12 \int^{\infty}_{-\infty} 
\tilde{f}_{\mathrm{L}*}^{\mathrm{X}}{}''(y) \log (1 - e^{-(z-y)}) dy
\right],
\cr
\mathcal{Y}_{\mathrm{R}}^{\mathrm{X}}{}(z) 
= &(-1)^N \exp \left[ (N-2) z 
+ \frac12 m_0 - 2^{N-4} m_{\mathrm{R}} 
\right]
\cr
& \qquad \qquad 
\times \exp 
\left[
\frac12 \int^{\infty}_{-\infty} 
\tilde{f}_{\mathrm{R}*}^{\mathrm{X}}{}''(y) \log (1 - e^{-(z-y)}) dy
\right].
\end{align}

In the following, we rewrite the conditions that $\tilde{f}_{\mathrm{L}*}^{\mathrm{X}}{}''(y)$ and $\tilde{f}_{\mathrm{R}*}^{\mathrm{X}}{}''(y)$ satisfy into the conditions for these amplitude functions.

First, $\tilde{f}_{\mathrm{L}*}^{\mathrm{X}}{}''(y)$ and $\tilde{f}_{\mathrm{R}*}^{\mathrm{X}}{}''(y)$ are related as in \eqref{eq:fLRSC}, from which we find the constraints
\begin{align}
&\tilde{f}^{\mathrm{X}}_{\mathrm{L}}(x) + \tilde{f}^{\mathrm{Y}}_{\mathrm{R}}(x)= 0 
\qquad x \in \mathcal{C}_i \quad (i=1,\cdots, N-1)
\cr
&\tilde{f}^{\mathrm{S}}_{\mathrm{L}}(x) + \tilde{f}^{\mathrm{S}}_{\mathrm{R}}(x)= 0 ,
\quad
\tilde{f}^{\mathrm{S}}_{\mathrm{L}}(x) + \tilde{f}^{\mathrm{C}}_{\mathrm{R}}(-x)= 0 
\qquad x \in \mathcal{C}_N
\cr
& \tilde{f}^{\mathrm{X}}_{\mathrm{L}}(x) + \tilde{f}^{\mathrm{X}}_{\mathrm{L}}(-x) = 0 \qquad x \in \mathcal{C}_{\mathrm{L}}
\cr
& \tilde{f}^{\mathrm{X}}_{\mathrm{R}}(x) + \tilde{f}^{\mathrm{X}}_{\mathrm{R}}(-x) = 0 \qquad x \in \mathcal{C}_{\mathrm{R}}
\end{align}
These constraints are written in terms of the amplitude functions as
\begin{align}
&\mathcal{Y}_{\mathrm{L}}^{\mathrm{X}}{}(x+i0) 
\mathcal{Y}_{\mathrm{R}}^{\mathrm{X}}{}(x+i0) 
= 
\mathcal{Y}_{\mathrm{L}}^{\mathrm{X}}{}(x-i0) 
\mathcal{Y}_{\mathrm{R}}^{\mathrm{X}}{}(x-i0) 
\quad 
x \in \mathcal{C}_i \quad (i=1,\cdots, N-1)
\cr
&\mathcal{Y}_{\mathrm{L}}^{\mathrm{S}}{}(x+i0) 
\mathcal{Y}_{\mathrm{R}}^{\mathrm{S}}{}(x+i0) 
= 
\mathcal{Y}_{\mathrm{L}}^{\mathrm{S}}{}(x-i0) 
\mathcal{Y}_{\mathrm{R}}^{\mathrm{S}}{}(x-i0) 
\quad 
x \in \mathcal{C}_N
\cr
&\mathcal{Y}_{\mathrm{L}}^{\mathrm{S}}{}(x+i0) 
\mathcal{Y}_{\mathrm{R}}^{\mathrm{C}}{}(-x+i0) 
= 
\mathcal{Y}_{\mathrm{L}}^{\mathrm{S}}{}(x-i0) 
\mathcal{Y}_{\mathrm{R}}^{\mathrm{C}}{}(-x-i0) 
\quad 
x \in \mathcal{C}_N
\cr
&\mathcal{Y}_{\mathrm{L}}^{\mathrm{X}}{}(x+i0) 
\mathcal{Y}_{\mathrm{L}}^{\mathrm{Y}}{}(-x+i0) 
= 
\mathcal{Y}_{\mathrm{L}}^{\mathrm{X}}{}(x-i0) 
\mathcal{Y}_{\mathrm{L}}^{\mathrm{Y}}{}(-x-i0) 
\quad x \in \mathcal{C}_{\mathrm{L}}
\cr
&\mathcal{Y}_{\mathrm{R}}^{\mathrm{X}}{}(x+i0) 
\mathcal{Y}_{\mathrm{R}}^{\mathrm{Y}}{}(-x+i0) 
= 
\mathcal{Y}_{\mathrm{R}}^{\mathrm{X}}{}(x-i0) 
\mathcal{Y}_{\mathrm{R}}^{\mathrm{Y}}{}(-x-i0) 
\quad x \in \mathcal{C}_{\mathrm{R}}
\end{align}
which means that a specific combination of amplitude functions does not have a branch cut in the indicated region.

Taking these constraints into account, the saddle point equations \eqref{eq:saddle-deriv-1-1}, \eqref{eq:saddle-deriv-1-2}, \eqref{eq:saddle-deriv-2}, and \eqref{eq:saddle-deriv-3} lead to 
\begin{align}\label{eq:Y-sol}
&\mathcal{Y}_{\mathrm{L}}^{\mathrm{X}}(x \pm i0) 
= \mathcal{Y}_{\mathrm{R}}^{\mathrm{Y}}(x \mp i0) ,
\qquad x \in \mathcal{C}_i, \quad (i=1,\cdots, N-1)
\cr 
&\mathcal{Y}_{\mathrm{L}}^{\mathrm{S}}(x \pm i0) 
= \mathcal{Y}_{\mathrm{R}}^{\mathrm{S}}(x \mp i0) ,
\qquad x \in \mathcal{C}_N
\cr 
&\mathcal{Y}_{\mathrm{L}}^{\mathrm{S}}(x \pm i0) 
= \mathcal{Y}_{\mathrm{R}}^{\mathrm{C}}(-x \mp i0) ,
\quad x \in \mathcal{C}_N
\cr 
&\mathcal{Y}_{\mathrm{L}}^{\mathrm{X}}(x \pm i0) 
= \mathcal{Y}_{\mathrm{L}}^{\mathrm{Y}}(-x \mp i0) ,
\quad x \in C_{\mathrm{L}},
\cr 
&\mathcal{Y}_{\mathrm{R}}^{\mathrm{X}}(x \pm i0) 
= \mathcal{Y}_{\mathrm{R}}^{\mathrm{Y}}(-x \mp i0) ,
\quad x \in C_{\mathrm{R}},
\end{align}
where the $\pm$ or $\mp$ sign are taken in the same order.

This indicates, for example, that for the case with two spinors, the four functions $\mathcal{Y}_{\mathrm{L}}^{\mathrm{S}}(z), \mathcal{Y}_{\mathrm{L}}^{\mathrm{S}}(-z),
\mathcal{Y}_{\mathrm{R}}^{\mathrm{S}}(z), \mathcal{Y}_{\mathrm{R}}^{\mathrm{S}}(-z)$ are exchanged with each other when we go across the branch cuts $\mathcal{C}_i, \mathcal{C}_{\mathrm{L}}, \mathcal{C}_{\mathrm{R}}$ as
\begin{align}\label{eq:Weyl}
\mathcal{Y}_{\mathrm{L}}^{\mathrm{S}}{}(z) 
&\leftrightarrow \mathcal{Y}_{\mathrm{R}}^{\mathrm{S}}{}(z) ,
\qquad z \in \mathcal{C}_i \quad (i=1,\cdots, N),
\cr 
\mathcal{Y}_{\mathrm{L}}^{\mathrm{S}}{}(z) 
&\leftrightarrow \mathcal{Y}_{\mathrm{L}}^{\mathrm{S}}{}(-z) ,
\qquad z \in \mathcal{C}_{\mathrm{L}}
\cr 
\mathcal{Y}_{\mathrm{R}}^{\mathrm{S}}{}(z) 
&\leftrightarrow \mathcal{Y}_{\mathrm{R}}^{\mathrm{S}}{}(-z) ,
\qquad z \in \mathcal{C}_{\mathrm{R}}
\end{align}
Analogously, for the case of one spinor and one conjugate spinor, the four functions $\mathcal{Y}_{\mathrm{L}}^{\mathrm{S}}(z), \mathcal{Y}_{\mathrm{L}}^{\mathrm{S}}(-z),
\mathcal{Y}_{\mathrm{R}}^{\mathrm{C}}(z), \mathcal{Y}_{\mathrm{R}}^{\mathrm{C}}(-z)$ are exchanged with each other as
\begin{align}\label{eq:Weyl2}
\mathcal{Y}_{\mathrm{L}}^{\mathrm{S}}{}(z) 
&\leftrightarrow \mathcal{Y}_{\mathrm{R}}^{\mathrm{C}}{}(z) ,
\qquad z \in \mathcal{C}_i \quad (i=1,\cdots, N-1),
\cr 
\mathcal{Y}_{\mathrm{L}}^{\mathrm{S}}{}(z) 
&\leftrightarrow \mathcal{Y}_{\mathrm{R}}^{\mathrm{C}}{}(-z) ,
\qquad z \in \mathcal{C}_N ,
\cr 
\mathcal{Y}_{\mathrm{L}}^{\mathrm{S}}{}(z) 
&\leftrightarrow \mathcal{Y}_{\mathrm{L}}^{\mathrm{S}}{}(-z) ,
\qquad z \in \mathcal{C}_{\mathrm{L}}
\cr 
\mathcal{Y}_{\mathrm{R}}^{\mathrm{C}}{}(z) 
&\leftrightarrow \mathcal{Y}_{\mathrm{R}}^{\mathrm{C}}{}(-z) ,
\qquad z \in \mathcal{C}_{\mathrm{R}}
\end{align}
These exchanges can be understood as analytic continuation.

Finally, we also note the cycle integral. The constraints \eqref{eq:constraint-fg} are translated as
\begin{align}\label{eq:A-cycle}
&\oint_{\mathcal{A}_i} z \, d \log \mathcal{Y}_{\mathrm{L}}^{\mathrm{X}} = a_i, 
\quad
\oint_{\mathcal{A}_i} d \log \mathcal{Y}_{\mathrm{L}}^{\mathrm{X}} = 1,
\cr
&\oint_{\mathcal{A}_{\mathrm{L}}} z \, d \log \mathcal{Y}_{\mathrm{L}}^{\mathrm{X}} = a_{\mathrm{L}}^{\mathrm{X}},
\quad
\oint_{\mathcal{A}_{\mathrm{L}}}  d \log \mathcal{Y}_{\mathrm{L}}^{\mathrm{X}} = 1,
\cr
&\oint_{\mathcal{A}_{\mathrm{R}}} z \, d \log \mathcal{Y}_{\mathrm{R}}^{\mathrm{X}} = a_{\mathrm{R}}^{\mathrm{X}},
\quad
\oint_{\mathcal{A}_{\mathrm{R}}} z \, d \log \mathcal{Y}_{\mathrm{R}}^{\mathrm{X}} = 1.
\end{align}
where $\mathcal{A}_i, \mathcal{A}_{\mathrm{L}}, \mathcal{A}_{\mathrm{R}}$ are cycles that goes around the region $\mathcal{C}_i$,  $\mathcal{C}_{\mathrm{L}}$,  $\mathcal{C}_{\mathrm{R}}$, respectively.  
We also note that by using the first derivative of the saddle point equations \eqref{eq:saddle1}, \eqref{eq:saddle1-2} that
\begin{align}\label{eq:B-cycle}
\frac{\partial \mathcal{F}^{\mathrm{XY}}}{\partial a_i} = \xi_i = \oint_{B_i} z\, d \log \mathcal{Y}_{\mathcal{B}_{i}},
\end{align} 
where the first equality follows from the fact that $a_i$ and $\xi_i$ are related by the Legendre transformation.
Here, $B_i$ is the non-compact cycle that goes from infinity to a point in $\mathcal{C}_i$.

\subsection{Cameral curve}

The following combinations are invariant under \eqref{eq:Weyl} and \eqref{eq:Weyl2}
\begin{align}\label{eq:def-character}
\mathcal{X}_1^{\mathrm{XY}} (z)
:= &
\mathcal{Y}_{\mathrm{L}}^{\mathrm{X}}{}(z)  
+ \mathcal{Y}_{\mathrm{L}}^{\mathrm{X}}{}(-z)  
+ \mathcal{Y}_{\mathrm{R}}^{\mathrm{Y}}{}(z)
+ \mathcal{Y}_{\mathrm{R}}^{\mathrm{Y}}{}(-z),
\cr
\mathcal{X}_2^{\mathrm{XY}} (z)
:= & \mathcal{Y}_{\mathrm{L}}^{\mathrm{X}}{}(z) \mathcal{Y}_{\mathrm{L}}^{\mathrm{X}}{}(-z)
+ \mathcal{Y}_{\mathrm{L}}^{\mathrm{X}}{}(z) \mathcal{Y}_{\mathrm{R}}^{\mathrm{Y}}{}(z)
+ \mathcal{Y}_{\mathrm{L}}^{\mathrm{X}}{}(z)  \mathcal{Y}_{\mathrm{R}}^{\mathrm{Y}}{}(-z)
\cr
& \quad
+ \mathcal{Y}_{\mathrm{L}}^{\mathrm{X}}{}(-z) \mathcal{Y}_{\mathrm{R}}^{\mathrm{Y}}{}(z)
+ \mathcal{Y}_{\mathrm{L}}^{\mathrm{X}}{}(-z) \mathcal{Y}_{\mathrm{R}}^{\mathrm{Y}}{}(-z)
+ \mathcal{Y}_{\mathrm{R}}^{\mathrm{Y}}{}(z)\mathcal{Y}_{\mathrm{R}}^{\mathrm{Y}}{}(-z)
\cr
\mathcal{X}_3^{\mathrm{XY}} (z)
:= &\mathcal{Y}_{\mathrm{L}}^{\mathrm{X}}{}(z)
\mathcal{Y}_{\mathrm{L}}^{\mathrm{X}}{}(-z) 
\mathcal{Y}_{\mathrm{R}}^{\mathrm{Y}}{}(z)
+ \mathcal{Y}_{\mathrm{L}}^{\mathrm{X}}{}(z)
\mathcal{Y}_{\mathrm{L}}^{\mathrm{X}}{}(-z)  
\mathcal{Y}_{\mathrm{R}}^{\mathrm{Y}}{}(-z)
\cr
&+ \mathcal{Y}_{\mathrm{L}}^{\mathrm{X}}{}(z) 
\mathcal{Y}_{\mathrm{R}}^{\mathrm{Y}}{}(z)
\mathcal{Y}_{\mathrm{R}}^{\mathrm{Y}}{}(-z)
+ \mathcal{Y}_{\mathrm{L}}^{\mathrm{X}}{}(-z)  
\mathcal{Y}_{\mathrm{R}}^{\mathrm{Y}}{}(z)
\mathcal{Y}_{\mathrm{R}}^{\mathrm{Y}}{}(-z),
\cr
\mathcal{X}_4^{\mathrm{XY}} (z)
:= &\mathcal{Y}_{\mathrm{L}}^{\mathrm{X}}{}(z)
\mathcal{Y}_{\mathrm{L}}^{\mathrm{X}}{}(-z) 
\mathcal{Y}_{\mathrm{R}}^{\mathrm{Y}}{}(z)
\mathcal{Y}_{\mathrm{R}}^{\mathrm{Y}}{}(-z) .
\end{align}
This means that these functions are continuous even at $\mathcal{C}_i, \mathcal{C}_{\mathrm{L}}, \mathcal{C}_{\mathrm{R}}$. 
Furthermore, from \eqref{eq:int-fL-fR}, \eqref{eq:def-amp-fn}, and \eqref{eq:def-character}, we find that these functions are periodic in the pure imaginary direction and invariant under the reflection $z \to -z$,
\begin{align}\label{eq:period-reflection}
\mathcal{X}_i^{\mathrm{XY}}(z + 2 \pi i) = \mathcal{X}_i^{\mathrm{XY}}(z),
\quad
\mathcal{X}_i^{\mathrm{XY}}(-z) = \mathcal{X}_i^{\mathrm{XY}}(z)
\qquad (i=1,2,3,4)
\end{align}
This indicates that the only possible singularity of these functions is at $z \to \pm \infty$ and that the singularity at $z \to + \infty$ is identical to that at $z \to - \infty$.

In the following, we consider the asymptotic behavior at $z \to - \infty$.
The asymptotic behaviors of the amplitude functions $\mathcal{Y}_{\mathrm{L}}^{\mathrm{X}}(z)$ and $\mathcal{Y}_{\mathrm{R}}^{\mathrm{Y}}(z)$ at $z \to - \infty$ can be again read off from \eqref{eq:def-amp-fn} that
\begin{align}
&\mathcal{Y}_{\mathrm{L}}^{\mathrm{X}}(z) \sim \exp  \left[ 2 z
- \frac12 m_0 + (2^{N-4}-2) m_{\mathrm{L}} 
\right]
\qquad \text{ as } \,\,z \to -\infty
\cr
&\mathcal{Y}_{\mathrm{R}}^{\mathrm{Y}}(z) \sim \exp  \left[ - 2 z
+ \frac12 m_0 - (2^{N-4}-2) m_{\mathrm{R}}
\right]
\quad \text{ as } \,\,z \to -\infty
\end{align}
while the asymptotic behaviors of $\mathcal{Y}_{\mathrm{L}}^{\mathrm{X}}(-z)$ and $\mathcal{Y}_{\mathrm{R}}^{\mathrm{X}}(-z)$ at $z \to - \infty$ can be again read off from \eqref{eq:def-amp-fn-2} that
\begin{align}
&\mathcal{Y}_{\mathrm{L}}^{\mathrm{X}}(-z) \sim (-1)^N \exp \left[ (N-2) z
- \frac12 m_0 + 2^{N-4} m_{\mathrm{L}}  
\right]
\qquad \text{ as } \,\,z \to -\infty
\cr
&\mathcal{Y}_{\mathrm{R}}^{\mathrm{Y}}(-z) \sim (-1)^N \exp  \left[ - (N-2) z
+ \frac12 m_0 - 2^{N-4} m_{\mathrm{R}}
\right]
\quad \text{ as } \,\,z \to -\infty
\end{align}

Using these, the asymptoric behavor of the function $\mathcal{X}_i^{\mathrm{XY}}$ at $z \to - \infty$ are given, respectively, as
\begin{align}
\mathcal{X}_1^{\mathrm{XY}} (z) \sim C_1 e^{-2z},
\quad
\mathcal{X}_2^{\mathrm{XY}} (z) \sim C_2 e^{-Nz},
\quad
\mathcal{X}_3^{\mathrm{XY}} (z) \sim C_3 e^{-2z},
\quad
\mathcal{X}_4^{\mathrm{XY}} (z) \sim C_4,
\end{align}
where we put
\begin{align}\label{eq:def-CCtilde}
C_1:=&
\left\{ 
\begin{array}{ll}
\exp \left( 
\frac12 m_0 - (2^{N-4}-2) m_{\mathrm{R}} 
\right)
& \text{ for } \,\, N =2,3
\\
\exp \left( \frac12 m_0 + m_{\mathrm{R}}\right) + \exp \left( \frac12 m_0 -  m_{\mathrm{R}} \right) 
& \text{ for } \,\, N =4
\end{array}
\right.
\cr
C_2 :=&
\left\{
\begin{array}{ll}
\exp \left( m_0 + \frac32  m_{\mathrm{R}} \right)
+ \exp \left( \frac14 m_{\mathrm{L}} + \frac74 m_{\mathrm{R}} \right)
& \text{ for } \quad N = 2
\cr
(-1)^N \exp \left( m_0 - (2^{N-3}-2)  m_{\mathrm{R}} \right)
&
\text{ for } \quad N = 3,4
\end{array}
\right.
\cr
C_3:=&
\left\{ 
\begin{array}{ll}
\exp \left( 
\frac12 m_0 + 2^{N-4} m_{\mathrm{L}} - (2^{N-3}-2) m_{\mathrm{R}}
\right)
& \text{ for } \,\, N =2,3
\\
\exp \left( \frac12 m_0 - m_{\mathrm{L}} \right) + \exp \left( \frac12 m_0 +  m_{\mathrm{L}} \right) 
& \text{ for } \,\, N =4
\end{array}
\right.
\cr
C_4 :=& \exp \left( 
(2^{N-3}-2) (m_{\mathrm{L}} - m_{\mathrm{R}} )
\right).
\end{align}

Taking all these things into account, we find that these are given as Laurent polynomials in the variable
\begin{align}
w := e^{-z}
\end{align}
of the form
\begin{align}\label{eq:cameral}
&\mathcal{X}_1^{\mathrm{XY}} (z) = \sum_{k=0}^{2} c_{1,k} (w^k+w^{-k})
\cr
&\mathcal{X}_2^{\mathrm{XY}} (z) = \sum_{k=0}^{N} c_{2,k} (w^k+w^{-k})
\cr
&\mathcal{X}_3^{\mathrm{XY}} (z) = \sum_{k=0}^{2} c_{3,k} (w^k+w^{-k})
\cr
&\mathcal{X}_4^{\mathrm{XY}} (z) = C_4.
\end{align}
with
\begin{align}\label{eq:c12c2Nc32}
c_{1,2} = C_1, \qquad 
c_{2,N} = C_2, \qquad
c_{3,2} = C_3, 
\end{align}
where $C_1$, $C_2$, $C_3$, and $C_4$ are given in \eqref{eq:def-CCtilde}. 
The constraints on the other coefficients $c_{i,j}$ in \eqref{eq:cameral} will be determined in a later subsection. The set of equations \eqref{eq:cameral} are interpreted as a Cameral curve as discussed in \cite{Nekrasov:2012xe}.

\subsection{Seiberg-Witten curve and boundary conditions}
We define the Riemann surface defined by 
\begin{align}\label{eq:SW-curve}
F_{\mathrm{SW}} (t,w) =0
\end{align}
with
\begin{align}\label{eq:SW-poly}
&F_{\mathrm{SW}} (t,w) := 
\left( t-\mathcal{Y}_{\mathrm{L}}^{\mathrm{X}}(z) \right) 
\left( t-\mathcal{Y}_{\mathrm{L}}^{\mathrm{X}}(-z) \right)
\left( t-\mathcal{Y}_{\mathrm{R}}^{\mathrm{Y}}(z) \right)
\left( t-\mathcal{Y}_{\mathrm{R}}^{\mathrm{Y}}(-z) \right).
\end{align}
and the 1-form defined on this Riemann surface
\begin{align}
\lambda_{\mathrm{SW}} = \log w \,d \log t.
\end{align}
Due to the cycle integral \eqref{eq:A-cycle} and \eqref{eq:B-cycle}, we find that these are the Seiberg-Witten curve and the Seiberg-Witten 1-form, respectively.

Expanding \eqref{eq:SW-poly} in terms of $t$, the coefficients are the combination introduced in \eqref{eq:def-character} as
\begin{align}\label{eq:FSWtw}
&F_{\mathrm{SW}} (t,w) = 
t^4 
- \mathcal{X}_1^{\mathrm{XY}} (z) t^3 
+ \mathcal{X}_2^{\mathrm{XY}} (z) t^2
- \mathcal{X}_3^{\mathrm{XY}} (z) t 
+ \mathcal{X}_4^{\mathrm{XY}} (z).
\end{align}
Thus, we can derive Seiberg-Witten curve from the Cameral curve \eqref{eq:cameral}, \eqref{eq:c12c2Nc32}.

In the following, we discuss the boundary conditions satisfied by the Seiberg-Witten curve to give constraints on the coefficients $c_{i,j}$ in \eqref{eq:cameral}. We observe that \eqref{eq:SW-poly} has a pair of double roots $\mathcal{Y}_{\mathrm{L}}^{\mathrm{X}}(0)$ and $\mathcal{Y}_{\mathrm{R}}^{\mathrm{Y}}(0)$ at $w=1$ as 
\begin{align}\label{eq:w1double}
F_{\mathrm{SW}} (t,w=1) 
= &
\left( t-\mathcal{Y}_{\mathrm{L}}^{\mathrm{X}}(0) \right) ^2
\left( t-\mathcal{Y}_{\mathrm{R}}^{\mathrm{Y}}(0) \right) ^2
\end{align}
Since 
\begin{align}
 \mathcal{Y}_{\mathrm{L}}^{\mathrm{X}}(0)^2 \mathcal{Y}_{\mathrm{R}}^{\mathrm{Y}}(0)^2 
= \mathcal{X}_4^{\mathrm{XY}} (0)
= C_4 
= \exp \left( 
(2^{N-3}-2) (m_{\mathrm{L}} - m_{\mathrm{R}} )
\right)
\end{align}
from \eqref{eq:cameral}, these two pairs of double roots are related as
\begin{align}\label{eq:sign1}
\mathcal{Y}_{\mathrm{L}}^{\mathrm{X}}(0) \mathcal{Y}_{\mathrm{R}}^{\mathrm{Y}}(0) 
= \pm M_{\mathrm{L}}^{-(2^{N-4}-1)} M_{\mathrm{R}}^{(2^{N-4}-1)},
\end{align}
where we introduce the notation
\begin{align}
M_{\mathrm{L}}:= \exp(-m_\mathrm{L}),
\quad
M_{\mathrm{R}}:= \exp(-m_\mathrm{R}).
\end{align}

Analogously, taking into account the property \eqref{eq:period-reflection}, \eqref{eq:SW-poly} has a pair of double roots $\mathcal{Y}_{\mathrm{L}}^{\mathrm{X}}(\pi)$ and $\mathcal{Y}_{\mathrm{R}}^{\mathrm{Y}}(\pi)$ also at $w=-1$ as
\begin{align}\label{eq:wm1double}
F_{\mathrm{SW}} (t,w=-1) 
= &
\left( t-\mathcal{Y}_{\mathrm{L}}^{\mathrm{X}}(\pi i) \right) ^2
\left( t-\mathcal{Y}_{\mathrm{R}}^{\mathrm{Y}}(\pi i) \right) ^2
\end{align}
with the two double roots related as 
\begin{align}\label{eq:sign2}
\mathcal{Y}_{\mathrm{L}}^{\mathrm{X}}(\pi i) \mathcal{Y}_{\mathrm{R}}^{\mathrm{Y}}(\pi i) 
= \pm M_{\mathrm{L}}^{-(2^{N-4}-1)} M_{\mathrm{R}}^{(2^{N-4}-1)},
\end{align}

What we find in this paper is that the relative sign between \eqref{eq:sign1} and \eqref{eq:sign2} is determined by the choice of spinor or conjugate spinor.
For the case of two spinors, we find 
\begin{align}
&\frac{\mathcal{Y}_{\mathrm{L}}^{\mathrm{S}}(\pi i) \mathcal{Y}_{\mathrm{R}}^{\mathrm{S}}(\pi i) }{\mathcal{Y}_{\mathrm{L}}^{\mathrm{S}}(0) \mathcal{Y}_{\mathrm{R}}^{\mathrm{S}}(0) }
\cr
& = 
\exp \left[
\frac12 \int^{\infty}_{-\infty} 
(\tilde{f}_{\mathrm{L}}^{\mathrm{S}}{}''(y)+ \tilde{f}_{\mathrm{R}}^{\mathrm{S}}{}''(y))
\log \left( \frac{1+e^{-\beta y}}{1-e^{-\beta y}} \right) dy
\right]
\cr
& = 
\exp \left[
\frac12 \int^{\infty}_{-\infty} 
(g_{\mathrm{L}}''(y) - g_{\mathrm{L}}''(-y) - g_{\mathrm{R}}''(y) + g_{\mathrm{R}}''(-y) )
\log \left( \frac{1+e^{-\beta y}}{1-e^{-\beta y}} \right) dy
\right]
\cr
& = 
\exp \left[
\frac12 \int^{\infty}_{-\infty} 
(g_{\mathrm{L}}''(y) - g_{\mathrm{R}}''(y) )
\left( \log \left( \frac{1+e^{-\beta y}}{1-e^{-\beta y}} \right) - \log \left( \frac{1+e^{\beta y}}{1-e^{\beta y}} \right)\right)dy
\right]
\cr
& = 1
\end{align}
where we used the constraints \eqref{eq:constraint-fg}. 
This leads to the constraints 
\begin{align}\label{eq:const1-bdySS}
F_{\mathrm{SW}} (t,w=1) 
= 
\left( t^2 + U_1 t + M_{\mathrm{L}}^{-(2^{N-4}-1)}  M_{\mathrm{R}}^{(2^{N-4}-1)}  \right) ^2,
\cr
F_{\mathrm{SW}} (t,w=-1) 
= 
\left( t^2 + U_2 t + M_{\mathrm{L}}^{-(2^{N-4}-1)}  M_{\mathrm{R}}^{(2^{N-4}-1)} \right) ^2,
\end{align}
or 
\begin{align}
F_{\mathrm{SW}} (t,w=1) 
= 
\left( t^2 + U_1 t - M_{\mathrm{L}}^{-(2^{N-4}-1)}  M_{\mathrm{R}}^{(2^{N-4}-1)} \right) ^2,
\cr
F_{\mathrm{SW}} (t,w=-1) 
= 
\left( t^2 + U_2 t - M_{\mathrm{L}}^{-(2^{N-4}-1)}  M_{\mathrm{R}}^{(2^{N-4}-1)} \right) ^2.
\end{align}
where we introduce
\begin{align}
U_1 &:= - \left( \mathcal{Y}_{\mathrm{L}}^{\mathrm{X}}(0) + \mathcal{Y}_{\mathrm{R}}^{\mathrm{Y}}(0) \right),
\cr
U_2 &:= - \left( \mathcal{Y}_{\mathrm{L}}^{\mathrm{X}}(\pi i) + \mathcal{Y}_{\mathrm{R}}^{\mathrm{Y}}(\pi i) \right).
\end{align}
Since they are related by the redefinition 
$M_{\mathrm{L}} \to e^{2^{-N+5} \pi i } M_{\mathrm{L}}$, for example,
they will lead to the equivalent Seiberg-Witten curves and thus, it would be enough to consider only the former case.

On the contrary, for the case with one spinor and one conjugate spinor, we find
\begin{align}
\frac{\mathcal{Y}_{\mathrm{L}}^{\mathrm{S}}(\pi i) \mathcal{Y}_{\mathrm{R}}^{\mathrm{C}}(\pi i) }{\mathcal{Y}_{\mathrm{L}}^{\mathrm{S}}( 0 ) \mathcal{Y}_{\mathrm{R}}^{\mathrm{C}}(0) }
& = 
\exp \left[
\frac12 \int^{\infty}_{-\infty} 
(\tilde{f}_{\mathrm{L}}^{\mathrm{S}}{}''(y)+ \tilde{f}_{\mathrm{R}}^{\mathrm{C}}{}''(y))
\log \left( \frac{1+e^{-\beta y}}{1-e^{-\beta y}} \right) dy
\right]
\cr
& = 
\exp \biggl[
\frac12 \int^{\infty}_{-\infty} \bigl( g_{\mathrm{L}}''(y) - g_{\mathrm{L}}''(-y) - g_{\mathrm{R}}''(y) + g_{\mathrm{R}}''(-y) 
\cr
& \qquad\qquad
- f_N(y) + f_N(-y) \bigr)
\log \left( \frac{1+e^{-\beta y}}{1-e^{-\beta y}} \right) dy
\biggr]
\cr
&= -1.
\end{align}
This leads to the constraints 
\begin{align}\label{eq:const1-bdySC}
F_{\mathrm{SW}} (t,w=1) 
= 
\left( t^2 + U_1 t + M_{\mathrm{L}}^{-(2^{N-4}-1)}  M_{\mathrm{R}}^{(2^{N-4}-1)} \right) ^2,
\cr
F_{\mathrm{SW}} (t,w=-1) 
=  
\left( t^2 + U_2 t - M_{\mathrm{L}}^{-(2^{N-4}-1)}  M_{\mathrm{R}}^{(2^{N-4}-1)} \right) ^2,
\end{align}
or 
\begin{align}
F_{\mathrm{SW}} (t,w=1) 
= 
\left( t^2 + U_1 t - M_{\mathrm{L}}^{-(2^{N-4}-1)}  M_{\mathrm{R}}^{(2^{N-4}-1)} \right) ^2,
\cr
F_{\mathrm{SW}} (t,w=-1) 
= 
\left( t^2 + U_2 t + M_{\mathrm{L}}^{-(2^{N-4}-1)}  M_{\mathrm{R}}^{(2^{N-4}-1)} \right) ^2.
\end{align}
Again, we consider only the former case. 
 
For the case of two spinors, by imposing the constraints \eqref{eq:const1-bdySS} to the Seiberg-Witten curve \eqref{eq:FSWtw}, we find that it is rewritten in the form
\begin{align}\label{eq:SW2S}
F_{\mathrm{SW}}^{\mathrm{SS}} (t,w)
= &(t^2 + U_1 t + M_{\mathrm{L}}^{-(2^{N-4}-1)}  M_{\mathrm{R}}^{(2^{N-4}-1)})^2 \frac{(w+1)^2}{4w}
\cr
&- (t^2 + U_2 t + M_{\mathrm{L}}^{-(2^{N-4}-1)}  M_{\mathrm{R}}^{(2^{N-4}-1)})^2 \frac{(w-1)^2}{4w}
\cr
&+ (w-w^{-1})^2 \left( p_2(w) t^2 + p_1 (w) t + p_0(w) \right) t.
\end{align}
The first two terms are introduced in such a way as to satisfy the constraints \eqref{eq:const1-bdySS} and to fix the coefficients of $t^4$ and $t^0$ to be consistent with \eqref{eq:FSWtw}. The last term vanishes at both $w=1$ and $w=-1$ and does not alter this structure.
For the case of one spinor and one conjugate spinor, we find from \eqref{eq:const1-bdySC} that the Seiberg-Witten curve is of the form
\begin{align}\label{eq:SW1S1C}
F_{\mathrm{SW}}^{\mathrm{SC}} (t,w)
= &(t^2 + U_1 t + M_{\mathrm{L}}^{-(2^{N-4}-1)}  M_{\mathrm{R}}^{(2^{N-4}-1)})^2 \frac{(w+1)^2}{4w}
\cr
& - (t^2 + U_2 t - M_{\mathrm{L}}^{-(2^{N-4}-1)}  M_{\mathrm{R}}^{(2^{N-4}-1)})^2 \frac{(w-1)^2}{4w}
\cr
&+ (w-w^{-1})^2 \left( p_2(w) t^2 + p_1 (w) + p_0(w) \right) t.
\end{align}

Comparing the expressions \eqref{eq:SW2S} and \eqref{eq:SW1S1C} with the one for the Seiberg-Witten curve \eqref{eq:FSWtw} and the Cameral curve \eqref{eq:cameral}, we find that $p_i(w)$ are respectively given in the following form:
\begin{align}
&p_0(w) = C_3, \qquad p_2(w) = C_1,
\cr
\quad 
&p_1(w) = \sum_{k=-(N-2)}^{N-2} d_{2,k} w^k
\quad \text{with} \quad d_{2,k}= d_{2,-k}, \,\, d_{2,N-2} = d_{2,-N+2} = C_2
\end{align}
where $C_1$, $C_2$, and $C_3$ are given in \eqref{eq:def-CCtilde}. 

Finally, we now write down the Seiberg-Witten curves for each theory more explicitly. We rescale the variable $t$ as $t \to M_{\mathrm{R}}^{(2^{N-4}-1)} t $ and denote $U_1' = M_{\mathrm{R}}^{(2^{N-4}-1)}U_1$, $U_2' = M_{\mathrm{R}}^{(2^{N-4}-1)}U_2$, and $q=e^{-m_0}$. The parameters $U_1'$, $U_2'$, $V$, $V'$ are interpreted as Coulomb moduli. The plus signs of $\pm$ for the second term in the following expressions correspond to the theories with two spinors (2S) and the minus signs correspond to the theories with one spinor and one conjugate spinor (1S+1C).
\begin{itemize}
\item SO(4)+2S / SO(4)+1S+1C
\begin{align}
&( M_{\mathrm{R}}^{\frac34} t^2 + U_1' t + M_{\mathrm{L}}^{\frac34}  )^2 \frac{(w+1)^2}{4w}
- (M_{\mathrm{R}}^{\frac34} t^2 + U_2' t \pm M_{\mathrm{L}}^{\frac34} )^2 \frac{(w-1)^2}{4w}
\cr
&+ q^{-\frac12} (w-w^{-1})^2 \left( M_{\mathrm{R}}^{-\frac14} t^2 + (q^{-
\frac12} +q^{
\frac12} M_{\mathrm{L}}^{-\frac14} M_{\mathrm{R}}^{-\frac14}) t +  M_{\mathrm{L}}^{-\frac14} \right) t = 0.
\end{align}
\item SO(6)+2S / SO(6)+1S+1C
\begin{align}
& ( M_{\mathrm{R}}^{\frac12} t^2 + U_1' t + M_{\mathrm{L}}^{\frac12}  )^2 \frac{(w+1)^2}{4w}
- (M_{\mathrm{R}}^{\frac12} t^2 + U_2' t \pm M_{\mathrm{L}}^{\frac12} )^2 \frac{(w-1)^2}{4w}
\cr
&+ q^{-\frac12} (w-w^{-1})^2 \left( M_{\mathrm{R}}^{-\frac12} t^2 - (q^{-
\frac12} (w+w^{-1}) + V)t + M_{\mathrm{L}}^{-\frac12} \right) t = 0.
\end{align}
\item SO(8)+2S / SO(8)+1S+1C
\begin{align}
& ( t^2 + U_1' t + 1)^2 \frac{(w+1)^2}{4w}
- (t^2 + U_2' t \pm 1)^2 \frac{(w-1)^2}{4w}
\cr
&+ q^{-\frac12} (w-w^{-1})^2 \Bigl( (M_{\mathrm{R}}^{-1}+M_{\mathrm{R}}) t^2 
\cr 
& \quad
+ (q^{-
\frac12} (w^{2}+w^{-2}) + V(w+w^{-1}) + V')t 
+ (M_{\mathrm{L}}^{-1}+M_{\mathrm{L}}) \Bigr) t = 0.
\end{align}
\end{itemize}

\section{Conclusion and Discussion}\label{sec:concl}

In this paper, we have systematically investigated five-dimensional \(\mathcal{N}=1\) supersymmetric \(SO(2N)\) gauge theories coupled to hypermultiplets in the spinor and conjugate spinor representations, based on 5-brane web constructions with O5-planes. Our analysis yields new results of the partition functions and the Seiberg-Witten geometry, each of which encodes the distinction between the spinor and conjugate spinor matters.

As for the partition function, we have derived new expressions for the unrefined topological string partition functions corresponding to \(SO(2N)\) gauge theories with two spinors(2S), two conjugate spinors(2C), and one spinor plus one conjugate spinor(1S+1C), respectively. These expressions are obtained by applying topological vertex formalism to the 5-brane web diagrams with an O5-plane.
Differences among the partition functions of these theories are understood as differences in the structure of the gluing of the left and right spinor and/or conjugate spinor strips.
The resulting partition functions
take a form that is particularly suitable for the thermodynamic limit analysis. We have verified the consistency of our new expressions by recovering the expected perturbative parts in the weak coupling limit \(m_0 \to \infty\).
This check provides strong evidence for the correctness of our topological vertex computation.

As for the Seiberg-Witten geometry, we have clarified how the distinction between different spinor matter contents, two spinors versus one spinor plus one conjugate spinor, manifests itself through the thermodynamic limit  \(\hbar \to 0\). By applying the saddle point approximation, we derived the saddle point equations for the profile functions, from which we constructed the amplitude functions and the associated Cameral curve. 
The key finding is that the two theories are characterized by distinct boundary conditions of the Seiberg-Witten curve at the O5-plane positions \(w = \pm 1\). 
This boundary condition difference precisely encodes the distinction between the two theories, confirming that the 5-brane web diagrams indeed represent inequivalent gauge theories even though they look identical at some phases.

Our results have several important implications. First, they demonstrate that the topological vertex formalism with O5-planes, originally developed for Sp($N$) gauge theories with fundamental matter \cite{Kim:2017jqn}, can be successfully extended to accommodate spinor representations of SO($2N$) gauge groups. 
Second, the new expressions for the partition functions provide an alternative to the ADHM-based approach in \cite{Chen:2023smd}, offering a complementary perspective that may be more suitable for certain applications, particularly in the context of the thermodynamic limit and Seiberg-Witten geometry. 
Third, our results imply a useful technique to obtain
Seiberg-Witten curves directly from a 5-brane web diagram by imposing proper boundary conditions at the O5-plane.

Several interesting directions remain for future investigation. A natural extension is to generalize our analysis to the refined topological string partition function, which would allow access to the full five-dimensional superconformal index and provide richer information about the BPS spectrum. Another direction is to explore the \(SO(2N+1)\) cases, which involve slightly different orientifold actions and may exhibit analogous but distinct boundary condition structures. It would also be valuable to investigate whether the Seiberg-Witten geometries for 4d $\mathcal{N}=2$ SO($N$) gauge theories with spinors, which are studied in \cite{Terashima:1998fx, Chacaltana:2011ze, Tachikawa:2011yr, Chacaltana:2013oka, Chacaltana:2014ica}, can be reproduced from our results or a generalization of our results. Although the relation to their results does not appear obvious from the expression of our curve, we expect them to be reproduced by taking the 4d limit of our curves, which shrinks the compactified circle of the 5d spacetime, probably combined with appropriate coordinate transformations.
Finally, a deeper investigation of the relation between our partition function expressions and those obtained via the ADHM approach in \cite{Chen:2023smd} would be illuminating; while we have argued that they are expansions in different parameters, an explicit equivalence at the level of the full instanton partition function remains to be established.

We believe that our work provides a useful framework for studying non-perturbative aspects of supersymmetric gauge theories with orthogonal and symplectic gauge groups, based on a 5-brane web construction with an orientifold plane, particularly for extracting partition function and Seiberg-Witten geometry. We hope that the techniques and results presented here will prove useful for future explorations of five-dimensional supersymmetric gauge theories.

\acknowledgments
We thank Mohammad Akhond, Federico Carta, Sidharth Dwivedi, Hirotaka Hayashi, Sung-Soo Kim, Kimyong Lee, and Rui-Dong Zhu for useful discussions.
XL is supported by NSFC grant No.11501470, No.11426187, No.11791240561, and partially supported by NSFC grant No.11671328, Chengdu Science and Technology Program with Grant No.2025-YF09-00007-SN and the Fundamental Research Funds for the Central Universities 2682021ZTPY043, 2682025ZTPY001, 2682025ZTPY057, 2682025ZTO002. Especially, XL would like to thank Bohui Chen, An-min Li, Guosong Zhao for their constant support and also thank all the friends met in different conferences. XL would like to express special thanks to the Mainz Institute for Theoretical Physics (MITP) of the Cluster of Excellence PRISMA$^{+}$ (Project ID $390831469$), for its hospitality and support. FY is supported by start-up research grant RK21121 from Huzhou Normal University. 

\bibliographystyle{JHEP}
\bibliography{ref}
\end{document}